\newif\ifWithComments
\WithCommentsfalse

\newif\ifAlwaysWithReviewfix  
\AlwaysWithReviewfixfalse

\newif\ifisestimation
\isestimationtrue

\newif\ifisanon
\isanonfalse

\ifisanon
   \documentclass[letterpaper,twocolumn,10pt,hyphens]{article}
\else
  \documentclass[letterpaper,twocolumn,10pt,hyphens]{article}
\fi
\usepackage{usenix2019_v3}

\newcommand{\RelaxFloats}{
	\renewcommand{\topfraction}{0.9}
	\renewcommand{\floatpagefraction}{0.9}
	\renewcommand{\textfraction}{0.1}
}
\RelaxFloats

\AtBeginDocument{%
  \providecommand\BibTeX{{%
    \normalfont B\kern-0.5em{\scshape i\kern-0.25em b}\kern-0.8em\TeX}}}

\usepackage{tikz}
\usepackage{amsmath}

\usepackage{multirow}
\usepackage{url}
\usepackage{multicol}
\usepackage{subcaption}
\usepackage{graphicx} 
\usepackage{hyperref}

\usepackage{xspace}
\usepackage{dblfloatfix}

\usepackage{colortbl}
\usepackage{xcolor}
\usepackage{booktabs}
\usepackage{amssymb}

\usetikzlibrary{arrows,automata}
\usepackage{comment}
\usepackage{amssymb}
\usepackage{pifont}
\usepackage[title]{appendix}
\usepackage{tabularx}

\hypersetup{colorlinks,linkcolor=[rgb]{.75,0,0},urlcolor=[rgb]{0.75,0,0.75},citecolor=blue,breaklinks=true} 

\newcommand\Vsmall[1]{{\mbox{\small \textit{#1}}}}

\makeatletter
  \newcommand\EatSpacesHack{\@bsphack\@esphack}
\makeatother

\ifWithComments
  \renewcommand\comment[1]{{\color{blue} \sffamily [xxx:  #1]}}
  \newcommand\cmt[1]{{\color{blue} \sffamily [xxx:  #1]}}
  \newcommand\ceron[1]{{\color{cyan} \sffamily [xxx:  #1 - João Ceron  ]}}
  \newcommand\leandro[1]{{\color{orange} \sffamily [#1 -- Leandro ]}}
  \newcommand\PostSubmission[1]{\EatSpacesHack}
  \newcommand\reviewfix[1]{{\color{blue}\sffamily [RF:#1]\EatSpacesHack}}
\else
  \renewcommand\comment[1]{\EatSpacesHack}
  \newcommand\cmt[1]{\EatSpacesHack}
  \newcommand\commentceron[1]{\EatSpacesHack}
  \newcommand\ceron[1]{\EatSpacesHack}
  \newcommand\leandro[1]{\EatSpacesHack}
  \newcommand\PostSubmission[1]{\EatSpacesHack}
  \ifAlwaysWithReviewfix
    \newcommand\reviewfix[1]{{\color{blue}\sffamily [RF:#1]\EatSpacesHack}}
  \else
    \newcommand\reviewfix[1]{\EatSpacesHack}
  \fi
\fi
\makeatother

\newcommand{\tier}{{\mbox{Tier-1}}\xspace}

\def\Snospace~{\S{}} %
\date{}

\newcommand{\ok}{\checkmark}
\ifisanon
  \newcommand{\peering}{\texttt{Testbed-AnonX}\xspace}
  \newcommand{\tangled}{\texttt{Testbed-AnonY}\xspace}
  \newcommand{\DutchScrubbingCenter}{a National Scrubbing Center\xspace}
  \newcommand{\broot}{\texttt{X-Root}\xspace}
  
  \newcommand\anonymize[1]{\textit{(anonymized for review)}}
\else
  \newcommand{\peering}{\texttt{Peering}\xspace}
  \newcommand{\tangled}{\texttt{Tangled}\xspace}
  \newcommand{\DutchScrubbingCenter}{the Dutch National Scrubbing Center\xspace}
  \newcommand{\broot}{\texttt{B-root}\xspace}

\fi

\newcommand*\circled[1]{\tikz[baseline=(char.base)]{
            \node[shape=circle, draw, inner sep=0.2ex, baseline=-0.5ex] (char) {#1};}}

\newcommand{\tmpframe}[1]{\fbox{#1}}

\makeatletter
  \ifx\ifisarxiv\undefined
      \csname newif\expandafter\endcsname \csname ifisarxiv\endcsname
  \fi
  \isarxivtrue
  \renewcommand\section{\@startsection {section}{1}{\z@}%
                                   {-1.5ex \@plus -1ex \@minus -.2ex}%
                                   {1ex \@plus.2ex}%
                                   {\normalfont\large\bfseries}}
  \renewcommand\subsection{\@startsection{subsection}{2}{\z@}%
                                     {-0.75ex\@plus -1ex \@minus -.2ex}%
                                     {0.5ex \@plus .2ex}%
                                     {\normalfont\large\bfseries}}

  \renewcommand\subsubsection{\@startsection{subsubsection}{3}{\z@}%
                                     {-0.5ex\@plus -1ex \@minus -.2ex}%
                                     {0.5ex \@plus .2ex}%
                                     {\normalfont\normalsize\bfseries}}
  \def\@maketitle{\newpage
   \vbox to 1.7in{
   \vspace*{\fill}
   \vskip 1em
   \begin{center}%
    {\Large\bf \@title \par}%
    \vskip 0.175in minus 0.300in
    {\large\it
     \lineskip .5em
     \begin{tabular}[t]{c}\@author
     \end{tabular}\par}%
   \end{center}%
   \par
   \vspace*{\fill}
   }
 }

\ifisarxiv
\else
\fi

\makeatother

\begin{document}

\date{}

%
%
\title{\Large \bf Anycast Agility: Network Playbooks to Fight DDoS \ifisarxiv \thanks{A shorter version of this paper will appear in the proceedings of the 31st
USENIX Security’22 Symposium. This is the full version.}\fi}

\ifisanon
  \author{Authors Anonymized for Review}
\else
\author{
{\rm A S M Rizvi\thanks{Shared first author}}\\
USC/ISI
\and
{\rm Leandro Bertholdo\ifisarxiv\footnotemark[2]\else\footnotemark[1]\fi}\\
University of Twente
\and
{\rm Jo\~ao Ceron}\\
SIDN Labs
\and
{\rm John Heidemann}\\
USC/ISI
} 
\fi

\maketitle

\begin{abstract}
IP anycast is used for services such as DNS and Content Delivery
Networks (CDN) to provide the capacity to handle Distributed
Denial-of-Service (DDoS) attacks. 
During a DDoS attack service operators 
  redistribute traffic between anycast sites
  to take advantage of sites with unused or greater capacity.
Depending on site traffic and attack size,
  operators may instead concentrate
  attackers in a few sites to preserve operation in others.
Operators use these actions during attacks,
  but how to do so has not been described systematically or publicly.
This paper 
  describes several methods to use BGP to shift traffic when under DDoS,
  and shows that a \emph{response playbook} can provide a menu of responses
  that are options during an attack.
To choose an appropriate response from this playbook,
  we also describe a new method to estimate true attack size,
  even though the operator's view during the attack is incomplete.
Finally, operator choices are constrained by distributed routing policies,
  and not all are helpful.
\reviewfix{SS22-D13}
We explore how specific anycast deployment can constrain options in this playbook,
  and are the first to measure how generally applicable they are across multiple
  anycast networks.
\end{abstract}

\section{Introduction}



Anycast routing is used by services like DNS or CDN where multiple
sites announce the same prefix from geographically distributed
locations.  Defined in 1993~\cite{rfc1546} anycast was
widely deployed
by DNS roots in the
early-2000s~\cite{Shaikh2001dnsselection,rfc3258,ballani2006measurement},
and today it is used by many DNS providers and Content Delivery
Networks~\cite{weiden2010anycast, fan2013evaluating,
flavel2015fastroute,cicalese2015characterizing,
cicalese2018longitudinal}.

In IP anycast, BGP routes each network to a particular anycast site,
  dividing the world into \emph{catchments}.  
BGP usually associates networks with nearby anycast sites,
  providing generally good latency~\cite{Schmidt17a}. 
Anycast also helps during
  Distributed-Denial-of-Services (DDoS) attacks,
  with each site adds to the aggregate capacity
  at lower cost than a single very large site.
Each site is independent,
  so should DDoS overwhelm one site,
  sites that are not overloaded are unaffected.

DDoS attacks are getting larger and more common.  Different root servers
and anycast services frequently report DDoS events~\cite{2015-event,
2016-event, 2018-memcrashed, cloudflare-event}.
Different automated
tools make it easier to generate attacks~\cite{wilson2012tools},
and some offer DDoS-as-a-Service, allowing attacks from unsophisticated users
for as little as US\$10~\cite{smith2017ddosaas}.
\comment{added VoIP.ms example. ---asmrizvi 2021-10-10}
\reviewfix{SS22-G4}
DDoS intensity is still growing,
with the 2020 CLDAP attack exceeding 2.3\,Tb/s in size \cite{aws2020cldap},
  and the 2021 VoIP.ms attack lasting for over 5 days~\cite{voipms2021, voipms12021}.
The reservoir of attack sources grow with millions of Internet-of-Things
devices whose vulnerabilities fuel
botnets~\cite{krebs2016krebsonsecurity}.

Operators depend on anycast during DDoS attacks
  to provide capacity to handle the attack and to isolate attackers
  in catchments.
Service operators would like to adapt to an ongoing attack,
  perhaps shifting load from overloaded sites to other sites with excess capacity.
Prior studies of DDoS events have shown that operators take these
actions but suggested that the best action to take depends on attack
size and location compared to anycast site capacity~\cite{Moura16b}.
While prior work suggested countermeasures,
and we know that operators alter routing during attacks,
to date there has been only limited evaluation of
how routing choices change traffic~\cite{quoitin2005performance,
gao2005interdomain, ballani2006measurement, kuipers17}.
Only very recent work examined path poisoning to avoid congested paths~\cite{smith2018routing};
  there is no
  specific public guidance on how to use routing during an attack.

The goal of this paper is to guide defenders
  in traffic engineering (TE) to balance traffic across anycast
  during DDoS.

Our first contribution is a system
  with novel mechanism to estimate true attack rate and plan responses.
First, we propose a new mechanism 
  to \emph{estimate the true offered load},
  even when loss happens upstream of the defender.
Estimating the relative load on each site 
  (\autoref{sec:estimating})
  is the first step of defense,
  so that the defender can match load to the capacities of different sites,
  or decide that some sites should absorb as much of the attack as possible.
Second, we develop a \emph{BGP playbook}:
  a guide that allows operators to anticipate how
  TE actions
  rebalance load across a multi-site anycast system.
Together, these two elements provide a system
  that can automate response to DDoS attacks
  by adjusting anycast routing according to the playbook,
  or recommend actions to a human operator.

The second contribution is to understand how well 
  routing options for multi-hop TE work:
  AS prepending, community strings and path poisoning.
\reviewfix{SS22-A12: fixed typo in many places.}
\reviewfix{SS22-D9}
While well known, it is not widely understood how 
  \emph{available} and \emph{effective} these mechanisms are.
In \autoref{sec:exploit_catchment}    
  we show that while AS prepending 
  is available almost anywhere,
  community strings and path poisoning support varies widely.
We also show that their effectiveness varies greatly,
  in part because today's ``flatter Internet''~\cite{Chiu15a} 
  means AS prepending often shifts either nearly all or nearly no traffic.
Community strings provide finer granularity control,
  but we show their support is uneven.
Path poisoning may provide control multiple hops away, but
  like community strings it is often filtered,
  particularly for \tier ASes.
When these factors combine with
  the interplay between multiple sites and an anycast system,
  a BGP playbook is important to guide defenders.
Since the effects of TE are often specific to 
  the peers and locations of a particular
  anycast deployment, we
  explore how sensitive our results are to different
  locations and numbers of anycast sites (\autoref{sec:constraints_anycast}).

Our final contribution is to 
  demonstrate successful defenses in practice.
We replay real-world attacks in a testbed
  and show TE can defend
  (\autoref{sec:fight_ddos}).
\reviewfix{SS22-B1}
Of course no single defense can protect against all attacks,
  these examples show our approach provides a successful defense to
  many volumetric and polymorphic DDoS attacks.
\reviewfix{SS22-D1, SS22-D5, SS22-D10}
They show that our algorithm and process contributions
  (attack size estimation and playbook construction)
  have practical application.

Our work uses publicly available datasets.
Datasets for the input and results from our experiments
  are available at no charge.
Because our data concerns services but
  not individuals, we see no privacy concerns.


%

\section{Related Work}
	\label{sec:related}


Anycast routing has been studied for a long time from the perspective
of routing, performance, and DDoS-prevention.

\textbf{BGP to steer traffic:}
Prior work showed BGP is effective 
  to steer traffic to balance load on links~\cite{quoitin2003interdomain,Caesar05a,gao2005interdomain}.
However, Ballani et al.~showed that anycast requires planning
  and care for effective load balancing~\cite{ballani2006measurement}.
Others proposed to manipulate BGP based on packet loss, latency and
  jitter~\cite{quoitin2005performance, nanda2009scalable}.
We build on Ballani's recommendation to plan anycast, proposing a BGP playbook,
  and studying how well it can work.

Chang \emph{\mbox{et al.}}~\cite{chang2005inbound} suggested
  using BGP Communities for traffic engineering~\cite{rfc1997,comm_list, caida-communities}.
Recent work has examined BGP communities for blackhole routing
  in IXPs and ISPs~\cite{dietzel2016blackholing,giotsas2017inferring}.
Smith and Glenn examined path poisoning
  to address link congestion~\cite{smith2018routing}.
\reviewfix{SS22-C1, SS22-C7, SS22-D2, SS22-D4}
While each of these are important options in routing for defense,
  we show a system that guides the operator to select between them.
A system with multiple choices is necessary because no single method works
  against all attacks.
For example,  
  we show path poisoning does not work when we poison a Tier-1 AS.

\textbf{Anycast performance:}
Most anycast research focused on efficient delivery
  and stability~\cite{sarat2006use, liu2007two, calder2015analyzing,Wei17b,li2018internet}.
Later studies explicitly investigate the proximity of the 
  clients~\cite{ballani2006measurement, calder2015analyzing,li2018internet}.

Some studies try to improve anycast 
  through topology changes~\cite{Schmidt17a,mcquistin2019taming}.
Anycast services for DDoS is already used in commercial
  solutions \emph{\mbox{e.g.,}} Amazon \cite{scholl2015methods}, Akamai\cite{swildens2011global} 
  and AT\&T~\cite{spatscheck2013multi}. 
However, none of them address how to use routing manipulations
  as a DDoS defense mechanism.

\textbf{Anycast catchment control as a DDoS mitigation tool:}
To our knowledge,
  the idea of handling DDoS attacks by
  absorbing or shifting load across anycast sites
  was first published in 2016~\cite{Moura16b}.
Kuipers \emph{\mbox{et al.}}~\cite{kuipers17}
  refined that work, defining the traffic shifting approaches that
  we review in \autoref{sec:defense_strategy} and explore through experiment.
\reviewfix{SS22-D13}
We develop the idea of a BGP playbook to guide responses,
  and describe a new approach to estimate attack size,
  and finally show that responses can be effective with real-world events.
%
%


\textbf{Commercial and automated solutions:}
Most published commercial anti-DDoS solutions use routing 
  to steer traffic towards a mitigation infrastructure~\cite{dousti2018automated}.
Sometimes there is a requirement for all the sites
  to be connected through a private backbone to support traffic analysis~\cite{scholl2015methods}.
Another defense uses BGP 
  to divert all traffic to a scrubbing center, then 
  tunnels good traffic to the destination~\cite{smith2016network}.
Other methods use DNS manipulation~\cite{carney2015method}, or anycast proxies~\cite{holloway2014mitigating}
  which cannot be used in DNS anycast deployments itself.
Rather than outsourcing the problem, we explore how one can defend it.
\reviewfix{SS22-C1, SS22-D2, SS22-D4}
Other automated defenses include responsive
  resource management~\cite{fayaz2015bohatei},
  client-server reassignment~\cite{jia2014catch}, and
  filtering approaches~\cite{Rizvi19a}.
Our method uses TE approaches to efficiently
  use available resources in anycast.
  




\section{Mechanisms to Defend Against DDoS}
\label{sec:methodololgy}

In this section we describe our BGP mitigation process; how we
pre-compute a BGP playbook, estimate the attack size and select a TE
response.

\subsection{Overview and Decision Support}
	\label{sec:system_overview}

\begin{figure}
  \begin{center}
    \includegraphics[width=0.85\linewidth]{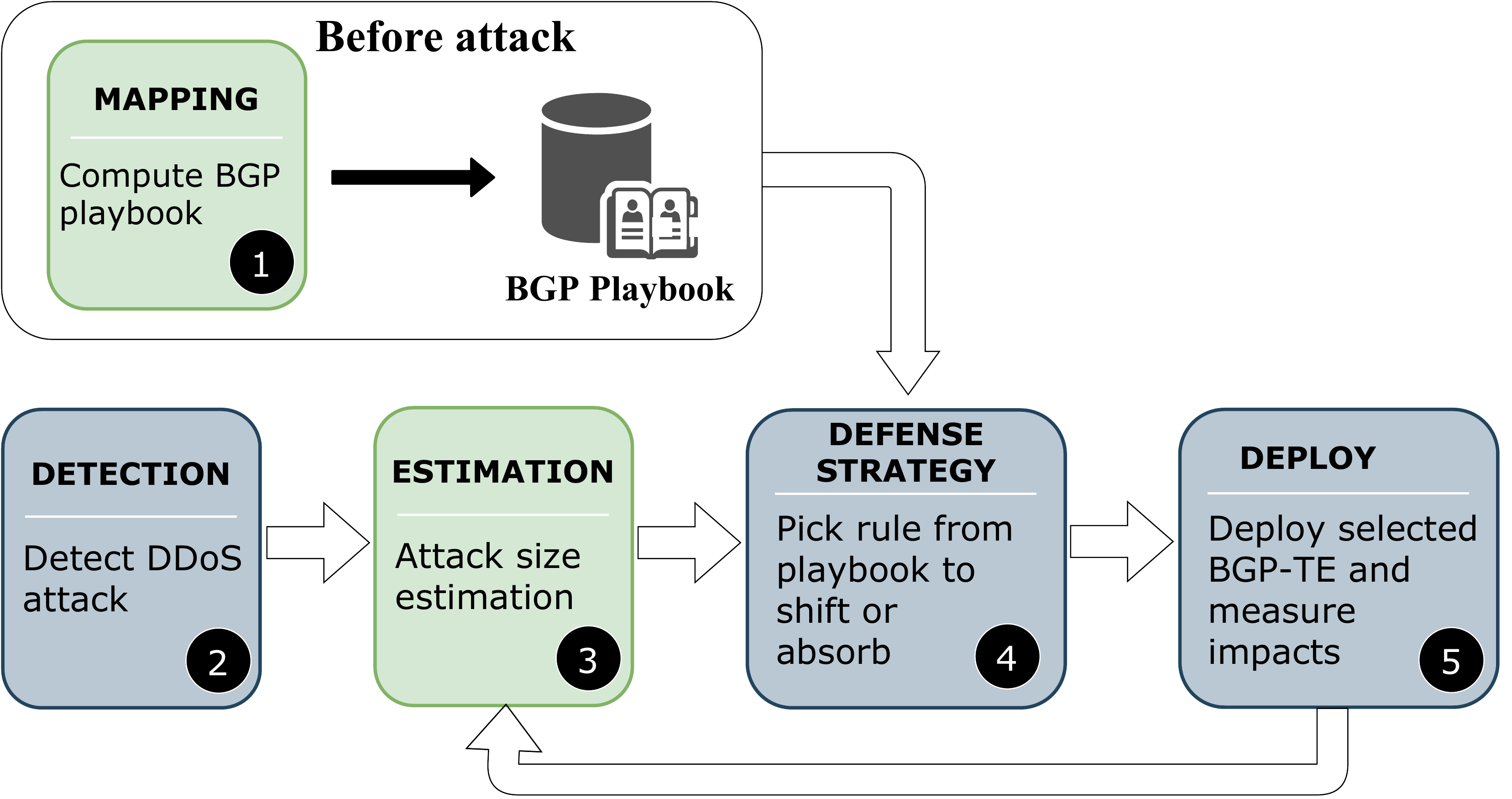}
  \end{center}
  \vspace*{-0.11in}
    \caption{Overview of the decision process.}
    \label{fig:overview-flow}
    \vspace*{-5mm}
\end{figure}
\setlength{\tabcolsep}{6pt} 

In \autoref{fig:overview-flow} we show how defense against DDoS works.
Defense against a DDoS begins with detection \circled{2},
  then defenders plan a defense \circled{4},
  carry it out \circled{5},
  and repeat this process until the attack is mitigated or it ends
  (bottom cycle in \autoref{fig:overview-flow}).
Detecting the attack is straightforward,
  since large attacks affect system performance.
The challenge is selecting the best response and quickly iterating.

We bring two new components to attack response (colored light green in \autoref{fig:overview-flow}):
  mapping before the attack,
  and estimating attack size when the attack begins.
Mapping \circled{1} (discussed in \autoref{sec:mapping})
  provides the defender with a \emph{playbook} of planned responses
  and the information about how they will change the traffic mix across
  their anycast system.
Size estimation \circled{3} (discussed in \autoref{sec:estimating})
  allows the defender to determine how much traffic should be moved
  and select a promising response from the playbook.
Together, these tools help to understand not only
  how to reduce traffic at a given site,
  but also the sites where that traffic will go.

These components come together
  in our automated response system (\autoref{sec:defense_strategy})
  that iterates between measurement and attack size estimation,
  defense selection, then deployment.
Defense uses the playbook built during mapping;
  we provide an example playbook in  \autoref{sec:playbook}.
We show how these defenses operate in
  testbed experiments in \autoref{sec:fight_ddos}.  

Our system is designed for services that
  operate with a fixed amount of infrastructure on specific anycast IP addresses
  and do not employ a third-party scrubbing service.
Operators of CDNs with multiple anycast services, DNS redirection, or scrubbing services
  may use our approach, but also have those other tools.
However, many operators cannot or prefer not to use scrubbing and DNS redirection:
  all operators of single-IP services (all DNS root servers),
  many ccTLDs who value national autonomy,
  and scrubbing services themselves.
Our approach defends against volumetric attacks
  where we have spare capacities in other sites.
\reviewfix{SS22-A4}
Since DDoS causes unavailability of services,
  suboptimal site selection during an attack is not a concern.

\subsection{Measurement: Mapping Anycast}
\label{sec:mapping}

We map the catchments of anycast service before an attack
  so that the defender can make an informed choice quickly during an attack,
  building a BGP playbook (\autoref{sec:playbook}).



To map anycast catchments we used Verfploeter~\cite{Vries17b}. As an
active prober (ICMP echo request), Verfploeter observes the responses
of all ping-responsive IPv4 /24s and maps which anycast site receives
the responses.
\reviewfix{SS22-A6}
\ifisarxiv
We provide a detailed description of anycast, BGP, and Verfploeter in \autoref{sec:anycast_bgp_background} and \autoref{sec:appendix_verfploeter}.
\else
We provide a detailed description of anycast and Verfploeter in \autoref{sec:anycast_bgp_verfploeter}.
\fi
Since mapping happens before the attack, mapping speed is not an issue.

Alternatively, we can map traffic by observing which customers are
seen at each site over time, or measuring from distributed
vantage points such as RIPE Atlas~\cite{staff2015ripe,
bajpai2015lessons}.
(Operators may already collect this information for optimization.)

Mapping should consider not only the current catchments but also
\emph{potential} shifts we might make during the attack.  This full
mapping is easy to do with Verfploeter, which can be continuously
running in an adjacent BGP prefix to map the possible shifts.
This mapping process is important to anticipate how traffic may shift.
We will show later that BGP control is limited by the
granularity of routing policy (\autoref{sec:exploit_catchment}) and by
the deployment of the anycast sites
(\autoref{sec:constraints_anycast}).  

A challenge in pre-computed maps with routing alternatives is that
routing is influenced by all ASes. Thus, the maps may shift over time
due to changes in the routing policies of other ASes. Fortunately,
prior work shows that anycast catchments are relatively slow to
change~\cite{Wei17b}.
We also show that our BGP playbook is stable over time (\autoref{sec:playbook} and \autoref{sec:appendix_stability}).

\ifisestimation
\subsection{Estimation of the Attack Size}
\label{sec:estimating}

After the detection of an attack, the first step in DDoS defense
  is to estimate the attack size,
  so we can then select a defense strategy of how much traffic to shift.
Our goal is to measure \emph{offered load},
  the traffic that is sent to (offered to) each site.
During DDoS offered load balloons with a mix of attack and legitimate traffic,
  and loss upstream of the service means we cannot directly
  observe true offered load.
We later evaluate our approach with real-world DDoS events (\autoref{sec:attack_size_estimate_accuracy}).

\textbf{Idea:}
Our insight is that we can \emph{estimate} true offered load
  based on changes in some known traffic that
  actually does arrive at the service, even when there is upstream loss.

To know how much offered load actually arrives at the service,
  we need to estimate some fraction of legitimate traffic.
We can then observe how much this traffic drops during the attack,
  inferring upstream loss.
Unfortunately, there is no general way to determine
  all legitimate traffic,
  since legitimate senders change their traffic rates,
  and attackers often make their traffic legitimate-appearing.
Our goal is to reliable estimate \emph{some} specific legitimate traffic;
  we describe several sources next.
  
\textbf{Traffic sources:}
There are several possible sources of known legitimate traffic---we
  consider known measurement traffic
  and regular traffic sources that are heavy hitters \cite{basat2016heavyhitters}.

For DNS, our demonstration application,
  RIPE Atlas provides a regular source of known-good traffic,
  sent from many places.
RIPE makes continuous traffic from around 10k publicly available
  vantage points~\cite{RIPE21a}.
Each RIPE vantage point queries every 240\,s,
  and there is enough traffic (about 2500\,queries/minute) to provide a good estimate
  of offered load.
(Although RIPE Atlas is specific to DNS,
  other commercial services often have similar types of known monitoring traffic.)

To find the known-good traffic at each site,
  we use the catchments of RIPE vantage points
  with pre-deployed RIPE DNS CHAOS queries
  (one exists for each root DNS IP, such as measurement ID 11309 for A-root). 
We can also use Verfploeter or captured traces in the anycast sites.
\reviewfix{SS22-A2, SS22-A13, SS22-G2}
An advantage of using RIPE traffic
  is that it does not place any new load on the service.

\emph{Heavy hitters} can provide
 an additional source of known-good traffic.
\reviewfix{SS22-G2} 
Many services have a few consistently large-volume users
  with regular traffic patterns,
  and while they vary over time, many are often stable.
For DNS, we find that most heavy hitters have a strong diurnal variation in rate;
  we model them with TBATS
  (Trigonometric seasonality, Box-Cox transformation, ARMA errors, Trend and Seasonal)~\cite{de2011forecasting}
  to factor out such known variation.
\reviewfix{SS22-A2, SS22-A13}
While an adversary could spoof heavy hitters,
  that requires a large and ongoing investment to succeed.

\textbf{Estimation:}
Our goal is to estimate offered load, $T_{\Vsmall{offered}}$.
We can measure the  observed traffic rate, $T_{\Vsmall{observed}}$,
  at the access link.
We define $\alpha$ as the \emph{access fraction}---the fraction
  of traffic that is not dropped.
Therefore $T_{\Vsmall{observed}} = \alpha \cdot T_{\Vsmall{offered}}$.

To estimate the access fraction ($\alpha$),
  we observe that known good traffic has the same loss on incoming
  links as does other good traffic and attack traffic.
We estimate the known traffic rate (from RIPE Atlas measurement traffic,
  or from heavy hitters, or both), as
  $T_{\Vsmall{known}}$.
Then $\alpha \cdot T_{\Vsmall{known,offered}} = T_{\Vsmall{known,observed}}$,
  and our estimate of offered load is
   $\hat{T}_{\Vsmall{offered}} = 
  T_{\Vsmall{observed}} \cdot T_{\Vsmall{known,offered}} /  {T_{\Vsmall{known,observed}}}$.

\fi  

\subsection{Traffic Engineering as a Defense Strategy}
	\label{sec:defense_strategy}

With knowledge of the offered load, the defender can select an overall defense
strategy that will drive traffic engineering decisions.
The defender first must determine if the attack exceeds overall capacity or not.

For attacks that exceed overall capacity,
  the defender's goal is to preserve successful
service at some sites, while allowing other sites to operate in
degraded mode as \emph{absorbers}~\cite{Moura16b}. The defender may
also choose to shift traffic away from some degraded sites to ease
their pain.
Unloading the overloaded sites is recognized as \emph{breakwaters}~\cite{kuipers17}.

For moderate-size attacks, the defender should try to serve all
traffic, \emph{rebalancing} to shift traffic from overloaded sites to less busy
sites.
In heterogeneous anycast networks, where some sites have more
capacity than others, the defense approach can be different. In these
cases, larger, ``super''-sites can attract traffic from smaller sites.
For moderate-size attacks, it may even be best for smaller sites to
shut down if the super-sites can handle the traffic.

Regardless of attack size, traffic engineering allows the defender
  to shift attack traffic to absorber or breakwater sites.
We next describe traffic engineering options,
  and then how one can automate response.
For operators unwilling to fully automate response,
  our system can still provide recommendations for possible actions
  and their consequences.

\subsubsection{Traffic Engineering to Manage an Attack}
	\label{sec:route_manipulation}

Given an overall defense strategy (absorb or rebalance),
  the defender will  use traffic engineering
  to shift traffic,
  either automatically (\autoref{sec:automatic-defense-selection})
  or as advice under operator supervision.
For anycast deployments connected by the public Internet,
  BGP~\cite{Caesar05a} will be the tool of choice
  to control routing and influence anycast catchments.
Organizations that operate their own wide-area networks
  may also be able to use SDN to manage traffic on
  their internal WAN~\cite{Schlinker17a,Hong18a}.
Fortunately, BGP has well established mechanisms to manage routing policy.
%
%
We use three BGP mechanisms in the paper:
  AS-Path prepending, BGP communities and Path Poisoning.

\textbf{AS-Path Prepending} is a way to de-prefer a
  routing path, send traffic to other catchments. 
BGP's AS-Path is the list of ASes back to the route originator.
The AS-Path both prevents routing loops and 
  also serves as a rough estimate for distance,
  with BGP preferring routes with shorter AS-Paths.
By artificially inserting extra ASes into the AS-Path,
  the route originator can de-prefer one site in favor of others.
Path prepending is known to be a coarse
  routing technique for traffic engineering.
We measure how fine the control AS-Path prepending provides to anycast
  in \autoref{sec:path_prepending}.

We define \textbf{Negative Prepending} as the use of AS-Path prepending
  to draw traffic towards a site,
  preferencing one site over others.
Prepending can only increase path lengths,
  but an anycast operator in control of all anycast sites can
  prepend at all sites except one, in effect giving that site a shorter
  AS-Path (relative to the other sites) than it had before.
``Negative prepending by one at site S'' is, therefore, shorthand for prepending
  by one at all sites other than S.

Long AS-Paths due to prepending can make prefixes more vulnerable to route
  hijacking~\cite{madory19a}.
\reviewfix{SS22-A10}
However, this issue has a small impact on anycast prefix, as always 
there is a site announcing without any prepend, keeping 
the path length limited.
We suggest that formal defenses to hijacking such as RPKI are needed
  even without prepending,
  and when they are in place, prepending can be an even
  more valuable tool for TE.

\textbf{BGP Communities} (or community strings)
  label specific BGP routes with 32 or 64 bits of information.
How this information is interpreted is up to the ASes.
While not officially standardized, a number of conventions exist
  where part of the information identifies an AS
  and the other part a policy such as blackholing, prepend, or set local-preference.
Community strings are widely supported to allow ISPs to delegate
  some control over routing policy to their customers~\cite{ampath,comm_list}.

\textbf{Path Poisoning} is another way to control
  the incoming traffic.
This technique consists of adding
  the AS of another carrier to the AS PATH\@.
Paths that repeat ASes in different parts of the AS PATH
  indicate routing loops and must be discarded by BGP\@.

When using path poisoning we announce a path
  with both the poisoned AS and own AS
  (otherwise neighbors may filter our announcement as not from us).
We must therefore \emph{also} prepend twice at all other anycast sites,
  otherwise poisoning also results in a longer AS path.

\begin{figure}
\centering

\includegraphics[width=0.8\linewidth]{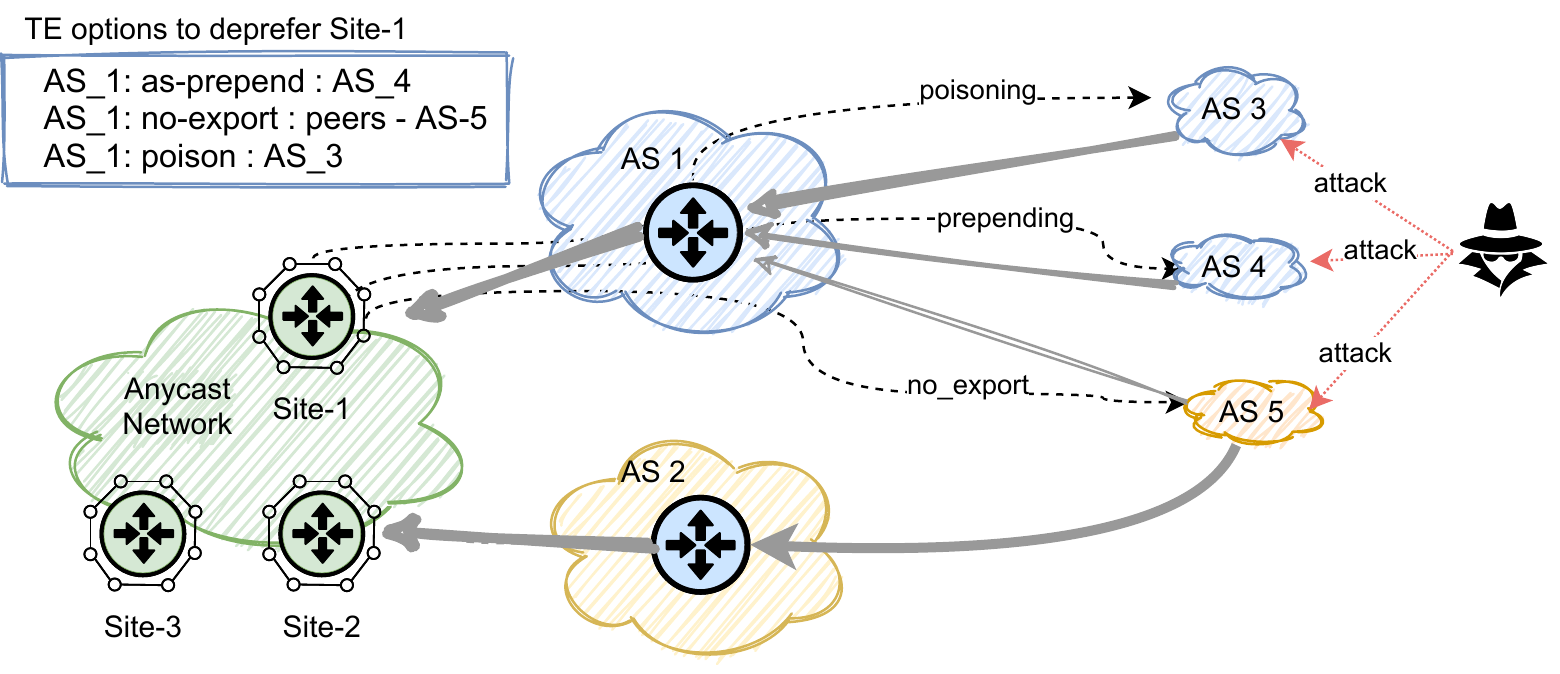}
    \caption{TE techniques to shift traffic from Site-1 to Site-2.}
\label{fig:ddos-bgp-solution}
  \vspace{-0.5em}
\end{figure}

\autoref{fig:ddos-bgp-solution} shows how traffic engineering
  can be applied to anycast systems in order to modify the catchment.
In this example, {\it site-1} \/ is overwhelmed by an attack.
Aiming to shift bins of traffic to {\it site-2} \/
  with spare capacity, we can make BGP announcements.
{\it site-1} \/ poisons AS3, prepends (only showing to AS4),
  and prevents announcement to AS5 using \textit{not-export} BGP community.
These changes decrease load in \emph{site-1}, shifting the traffic to
\emph{site-2}.

\subsubsection{Automatic Defense Selection}
    \label{sec:automatic-defense-selection}

\reviewfix{SS22-A8}

To automate defense
  we use a centralized \emph{controller}.
The controller collects observations for all sites
  (from external measurements, or assuming the site is saturated if it cannot reach the site),
    then takes action, if required (\circled{4} of \autoref{fig:overview-flow}):
(1) The controller identifies sites that are overcapacity
  by comparing estimated load to expected capacity and observed resources at each site.
(2) The controller identifies all playbook options
  that will reduce load at any impacted sites
  without overloading currently acceptable sites.
(3) It selects from any viable options,
  favoring a uniform distribution and smallest change (or selecting arbitrarily if necessary).
If all changes leave some sites overwhelmed, it can choose the ``least bad'' scenario,
  or request operator intervention.

After deploying a new routing policy,
  the decision machine continues to evaluate
  the traffic level at each site (\circled{5} of \autoref{fig:overview-flow}).
If any site is still overwhelmed
  after 5 minutes, 
  we try again, repeating size estimation, decision, and action.
In the subsequent iterations, the controller
  only considers the routing options that
  were considered in the previous iteration (from step (2) of this decision process).
We allow time between attempts so announcements can propagate~\cite{labovitz2000delayed}.
To avoid oscillation or interference with route flap dampening,
  after three attempts
  we escalate the problem to the human operator.
\reviewfix{SS22-A11}
We choose these values for timer duration and number of retries
  from recommendations of operators to avoid oscillation,
  other options are possible.
Explore other options is possible as future work.

\emph{Return to service:}
\reviewfix{SS22-G3}
After a period with no overloaded sites,
  we can automatically revert any interventions,
  on the assumption that default routing provides users best service.
Leaving interventions in place for some time can help with
  polymorphic attacks (\autoref{sec:fight_ddos}).

\subsubsection{Operator Assistance System}
    \label{sec:op-assistance-system}
\reviewfix{SS22-F4, SS22-E2, SS22-E5, SS22-C2, SS22-C6}
We discussed our approach with operators of root DNS and cloud services
  to get feedback on the approach.
While they were enthusiastic about automated defenses
  to deal with common DDoS events,
  and to handle events during non-business hours,
  some operators prefer human-supervised (non-automated) response,
  and all expected human supervision of response
  during initial deployment to build trust before full automation.

\begin{figure}[]
    \centering
    \tmpframe{\includegraphics[scale=0.11]{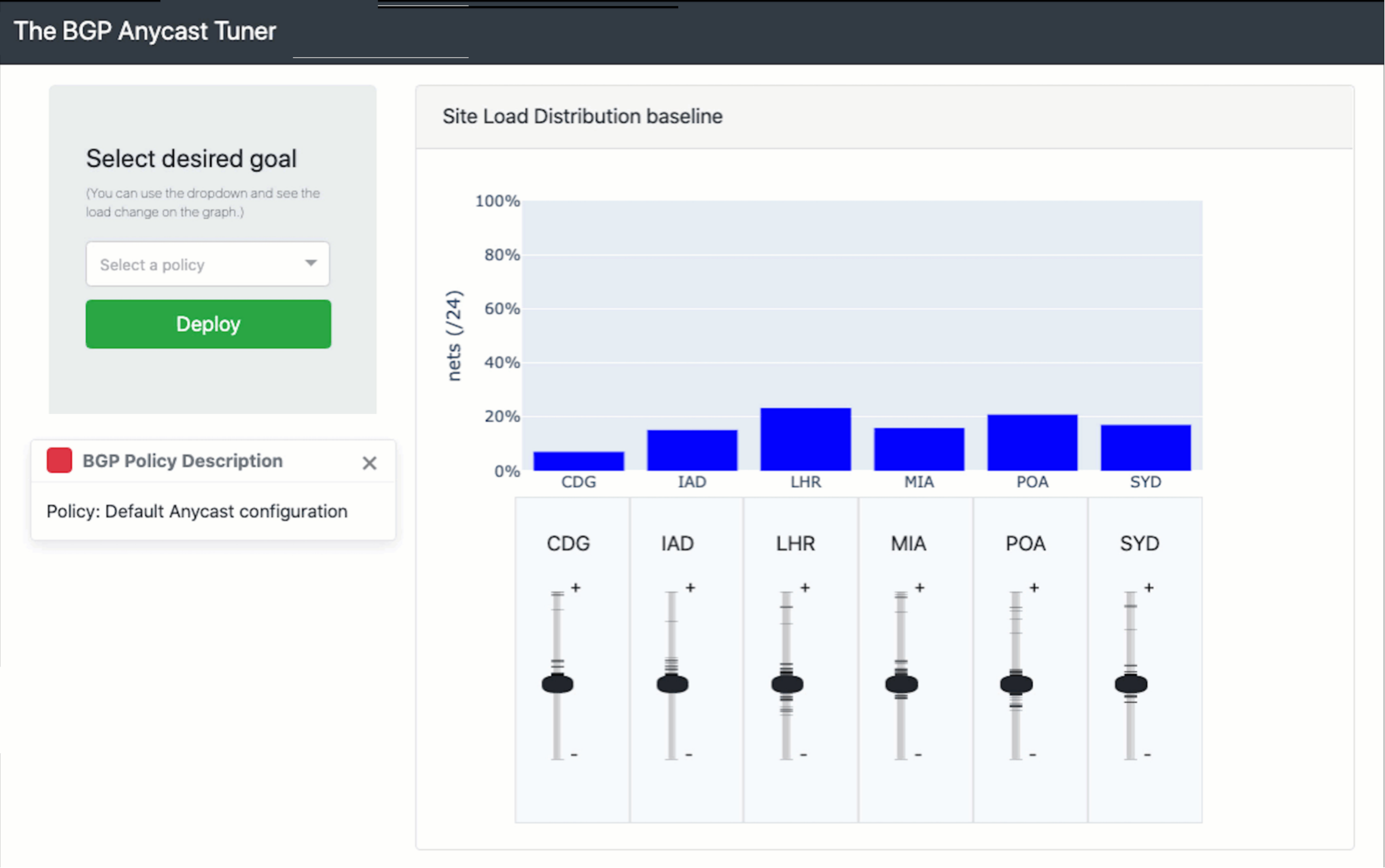}}
    \caption{Operator assistance system.}                
    \label{fig:user-interface}
    \vspace{-0.18in}
\end{figure}

To support human-supervised response,
  we design an operator assistance system
  as an alternative (or precursor) to automation. 
This system provides a web-based interface that activates route changes,
  coupled with playbook lookup that
  recommends good options based on current sensor status
  (\autoref{fig:user-interface}). Details are described in
  \autoref{sec:operator_assistance}.


\ifisestimation
\section{Evaluation of Offered Load Estimation}
	\label{sec:attack_size_estimate_accuracy}

We next evaluate estimating offered load
  with real-world events;
  a testbed evaluation is in \autoref{sec:testbed-experiment-details}.


\begin{table*}
\centering
\small
\begin{small}
\begin{tabular}{lr|rrr|rrrrr|r}
\textbf{Scenario/}  & & \multicolumn{3}{c|}{\textbf{known-good traffic}} &  \multicolumn{5}{c|}{\textbf{offered load during attack}} & \textbf{estimated/}\\
\textbf{Date} & \textbf{Dur.} & \textbf{normal} & \textbf{observed} & $\alpha$ & \textbf{normal} & \textbf{observed} & \textbf{reported} &  \textbf{estimated} & $\hat{\alpha}$ & \textbf{reported} \\
  \hline
 \rowcolor{gray!20} 

	2015-11-30 & 3h& 33.08& 1.85& 0.0559  & 0.03\,M & 0.37\,M & 5.1\,M & 6.6\,M & 0.07 & 1.3\\
\rowcolor{gray!20}
	2016-06-25 & 3h & 36.58& 0.33& 0.0091 & 0.03\,M & 0.10\,M & 10\,M & 11\,M & 0.01 & 1.1 \\

	Testbed & 5min & 425.2& 207.0& 0.4900 & 8.5\,k & 16.3\,k & 29.2\,k & 33.2\,k & 0.56 & 1.1 \\

\end{tabular}
\end{small}
  \caption{Estimating sizes of offered load (second from right) 
  based on known-good traffic (second from left) with real-world attacks at \broot and testbed experiment.
    Traffic rates are in queries/second (reporting only the peaks).}
\label{tab:events-estimation}
\vspace{-0.15in}
\end{table*}

\subsection{Case Studies}
	\label{sec:estimation_case_studies}

We test our approach with two large DNS DDoS events
  from
  2015-11-30 and 2016-06-25.
The  November~2015 event was a DNS flood, and
  the June~2016 was a SYN and ICMP flood attack.
\broot exhibited significant upstream loss
  in both these events,
  so we estimate true offered load to \broot 
  and compare to observations at other roots for ground truth.

To apply our system we measure the access fraction ($\alpha$)
  using the known-good traffic.
\autoref{tab:events-estimation} shows the expected typical known-good traffic (``normal''),
  the observed rate under attack (``observed'') and the computed $\alpha$.
Here we use RIPE Atlas as known good\ifisanon~\cite{anon-b-ripe}\else~\cite{ripe-dns-built-in}\fi.
We see similar results (omitted due to space) when using the top 100 heavy hitters.

\begin{figure}
\centering
\includegraphics[width=0.75\linewidth]{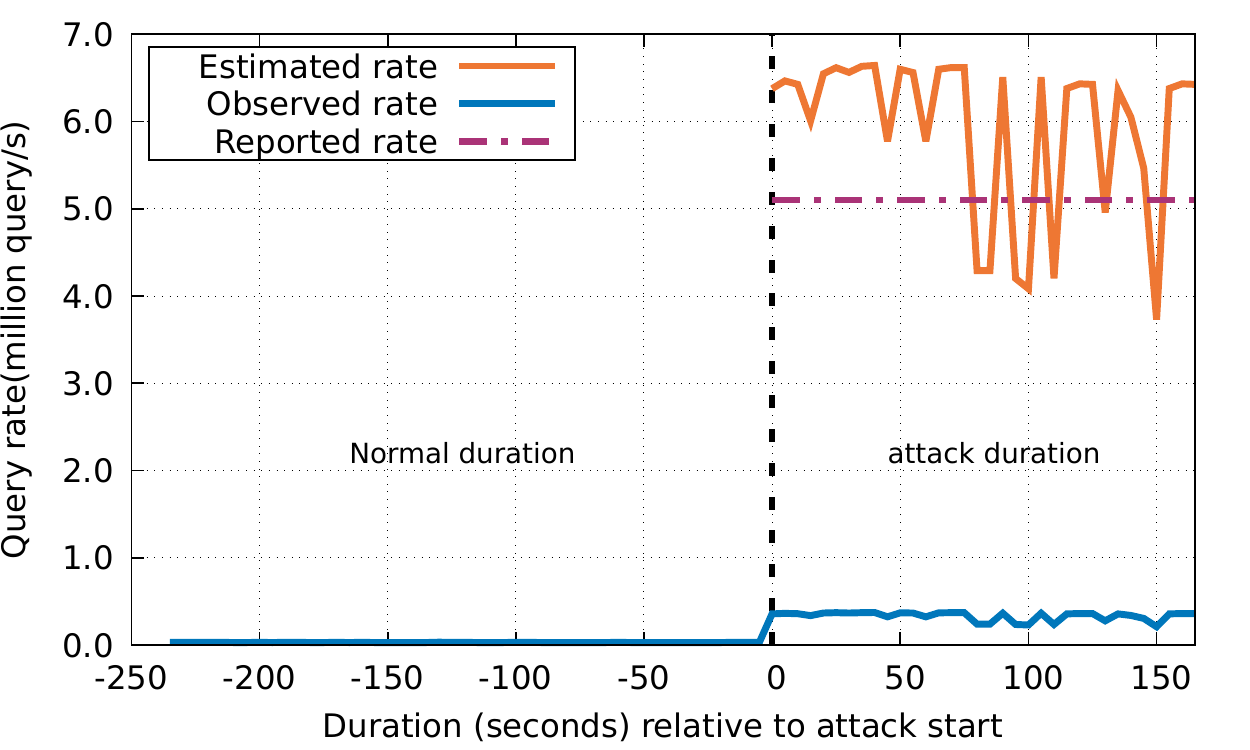}
\vspace{-0.07in}
    \caption{Estimating real-world attack events: estimating Nov. 2015 event with 5.59\% access fraction.}
\label{fig:show-estimation}
    \vspace{-0.17in}
\end{figure}

\autoref{fig:show-estimation} compares 
  the observed load (the bottom blue line)
  with the estimated offered load (the middle, varying, orange line)
  from our system,
  as compared to the attack rate reported from other roots (the dashed purple line).
The offered load columns of \autoref{tab:events-estimation}
  give numeric values.

Even though the attack was large, we see that the
  estimated attack size of the 2015 event of 4--6.5\,Mq/s
  is close to the reported 5.1\,Mq/s~\cite{Moura16b, 2015-event}.
We also see similar results from the 2016 event~\cite{2016-event},
  where we estimate 8--11\,Mq/s of total traffic,
  compared to the 10\,Mq/s reported rate (details with figure in \autoref{sec:case-study-2016}).
We also add the result from Testbed experiment
  which also shows a good accuracy (details in \autoref{sec:testbed-experiment}).

These two events show that even with high rates of upstream loss
  we are able to get reasonable estimates of total offered load.
Our results provide good accuracy when the known-good traffic
  has 2500 queries/minute with RIPE, and additional known-good traffic
  can improve accuracy.
Use of additional known-good traffic (such as heavy hitters)
  improves accuracy in these cases by providing a larger signal.
However, in practice, even a rough estimation
  allows a far better response than using directly observed load.

We conclude that attack size estimation
  is close enough to help plan response to DDoS events.

\fi

\section{Evaluation Approach}
    \label{sec:measurement_methodology}


We next describe how we will evaluate
  the effectiveness of TE (\autoref{sec:exploit_catchment})
  and that results generalize to different deployments
  (\autoref{sec:constraints_anycast}).
Traffic engineering in response to DDoS depends on
  the anycast deployment---where sites are and with whom they peer.
We evaluate on two different testbeds.
\reviewfix{SS22-A1, SS22-A5, SS22-B2}
Our approach (estimation, TE, and playbook construction)
  can be applied anywhere with different anycast setups.
\reviewfix{SS22-A4} We expect network operators will execute our approaches
  on a test prefix (in parallel with their operational network) prior to an event so that no service interruption happens.

\subsection{Anycast Testbeds}
	\label{sec:anycast_testbeds}

We evaluate our ideas on testbeds
  to see the constraints of real-world peering and deployments.
We use two independent testbeds: \peering\ifisanon~\cite{schlinker19peeringAnon} \else~\cite{schlinker19peering} \fi 
  and
  \tangled\ifisanon~\cite{le2020tangledAnon}\else~\cite{le2020tangled}\fi.
\autoref{tab:testbeds} summarizes information about each testbed
  with their own set of geographically distributed sites 
  along with their locations (\peering supports more sites but we used 8 sites).
These sites show different connectivity, and have
  one or more transits and IXP peers.
Most \peering sites have
  academic transits while \tangled has more commercial providers.
Our testbed is about the
  same size as many operational networks, since nearly half of
  real-world networks have five or fewer sites~\cite{cicalese2015characterizing}.


\begin{table}
\centering
\footnotesize
\begin{tabular}{lll}
    \textbf{Testbed}  & \textbf{Used Sites} & \textbf{\#} \\
  \hline
    {\footnotesize \peering }  & \begin{tabular}[c]{@{}l@{}}Amsterdam*\dag (AMS), Boston* (BOS), \\
                           Belo Horizonte*\dag (CNF), Seattle* (SEA)\\
                          Athens* (ATH), Atlanta* (ATL), \\Salt Lake City* (SLC), Wisconsin* (MSN)  
                          \end{tabular} & 8\\
  \hline

    {\footnotesize \tangled } & \begin{tabular}[c]{@{}l@{}} 
  Miami (MIA)*, London (LHR)*, \\
        Sydney (SYD)*, Paris (CDG)*, \\ 
        Los Angeles (LAX)*, Enschede (ENS)*, \\
        Washington (IAD)*, Porto Alegre (POA)*\dag\\
  \end{tabular} & 8 \\
    \hline
\end{tabular}
    \caption{Testbed and respective sites used in our experiments. Transit
    providers (*) and IXP (\dag).}
\label{tab:testbeds}
\vspace{-0.12in}
\end{table}




\subsection{Measuring Routing Changes}

To measure the effect of a BGP change, 
  we first change the routing announcement at a site,
  give some time to propagate, confirm that the announcement is accepted,
  and finally start the anycast measurement.

\textbf{Route convergence}: 
After a change, we allow some time for \emph{BGP route propagation}.
We know that routing and forwarding tables can be inconsistent (resulting in loops or
black holes) while prefix is
updating~\cite{labovitz2000delayed,teixeira2007bgp,silva2017bgpdelay}.
Although routing updates are usually stable within 5 minutes~\cite{silva2017bgpdelay},
  we wait 15 minutes for routing to settle when building our playbook since
  it is a non-attack period.
\reviewfix{SS22-A9} When the attack is not mitigated after deploying a routing policy, our system moves to a different approach after 5 minutes.


\textbf{Propagation of BGP policies}: 
Policy filtering could limit the acceptance of announced routes,
   although in practice these limits do not affect our traffic engineering.
Best practices for networks at the edge to filter out
  AS-Paths longer than 10 hops,
  and ASes in the middle often accept up to 50 hops,
  both more prepends than we need.
Based on routing observations from multiple global locations
  using RIPE RIS, we confirm that configurations in
  our experiments are never blocked due to route filtering
  in multi-hops away from our anycast sites. 

\section{Traffic Engineering Coverage and Control}
	\label{sec:exploit_catchment}

\ifisestimation
From an estimate of attack load,
  operators use BGP to 
  shift traffic.
\else
The operators want to 
  shift traffic with BGP mechanisms.
\fi
We next evaluate three TE mechanisms:
  AS-Path prepending, community strings and
  path poisoning.
For each we consider when it works
  and what degree of control it provides.
\autoref{tab:experiment-and-takeaways} summarizes our key results
  from tests on two testbeds (\autoref{sec:anycast_testbeds});
  in \autoref{sec:constraints_anycast}
  we evaluate generalizability.

\begin{table}[t]
   \centering
   \footnotesize
   \begin{tabularx}{\columnwidth}{p{20mm}p{60mm}}
   \textbf{Experiment} &  \textbf{Key Takeaways} \\
   \hline
  Path prepending 
   &  Works everywhere to effectively de-prefer a site
     (\autoref{sec:prepend_works}),
       but shifts traffic in large amounts
       (\autoref{sub:granularity}),
       and has few traffic levels (\autoref{fig:selfprepending}).
    \\ \hline
 \rowcolor{gray!10}
       Neg.~Prepending&
          Works everywhere to prefer a site 
          (\autoref{sec:prepend_works}). 
  \\ \hline
   BGP communities & Although widely implemented,
   well-known communities are not universal
   (\autoref{sec:communities_coverage}). \\
 \rowcolor{gray!10}
&   When supported, they provide
   finer-granularity control than prepending (\autoref{sec:granu}).
  \\ \hline
   BGP path poisoning & Many Tier-1 ASes drop the announcements when it sees Tier-1 ASes in the paths. (\autoref{sec:coverage-poisoning}) \\
    \rowcolor{gray!10}
   & Control over traffic is limited by the filters from other ASes. (\autoref{sec:granularity-poison}).  \\ \hline
  
    \end{tabularx}

    \caption{Experiment summarization and findings.}
    \label{tab:experiment-and-takeaways}
        \vspace{-0.15in}
\end{table}

\subsection{Control With Path Prepending}
	\label{sec:path_prepending}

First we consider AS-Path prepending as a defense strategy.

\subsubsection{Prepending coverage}

Support for AS-Path prepending is quite complete---it requires no
explicit support from the upstream provider, so we found prepending
worked at all sites in both of our testbeds.  In \peering, we are
allowed to use a maximum of three prepends, and in \tangled we use up
to five prepends.  Previous study \cite{chang2005inbound} shows a
maximum of 5 prepends is sufficient because 90\% of active ASes are
located less than six AS hops away.  We use RIPE RIS~\cite{riperis} to
check the routing visibility when prepends are in place, and we do not
observe changes in the routing propagation for both testbeds. Otherwise,
this might reveal the existence of AS path length
filters~\cite{cymru-secure-bgp-template,Huston2018BGP}.

%



\subsubsection{Does prepending work?}
	\label{sec:prepend_works}

Since AS-Path prepending is widely supported,
  we next evaluate this attractive TE method.

We explore this question 
  for a representative scenario using \peering
  using three sites 
  from three continents---Europe (Amsterdam-AMS), North America (Boston-BOS)
  and South America (Brazil-CNF).
In \autoref{sec:constraints_anycast} we generalize to other configurations.
We estimate load by counting /24 blocks in catchments,
  then compare the baseline with TE options.
\reviewfix{SS22-A7}
(We also explored traffic weighted by traffic loads instead of blocks,
  getting the same qualitative results and shapes with different constants, \autoref{sec:appendix_playbook_load}.)

\PostSubmission{
where do we compare to \tangled? ---johnh 2020-05-25.
joao: We do have this result. However I'm not sure they are meanfull here since we summarize them on \autoref{fig:selfprepending}}

\begin{figure*}
        \begin{subfigure}[t]{0.33\textwidth}
                \centering
                \includegraphics[width=.98\linewidth]{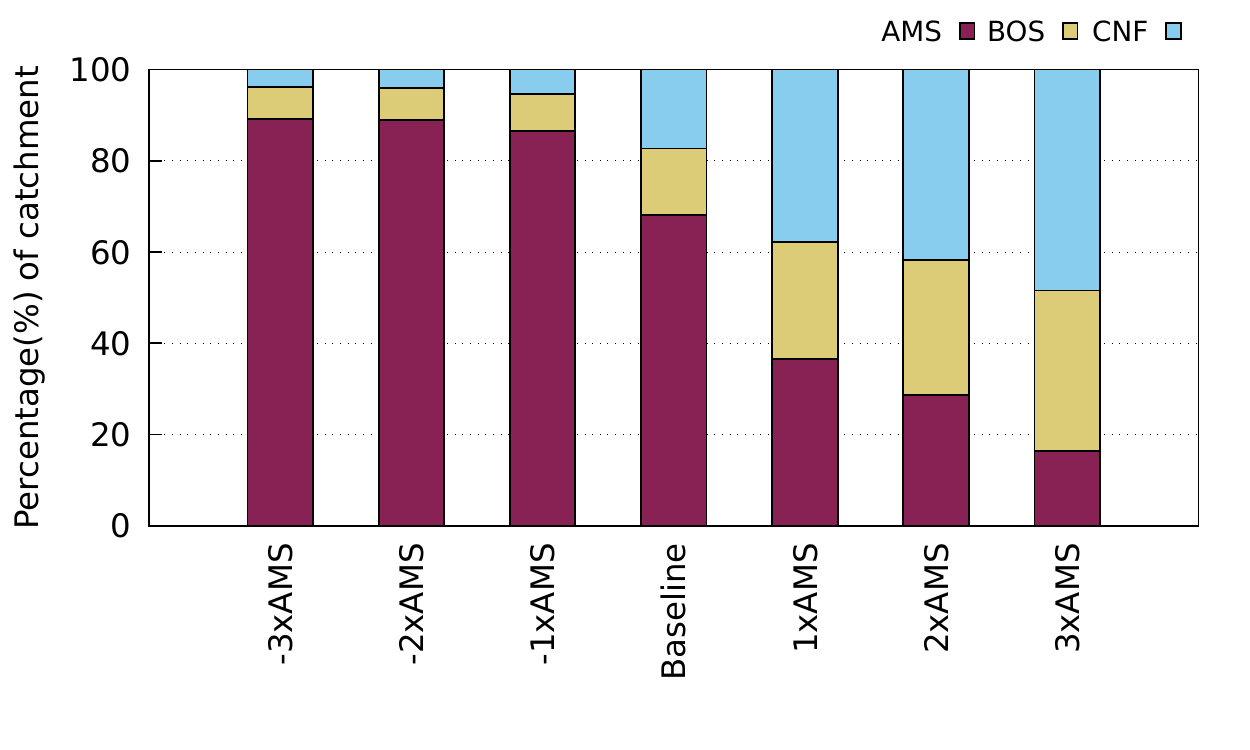}
                \vspace{-0.15in}
                \caption{AMS site.}
                \label{fig:ams-only-site-2020-02-24}
        \end{subfigure}
	\begin{subfigure}[t]{0.33\textwidth}
                \centering 
                \includegraphics[width=.98\linewidth]{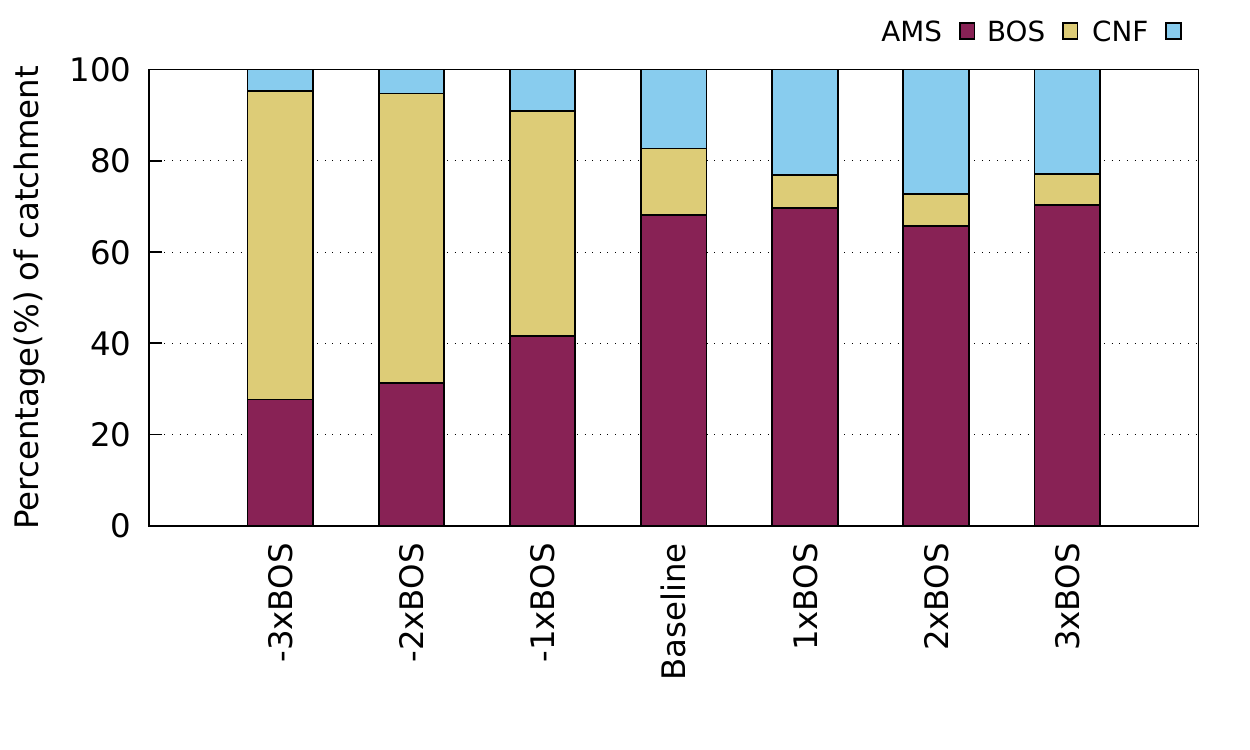}
                \vspace{-0.15in}
                \caption{BOS site.}
                \label{fig:bos-only-site-2020-02-24}
        \end{subfigure}
        \begin{subfigure}[t]{0.33\textwidth}
                \centering
                \includegraphics[width=.98\linewidth]{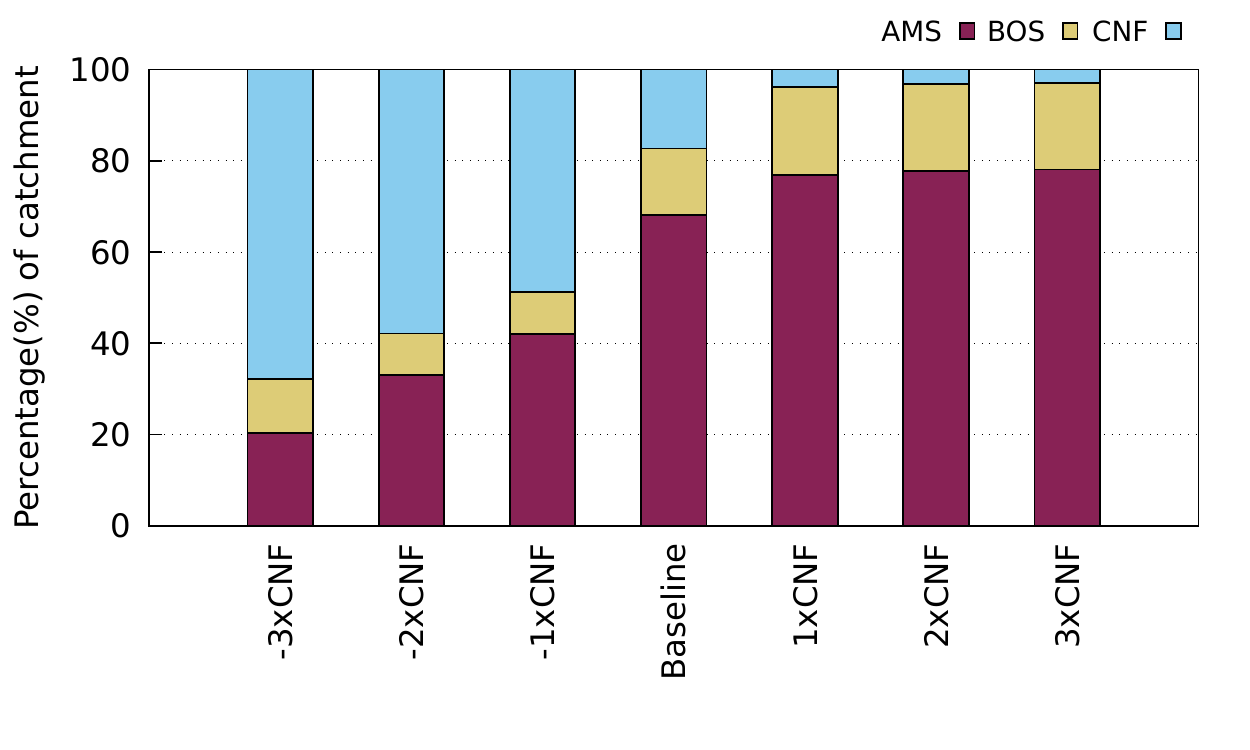}
                \vspace{-0.15in}
                \caption{CNF site.}
                \label{fig:cnf-only-site-2020-02-24}
        \end{subfigure}
        \vspace{-0.2in}
       \caption{\peering: Impact of path prepending in catchment distribution with AMS, BOS and CNF sites on 2020-02-24.}
       \label{fig:ams-bos-cnf-only-site-2020-02-24}
       \vspace{-0.15in}
\end{figure*}

\autoref{fig:ams-bos-cnf-only-site-2020-02-24}
  shows the traffic from each site under different conditions.
The middle bar in each graph is the baseline,
  the default condition with no prepending.
We then add prepending at each site,
  with one, two or three prepends in each bar going to the right of center.
We also consider negative prepending (~\autoref{sec:route_manipulation})
  in one to three steps, with bars going left of center.

\cmt{POSSIBLE TO CUT: next two paragraphs -- 2021-10-11 Leandro}

We first consider the baseline (the middle bar) of all three graphs in 
  \autoref{fig:ams-bos-cnf-only-site-2020-02-24}.
Amsterdam (AMS, the bottom, maroon part of each bar)
  gets about 68\% of the traffic.
AMS receives more traffic than BOS and CNF because
  that site
  has two transit providers and several peers,
  and Amsterdam is very well connected with the rest of the world.

We next consider prepending at each site (the bars to the right of center).
In each case, \emph{prepending succeeds at pushing traffic away from the site},
  as expected.
For AMS, each prepend shifts more traffic away,
  with the first prepend cutting traffic from 68\% to 37\%,
  then to 29\%, then to about 16\%.
BOS and CNF start with less traffic
  and prepending has a stronger effect, with one prepend
  sending most traffic away (at BOS, from 15\% to 7\%) and additional prepends showing little
  further change.
These non-linear changes are because changing BGP routing with prepending
  is based on path length, and the Internet's AS-graph is relatively flat~\cite{APNIC20a,Chiu15a}.

The bar graphs also show that when prepending pushed traffic away from a site,
  it all goes to some other site.
Where it goes depends on routing and is not necessarily proportional
  to the split in other configurations.
For example, after one prepend to AMS, 
  more traffic goes to CNF (the top sky blue bar) than to BOS (the middle yellowish bar).
These unexpected shifts are why we suggest
  pre-computing a ``playbook'' of routing options before an attack
  (\autoref{sec:mapping}) to guide decisions during an attack
  and anticipate the consequences of a change.

We also see that negative prepending succeeds at drawing traffic towards
  the site---in each case the bars to the left of center see more
  traffic in the site that is not prepending while the others prepend.
AMS sees relatively little change (68\% to 89\%) since it already has most traffic,
  while BOS and CNF each gain up to 68\% of traffic.

All three sites show some networks that are ``stuck'' on that site,
  regardless of prepending.
One reason for this stickiness is when some networks are only routable
  through one site because they are downstream of that exchange.
We confirm this by taking traceroute to two randomly chosen blocks
  that are stuck at BOS.  
Traceroutes and geolocation (with Maxmind)
  confirm they are in Boston, at MIT
  and a Comcast network (based on the penultimate traceroute hop).
We have used the local-preference BGP attribute to move such stuck blocks,
  but a systematic exploration of that option is future work.

\begin{figure*}
\vspace{0.1in}
\centering
  \begin{minipage}{0.3\textwidth}
\begin{center}
   \vspace{-0.1in}
    \includegraphics[width=\textwidth]{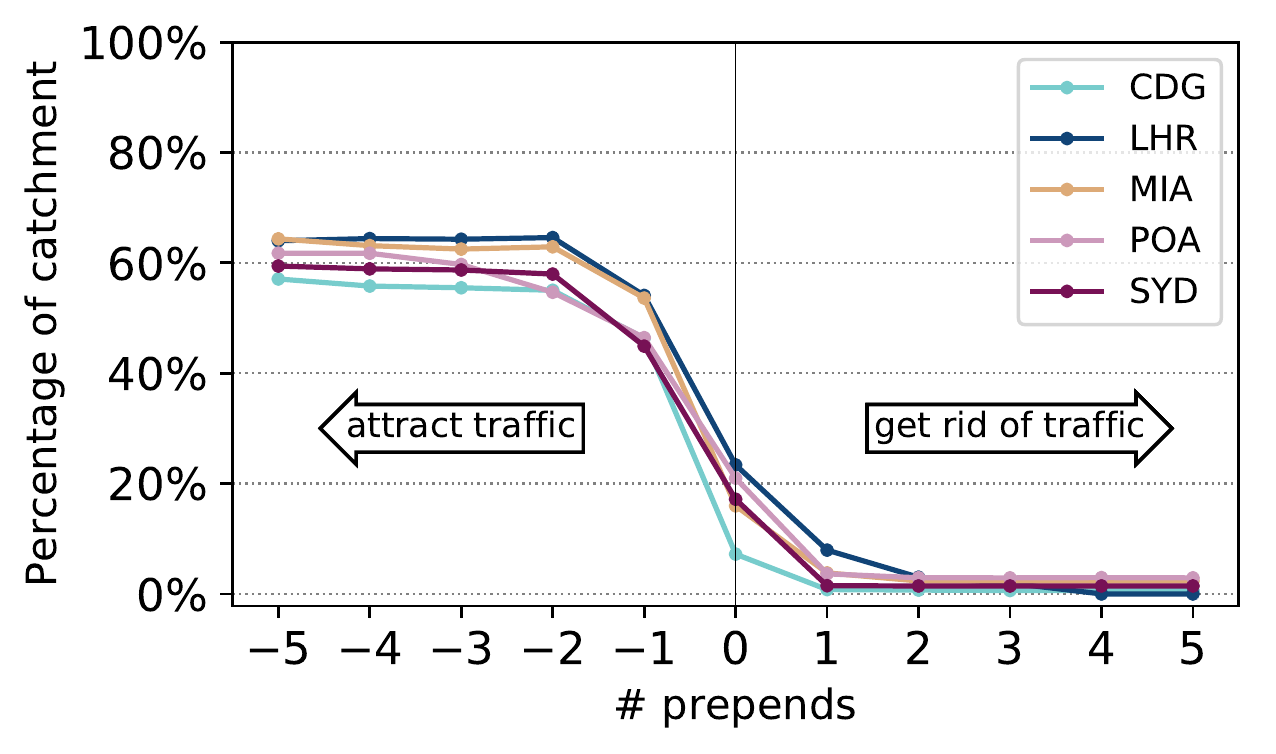}
\end{center}
     \vspace{-0.18in}
  \caption{\tangled: Effect of path prepending on catchments.}
          \label{fig:selfprepending}
  \end{minipage}\hfill
  \begin{minipage}{0.3\textwidth}
\begin{center}
    \includegraphics[width=\textwidth]{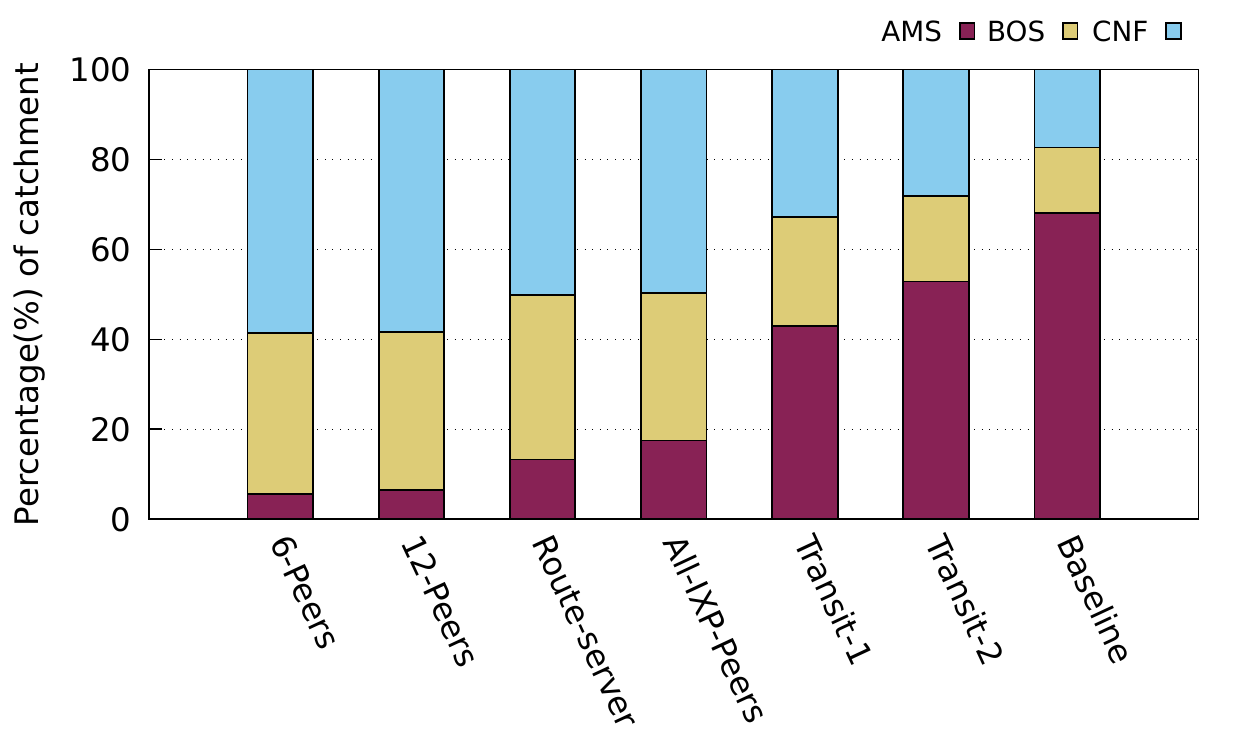}
\end{center}
       \vspace{-0.18in}
  \caption{\peering: Community strings (at AMS) on catchments for AMS, BOS, CNF on 2020-02-25.}
          \label{fig:community-2020-02-26}
  \end{minipage}\hfill
  \begin{minipage}{0.3\textwidth}
\begin{center}
\includegraphics[width=\textwidth]{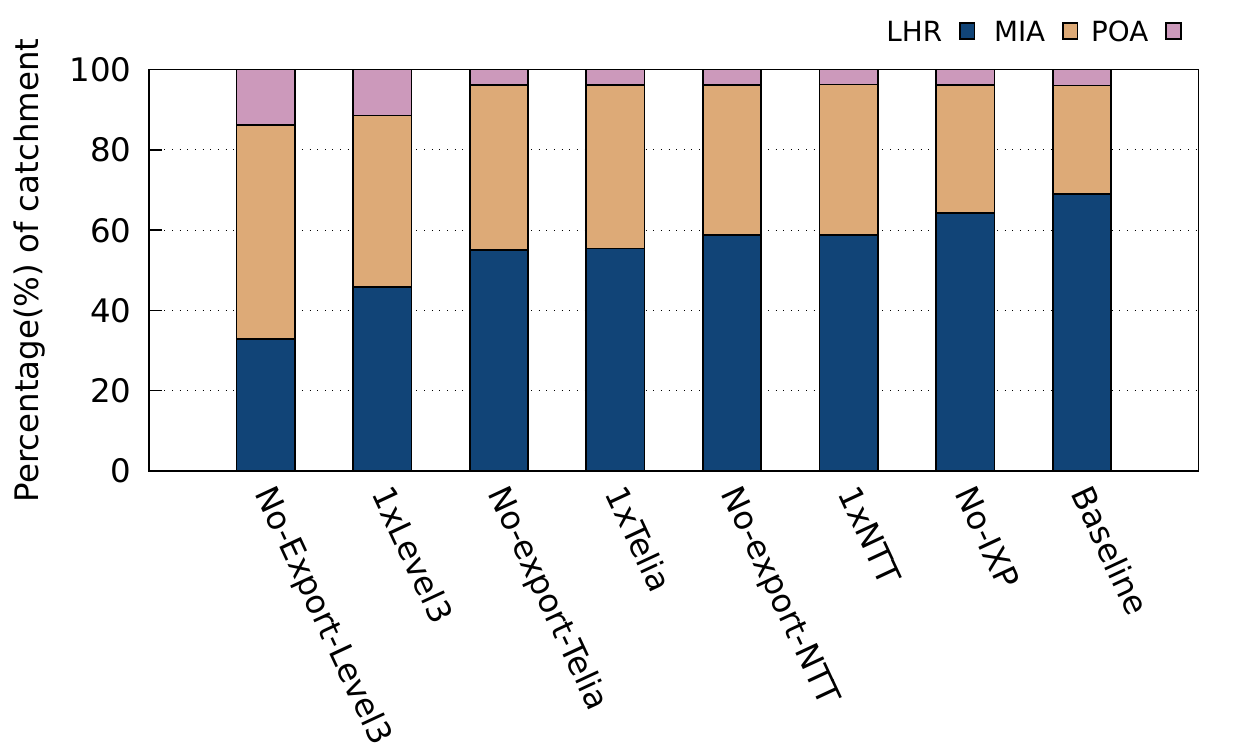}
\end{center}

     \vspace{-0.18in}
\caption{\tangled: using different communities to shift
    traffic on site LHR on 2020-04-05.}
        \label{fig:community-tangled}
  \end{minipage}
\end{figure*}


In summary, the experiment shows that AS prepend does work and can
shift traffic among sites, however, this traffic shift is not uniform.

%
%

\subsubsection{What granularity does prepending provide?}
\label{sub:granularity}
Having established that prepending can shift traffic,
  we next ask: how much control does it provide?
This question has two facets:
  how much traffic can we push away from a site or attract to it,
  and how many different levels are there between minimum and maximum.

\textbf{Limits:}
\autoref{fig:ams-bos-cnf-only-site-2020-02-24} suggested that
  in \peering, with those three sites,
  there is a limit to the traffic that can shift.
AMS, BOS, and CNF always get about 16\%, 7\% and 3\% of blocks,
  regardless of prepending.

\autoref{fig:selfprepending} confirms this result with a 5-site deployment (two from Europe, one from North America, one from South America and one from Australia)
  in our other testbed (\tangled).
X axis is presented with the number of
  prepends applied to each site.  The number zero (0) represents the
  \texttt{baseline}, the positive numbers (1-5) are the number of
  prepending applied and the negative numbers represent negative
  prepends. As depicted, each site can capture at most 55--65\%
  of blocks, and can shed at most 95\% of blocks, even with up to 5
  prepends.
We can also see that we do not get a granular control
  as only three points are between the minimum and maximum.

We conclude that while prepending can be a useful tool to shift traffic,
  it provides relatively limited control.

\subsection{Control with BGP Communities}

We next show that 
  BGP community strings have the opposite trade-off:
  what options they support vary from site to site,
  but when available, they provide more granular control over traffic.
We use whatever community strings that can be supported at each site.
Specific values for the same concept often vary.

\subsubsection{Community string coverage}
	\label{sec:communities_coverage}

ASes must opt-in to exchange community strings with peers,
  as opposed to prepending's near-universal support
  (since AS paths are used for loop detection,
  prepending works unless it is explicitly filtered out).
Explicit support is required because communities are only a tagging mechanism;
  the actions they trigger are at the discretion of peering AS\@.
Prior work has studied the diverse options supported by community strings~\cite{giotsas2017inferring}.

To evaluate coverage, we review support for BGP communities in
  the testbeds we use.
The testbeds provide information about two dozen locations
  with diverse peers. Each one of these peers has been evaluated 
  about its support to this feature.

In \autoref{tab:testbed-comunity} we
  describe path prepending and poisoning support
  and what types of community strings are supported at each site.
We group communities by class:
  advertisement options (no-peer, no-export to customers, and no export to anyone),
  selective prepending,
  and peers and transits that support selective advertisement.
We also show the number of non-transit peers and transits.

\comment{modified paragraph. ---asmrizvi 2022-02-24}
\comment{I think the paragraph now goes into details and I've lost the point.
(For example: what does the topic sen mean?!?!)
And intable 4, why doesn't it show the numbers we have here?
---johnh 2022-02-24}
\comment{The topic sentence is the same that we had with slightly different wordings. Added the number in the table. Rewritten. ---asmrizvi 2022-02-24}

\peering allows selective announcement to
  the transits and peers at each site,
  although the number of peers and transits varies.
Many sites with one transit provide no alternatives.
We considered selective announcement options at AMS,
  with 854 peers (106 bilateral peers including 2 route servers with 748 peers), and 2 transit providers~\cite{schlinker19peering}.
CNF has one transit provider and 129 peers (with only 6 bilateral peers, other peers are connected through 2 route servers).
For our Verfploeter measurement,
  we consider the peers and route servers
  with bilateral BGP sessions.
A single peer covers a small fraction of the address space
  in our Verfploeter measurement.
For some peers,
  we observed no coverage at all
  which requires further investigation with the peers
  to confirm our observation.
Hence, all the selective announcement options
  do not make difference in the catchment distribution (see the catchment in AMS with 12 peers compared to the transit-1 in \autoref{fig:community-2020-02-26}).
The \emph{options} column of \autoref{tab:testbed-comunity} 
  summarizes these results,
  showing how many routing options we have using community strings.

We evaluate \tangled to provide a second deployment with different peers.
\tangled built its anycast network over cloud providers, crowd-sourced  transit
providers and IXPs. 
All transit providers and IXPs sites support communities as
described in \autoref{tab:testbed-comunity}.
With \tangled, the POA site has 250 peers
  and most of them support communities strings. 

\begin{table*}
  \small
\newcommand{\psnt}{$\vartriangle$}
\scalebox{0.84}{%
  \begin{tabularx}{\textwidth}{l|cccccccc|cccccccc}
\multicolumn{1}{r|}{\textbf{Site:}}  & \multicolumn{8}{c}{\textbf{\peering}} & \multicolumn{8}{c}{\textbf{\tangled}}  \\
\multicolumn{1}{l|}{\textbf{Routing policy}}         & AMS & BOS & CNF  & SEA  & ATH & ATL & SLC   & MSN  
            & MIA & LHR & IAD  & CDG  & LAX & ENS & SYD   & POA \\ 
\hline
  \rowcolor{gray!20}
AS-path prepend     & \ok & \ok & \ok  & \ok & \ok & \ok & \ok  & \ok    & \ok & \ok  & \ok  & \ok  & \ok  & \ok  & \ok  & \ok \\ 
no-peer              &  \ok &  -- &  \ok  & --  & --  & --  &  --  & --    & \ok & \ok  & --   & \ok  & --   & --   & \ok  & \ok  \\
no-export            & \psnt  & --  & --   & --  & \psnt  & --  &  --  & --    & \ok & \ok  & --   & \ok  & --   & --   & \ok  & \ok \\
no-client            & --  & --  & --   & --  & --  & --  &  --  & --    & \ok & --   & --   &  --  & --   & --   & --   & --  \\
\rowcolor{gray!20}
Selective 
prepend             & \ok  & \ok  & \ok   & \ok  & \ok  & \ok  &  \ok  & \ok    & \ok & \ok  & --   & \ok  & --   & --   & \ok  & \ok \\
\rowcolor{gray!20}
Selective
announcement    & \ok & \ok & \ok & \ok & \ok & \ok & \ok & \ok & \ok & \ok & - & \ok & - & - & \ok & \ok \\
Path poisoning      & \ok  & \ok  & \ok   & \ok  & \ok    & \ok   & \ok     & \ok    & \ok & --   & --   & --   & --   & --   & --   & \ok  \\
\rowcolor{gray!20}
\# non-transit peers       & 854  & 0  & 129   & 0  & 0    & 0    &  0    & 0    &   0 &  0   &  0   & 0    &  0   & 0    & 0    & 250   \\
\rowcolor{gray!20}
\# transits    & 2  & 1  & 1   & 1  & 1    & 1    & 1     & 1    &   1 &  1   & 1    & 1    &  1   & 1    & 1    & 2   \\
\# options     & 856 & 1  & 130   & 1  & 1    & 1    & 1     & 1    &   1 &  1   & 1    & 1    &  1   & 1    & 1    & 252   \\
\end{tabularx} }
\vspace{-1ex}
  \caption{Traffic engineering options on each testbed sites. \ok: supported, -: not supported, \psnt: not tested. \comment{modified. ---2022-02-24}}
  \label{tab:testbed-comunity}
  \vspace{-0.6em}
\end{table*}

\comment{modified paragraph. ---asmrizvi 2022-02-24}
We conclude that the number of options at each anycast
  site may vary depending on the number of connections
  with peers and transits.
This uncertainty shows the need for a playbook
  that shows the possible options.

%

\subsubsection{At what granularity do community strings work?}
\label{sec:granu}
We next examine how well community strings work
  and what granularity of control they provide.
We use community strings to make BGP selective announcements,
  where we propagate our route only to 
  specific transit providers or IXP peers.

For our experiment, we use \peering,
  varying announcements at  AMS
  and observing traffic when anycast is provided from AMS, BOS and CNF
  (the same topology as \autoref{sec:prepend_works}).
\comment{modified line. ---asmrizvi 2022-02-25}
As described in \autoref{sec:communities_coverage}
  selective announcement community strings are provided only at AMS and CNF,
  and they affect our Verfploeter measurement only at AMS
  with several peers together, two transits one by one, and route servers.

To select the target ASes for selective announcement,
  we sort all the working peers of AMS site, based on the size of their customer cone
  using CAIDA's AS rank list~\cite{caida_rank}.
We then choose the 6 largest IXP peers and the 12 largest,
  as the left two bars in 
  \autoref{fig:community-2020-02-26}.
We then examine the route server, announced separately 
  (the next bar),
  and then all IXP peers including route servers.
Finally, we see the coverage with each of the two transit providers,
  announced separately.

First, we see that selective announcement provides 
  more control than prepending,
  as AMS shifts from baseline
   68\% of blocks
   to other configurations from 53 to 6\% of blocks.

Second, we see that there is some overlap in some combinations.
For example, each transit reaches more than half of all blocks
  reachable from AMS, so we know some blocks are reachable from
  both transit providers.
Thus, while there is some control over how many blocks to route to AMS,
  some peers are very ``strong'' and will pick up many blocks 
  if they are allowed to announce our prefix.

Third, we see the important role of route servers.
While direct coordination with 12 IXP peers
  brings only 7\% blocks at AMS,
  a route server lets AMS reach more ASes and 14\% of the blocks alone.

Finally, we see that transit providers play an important role.
AMS site has two transit providers---BIT BV (AS12859) 
  and Netwerkvereniging Coloclue (AS8283). 
Announcing to AS8283 attracts more traffic to AMS
  than announcing to AS12859.
Different AS relationship of these two transits with their upstream
  provides us a different traffic distribution.

As shown in our experiments, when compared to AS path prepending, 
  BGP communities provide way more better control over traffic distribution.

To investigate if the results found on \peering can be generalized, we
made a set of experiments on \tangled. Like \peering, we select 3
sites from three continents---London(LHR), Miami (MIA) and Porto
Alegre (POA), and use communities for selective prepending and
selective announcement from LHR. In \autoref{fig:community-tangled}, we show
the catchment distribution after using the community strings from LHR\@.
In the baseline, when no communities are used, LHR handles 69\% of
traffic.  From right to left, we see a gradual decrease in the
catchment distribution from 69\% to 33\%.  Stop announcing to IXP
peers reduces traffic from 69\% to 64\%.  But using prepending and no
export communities in AS2914 (NTT America), AS1299 (Telia Company) and
AS3356 (Level 3), we can get 30-60\% of the catchments in LHR.

Both testbeds show that community strings
  are not widely available in all sites, 
  and that even well-known
  communities are not fully adopted.
However,
  community strings can provide finer-grained control.
Selective announcement mostly provides more ``flexibility''
  depending on how many IXP peers and transits are connected.
We also find that
  some sites do not provide the support that we expect
  which means community strings require an extra step
  like contacting the transit provider for an explicit agreement.

\subsection{Control with Path Poisoning}
    \label{sec:poisoning}

We next turn to path poisoning,
  and show that like community strings,
  coverage and granularity are
  limited by routing filters deployed in upstream peers.

\subsubsection{Poisoning coverage}
    \label{sec:coverage-poisoning}

Support for path poisoning is dependent
  on the ASes we are poisoning and
  on route filters deployed by our upstream ASes.

We find that many ISPs, especially \tier ASes,
  filter out AS paths that poison \emph{any} \tier AS.
Tier-1 ASes deploy these filters to block BGP announcements
  from customers that contain other \tier ASes
  in the path to prevent route leaks~\cite{snijders2016practical, mcdaniel2020flexsealing}.
This filtering often makes path poisoning ineffective to control traffic.

To verify that poisoning \tier ASes is often ineffective from filtering,
  we poison \tier ASes announcing only from AMS in \peering,
  a unicast set-up blocking the impacts of other sites, and 
  make traceroutes from 1000 RIPE vantage points to our prefix.
Our measurement shows the evidence of filters
  when we poison \tier ASes---AS7018 (AT\&T), AS6453 (Tata Communications America), and AS1299 (Telia Company).
We observe many vantage points fail to reach our prefix as
  they are dependent on \tier ASes for their routes.
Some others change their paths avoiding \tier ASes.
We also validate route disappearance via most \tier ASes
  using RouteViews telescopes~\cite{routeviewsproject}.

Although poisoning \tier ASes is often ineffective,
  poisoning is effective with most non-\tier ASes.
Unfortunately, these ASes carry little traffic
   when they are not immediate upstreams.
Poisoning these small ASes only has little impact on traffic.
We again traceroute after poisoning a non-\tier AS (AS57866), and
  observe that \tier ASes propagate the poisoned path.
This proves poisoned paths with \tier and non-\tier
  ASes are treated differently by other ASes.

\subsubsection{What granularity does poisoning provide?}
    \label{sec:granularity-poison}

Path poisoning coverage is limited because one cannot usually poison a Tier-1 AS.
This same filtering limits the granularity that poisoning allows:
  poisoning Tier-1 ASes is not allowed,
  poisoning non Tier-1 ASes has little impact when they are multiple hops away
    because they represent little traffic.
Poisoning immediate neighbors may shift traffic,
  but is more complex than just not announcing to them.
We confirm these observations with detailed experiments in \autoref{sec:granularity-poison-detail},
  but we conclude that path poisoning is not generally an effective tool for traffic engineering.

\subsection{Playbook Construction}
    \label{sec:playbook}

\colorlet{green}{green!40}
\colorlet{red}{red!90}

\begin{table}
\begin{center}
{\footnotesize
\begin{tabular}{l|lll|lll}

	& \multicolumn{3}{c}{{\centering}{\bf Traffic to Site (\%) }}  \\
\textbf{Routing Policy} 
	& \bf AMS  & \bf BOS & \bf CNF	\\
\toprule
 \rowcolor{green!60}
  (a) 6peers, 12peers 		 & $\sim$5 	    & \cellcolor{green!30}$\sim$35 & \cellcolor{green!10} $\sim$55    \\
 \rowcolor{green!50}
  (b) Route-server       & 15 	    &\cellcolor{green!30}35 		& \cellcolor{green!10}55     \\
 \rowcolor{green!50}
  (c) All-IXP-Peers/Poison transits       & 15 	    & \cellcolor{green!30}35 		& \cellcolor{green!20}45     \\
\hline
 \rowcolor{green!50}
  (d) 3xPrepend AMS         & 15 	    &\cellcolor{green!30}35 		& \cellcolor{green!20}45     \\
 \rowcolor{green!40}
  (e) 2xPrepend AMS         & 25 	    & \cellcolor{green!30}35 		& \cellcolor{green!20}45     \\
 \rowcolor{green!30}
  (f) 1xPrepend AMS         & 35 	    & \cellcolor{green!40}25 		& \cellcolor{green!30}35 \\
\hline
 \rowcolor{green!40}
  (g) -3xPrepend BOS       & 25 	    & \cellcolor{white!0}65 		& \cellcolor{green!60}5     \\
 \rowcolor{green!30}
  (h) -2xPrepend BOS       & 35 	    &  \cellcolor{white!0}65 		& \cellcolor{green!60}5     \\
 \rowcolor{green!20}
  (i) -1xPrepend BOS       & 45 	    & \cellcolor{green!20}45 		&  \cellcolor{green!50}15     \\
\hline
 \rowcolor{green!30}
  (j) -3xPrepend CNF       & 25 	    &  \cellcolor{green!50}15 		& \cellcolor{white!0}65     \\
 \rowcolor{green!30}
  (k) -2xPrepend CNF       & 35 	    &  \cellcolor{green!60}5 		&  \cellcolor{green!10}55     \\
 \rowcolor{green!20}
  (l) -1xPrepend CNF       & 45 	    & \cellcolor{green!60}5 		& \cellcolor{green!20}45    \\
  \hline
 \rowcolor{green!20}
  (m)Transit-1       & 45 	    & \cellcolor{green!40}25 		&  \cellcolor{green!30}35     \\  
 \rowcolor{green!10}
  (n) Transit-2       & 55 	    &  \cellcolor{green!50}15 		&  \cellcolor{green!40}25     \\
\hline
 \rowcolor{green!30}
  (o) Poison Tier-1/Transit-2       & 35 	    &  \cellcolor{green!40}25 		&  \cellcolor{green!30}35     \\
  \rowcolor{green!10}
  (p) Poison Transit-1       & 55 	    &  \cellcolor{green!40}25 		&  \cellcolor{green!40}25     \\
\hline
  \textbf{(q) Baseline} 	            & 65 		&  \cellcolor{green!50}15 		&  \cellcolor{green!50}15 		\\
\hline
  (r) 1,2xPrepend BOS         & 65 	    & \cellcolor{green!60}5 		&  \cellcolor{green!40}25     \\
 \rowcolor{red!10}
  (s) 3xPrepend BOS         & 75 	    &  \cellcolor{green!60}5 		& \cellcolor{green!40}25     \\
\hline
 \rowcolor{red!10}
  (t) 1,2,3xPrepend CNF         & 75  	&  \cellcolor{green!50}15 	    &  \cellcolor{green!60}5  	\\
\hline
 \rowcolor{red!20}
  (u) -1,-2,-3xPrepend AMS       & 85 	    &  \cellcolor{green!60}5 		&  \cellcolor{green!60}5     \\

\end{tabular}
} 
\end{center}
    \vspace{-0.11in}
    \caption{Policies and traffic distribution (in 10\% bins); groups sorted by rough fraction of traffic to AMS, and colors showing the traffic compared to the baseline distribution.}
 \label{tab:policies_and_distribution}
    \vspace{-0.12in}
\end{table}

%

Based on our understanding of prepending, communities and poisoning,
  we can now build a playbook of
  possible traffic configurations for this anycast network.
\reviewfix{SS22-A4, SS22-C5, SS22-E4}
In practice, we build the playbook automatically using scripts
  that connect to BGP,
  then iterate through different BGP configurations,
  then run Verfploeter~\cite{Vries17b} to measure new catchments.
\reviewfix{SS22-A1}
Playbooks are necessarily specific to each anycast deployment,
  but we show in \autoref{sec:constraints_anycast}
  that the process generalizes.
\reviewfix{SS22-B1, SS22-B6, SS22-B7}
Using a playbook, an operator does not need a single ``best'' approach,
  rather a combination of approaches in the playbook
  ensures a greater control over traffic distribution.

A playbook is a list of variations of routing policy
  and the resulting traffic distributions.
\autoref{tab:policies_and_distribution} shows the playbook for our testbed,
  with the baseline of 65\% blocks to a site shown in white.
We group different levels of prepending (positive or negative) at each site,
  and show selected community string and poisoning configurations.

To summarize the many configurations from
  \autoref{tab:policies_and_distribution},
  \autoref{tab:granular-ams-bos-cnf} identifies which combinations
  result in specific traffic ratios at each site.
Each letter in this table
  refers back to a specific configuration from \autoref{tab:policies_and_distribution}.
During an attack, if the anycast system begins at
  the baseline configuration (q),
   if AMS is overloaded, the operator could select a TE configuration
   higher in the table (perhaps `e', `g', or `j').
The operator can then see the implications of that TE choice
  on other site (for example, `e' increases load on both other sites,
    with `g' increases load on BOS but decreases it at CNF). 

\begin{table}
\begin{center}
{\footnotesize
\begin{tabular}{l|lll}

\textbf{Traffic to Site (\%)}       & \multicolumn{1}{c}{\textbf{AMS}} & \multicolumn{1}{c}{\textbf{BOS}} & \multicolumn{1}{c}{\textbf{CNF }} \\ \hline
0-10                  & a                & \begin{tabular}[c]{@{}l@{}} k, l, r, s, u \end{tabular}        & \begin{tabular}[c]{@{}l@{}} g, h, t, u \end{tabular}        \\ 
10-20                 &    \begin{tabular}[c]{@{}l@{}}b, c, d          \end{tabular} & \begin{tabular}[c]{@{}l@{}}j, n, q, t   \end{tabular}             & i, q                                \\ 
20-30                 & e, g, j &\begin{tabular}[c]{@{}l@{}} f, m, o, p     \end{tabular}   &  n, r, p, s\\

30-40                 & f, h, k, o       & \begin{tabular}[c]{@{}l@{}}a, b, c, d, e  \end{tabular}     &  f, m, o                \\ 
40-50                 & \begin{tabular}[c]{@{}l@{}}i, l, m \end{tabular}    &   i        & \begin{tabular}[c]{@{}l@{}} c, d, e, l \end{tabular} \\ 
50-60                 &  n, p                               & --                               & \begin{tabular}[c]{@{}l@{}} a, b, k \end{tabular} \\ 
60-70                 & q, r                       & g, h                                & j                                               \\ 
70-80                 &  s, t      & --                               & -- \\ 
80-90                 & \begin{tabular}[c]{@{}l@{}} u  \end{tabular}         & --                               & --                               \\ 
90-100                &--                              & --                               & --                               \\ \hline
Traffic options    & 9                               & 6                                & 7                                \\ 
\end{tabular}
}
\end{center}
     \vspace{-0.11in}
\caption{ \peering playbook (AMS, BOS, and CNF)}
	\label{tab:granular-ams-bos-cnf}
      \vspace{-0.15in}
\end{table}

\reviewfix{SS22-F2 + SS22-A7}
An operator may also use
  a playbook with traffic load for two reasons.
First, loads in most interesting services have diurnal pattern.
Second, loads from each /24 prefix may vary because of
  the number of clients behind each prefix (more on \autoref{sec:appendix_playbook_load}).
Building the playbook with load is computationally simple;
  an operator can just use the same catchment mapping
  along with the per prefix load. 

Even with attack size estimation,
  attacks are accompanied by uncertainty,
  and attacker locations may be uneven.
However, the playbook provides a much better response than ``just relying on informal prior experience'' in two ways:
  the defender can anticipate the consequences of the TE action
    (that traffic will go somewhere!),
  and the defender can choose between different possible outcomes
    if the first is incomplete.


\textbf{Playbook flexibility and completeness:}
\autoref{tab:granular-ams-bos-cnf}
  helps quantify the ``flexibility'' that
traffic engineering allows us in this anycast deployment.
Using these 10\% traffic bins, we see that AMS has 9 options,
CNF 7, and BOS only 6.  Because AMS and CNF
mostly swap traffic after TE changes,
and because BOS is less well connected,
no configuration with three sites allows BOS to take traffic
within 50-60\% range, and no 3-site configuration can drive BOS or CNF over 70\%.

This analysis shows more central sites like AMS,
  and it may suggest the need for topology changes
  (perhaps adding another site in Europe or Asia to share AMS' load).

%
\section{Deployment Stability and Constraints}
	\label{sec:constraints_anycast}

\begin{figure*}
        \begin{subfigure}[t]{0.33\textwidth}
                \centering
                \includegraphics[width=.98\linewidth]{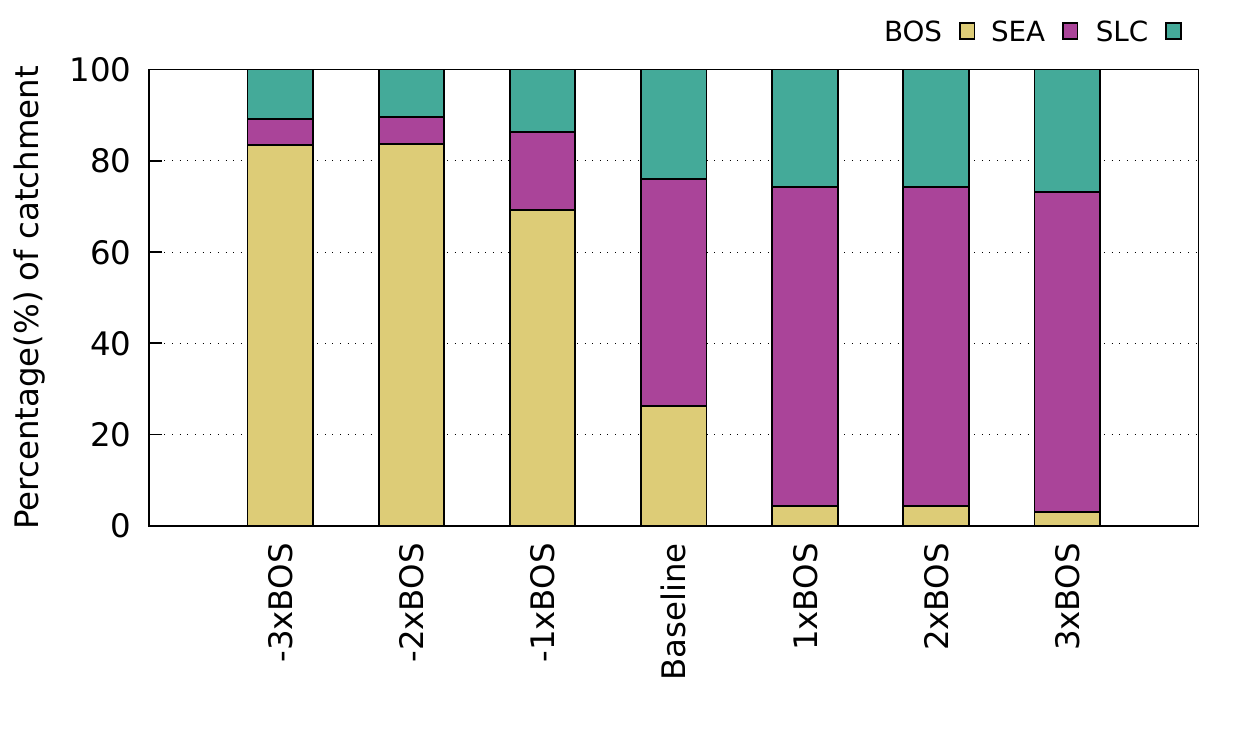}
                \vspace{-0.15in} 
                \caption{BOS site.}
                \label{fig:bos-2020-02-28}
        \end{subfigure}
	\begin{subfigure}[t]{0.33\textwidth}
                \centering 
                \includegraphics[width=.98\linewidth]{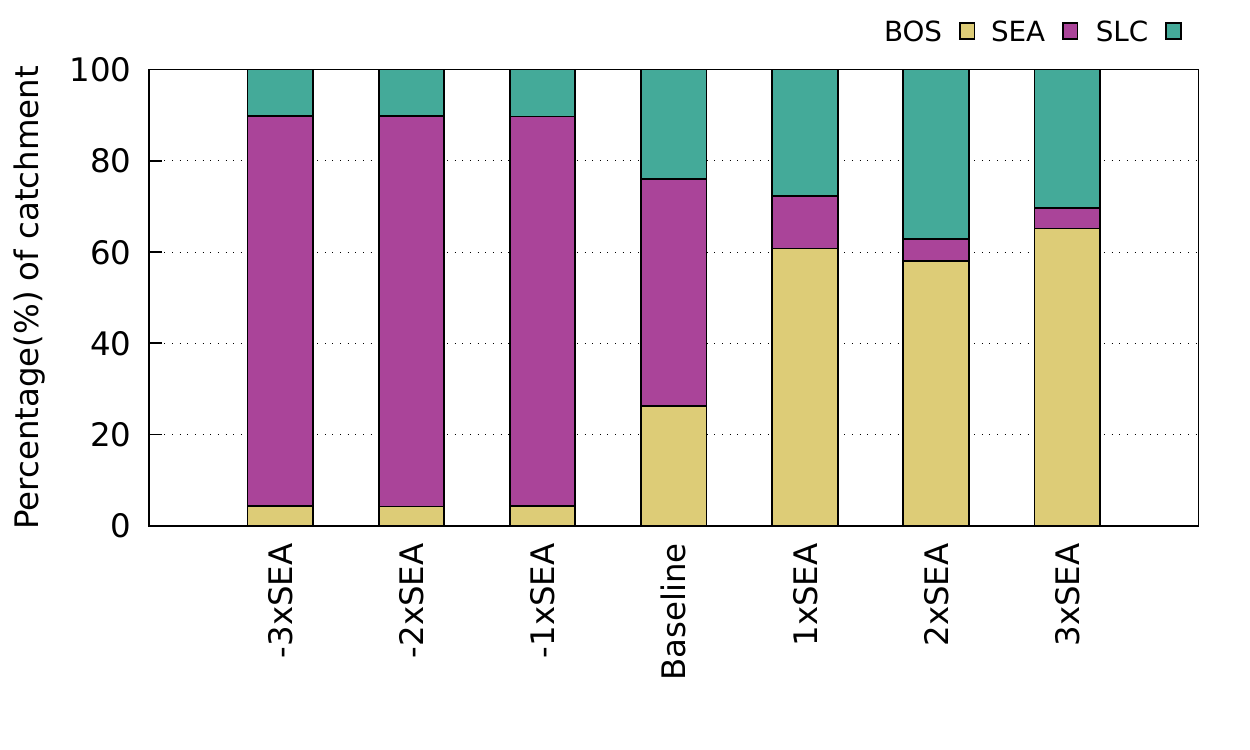}
                \vspace{-0.15in} 
                \caption{SEA site.}
                \label{fig:sea-2020-02-28}
        \end{subfigure}
        \begin{subfigure}[t]{0.33\textwidth}
                \centering
                \includegraphics[width=.98\linewidth]{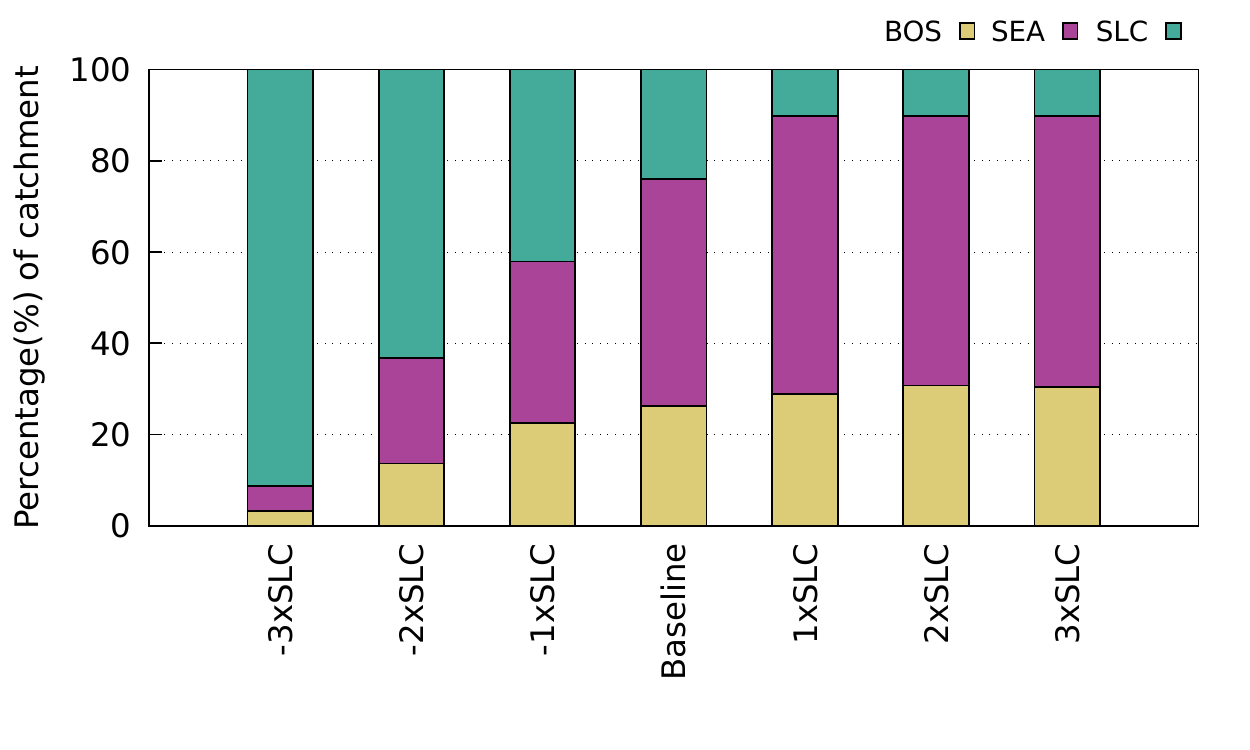}
                \vspace{-0.15in} 
                \caption{SLC site.}
                \label{fig:slc-2020-02-28}
        \end{subfigure}
       \vspace{-0.17in} 
       \caption{\peering: Impact of choosing BOS, SEA and SLC sites on 2020-02-28}
       \label{fig:sea-slc-bos-2020-02-28}
      \vspace{-0.15in}
\end{figure*}

In \autoref{sec:exploit_catchment}  we showed BGP-based TE provides
considerable flexibility.
\reviewfix{SS22-B2, SS22-B6, SS22-B7}
Building playbooks supports defenders
  by allowing them to explore how 
  transit providers, prepending, community strings, and poisoning
  affect their specific deployment.
\reviewfix{SS22-G1}
We next look at how stable the results are
  depending on choice of sites and the number of sites.
While the details of the playbook vary for each deployment,
  and we do not claim our testbeds represent all possible deployments,
  we show our approach is flexible and can respond to attacks in different deployments---our approach generalizes.


\subsection{Effects of Choice of Anycast Sites}
	\label{sec:location_effects}

First we see how sites affect our playbook.
New sites change catchments because they depend on location and peering,

In \autoref{sec:path_prepending}, we studied catchments with three specific
   \peering sites on three continents: 
   AMS, at a large, commercial IXP in Europe;
  CNF with an academic backbone transit in Brazil; and BOS, an academic site in the U.S\@.
We now switch to three educational sites all in the United States:
  SEA, at University of Washington on the west coast;
  SLC, at the University of Utah in the Rockies;
  and BOS, at
  Northeastern University in Boston on the east coast.

More important than just geographic location, site connectivity is
the most important factor in choosing sites. Multiple transit
providers increase the chance of having more BGP options to affect
traffic control and granularity. While a poorly connected site inside
a university network tends to provide less traffic control options.



\textbf{Prepending baseline:}
\autoref{fig:sea-slc-bos-2020-02-28}
  shows catchment sizes for the three North American sites
  with positive and negative prepending.
Now the baseline distribution is unbalanced, but less so than before,
  with SEA capturing 50\% of blocks.
We discussed SEA's heavy traffic with the \peering operators.
They suspect
  that SEA is near
  to the Seattle IXP,
  making its paths one hop from many commercial providers.
Which site has the greatest visibility depends on its peering
  and will vary from deployment to deployment.


\textbf{Prepending coverage and granularity:}
As with our prior experiments,
  we can adjust prepending to see how traffic shifts.
With these three sites,
  traffic shifts very quickly
  for BOS and SEA after one positive or negative prepend.
SLC has more flexibility, perhaps because it has the smallest
  catchment at the baseline,
  and gains more coverage with each step of negative prepending,
  to 42\%, 63\%, and 91\% of blocks.
Often (but not always), we see that academic sites exhibit less granularity because
  either they have few peers, 
  or their peers are academic networks with similar connectivity.
As a result,
  minor changes in AS-Path length place one site further from the others.
In addition, this less granular control shows the importance
  of building a playbook that is specific to a given deployment,
  or when the anycast topology changes.

\textbf{Community coverage:} 
While communities are common at IXPs and transit providers,
  academic networks (NRENs) have a more simple set of communities.
None of those academic sites provide community strings.

This observation confirms our prior coverage observation:
  community string support is not uniformly available.
We also looked at other combinations of sites in \peering
   and found similar results (\ifisarxiv\autoref{sec:small-eu} and \autoref{sec:nearby-sites}\else in the extended version of the paper~\cite{rizvi2020anycast}\fi).

\textbf{Path poisoning:}
We repeated our path poisoning experiments with three sites
  in Boston, Salt Lake City and Seattle.
We confirm that \tier ASes typically
  cannot be poisoned  (\autoref{sec:coverage-poisoning}).
We also see filters designed to prevent route leaks~\cite{snijders2016practical}  also interfere with poisoning.

Our experiments confirm that
  while catchments are deployment-specific,
  our qualitative results hold---prepending works but is coarse,
  and
  community strings and poisoning are not supported everywhere.

\subsection{Effects of Number of Anycast Sites}
	\label{sec:effects_no_sites}

\begin{figure}
\centering
\includegraphics[width=0.8\linewidth]{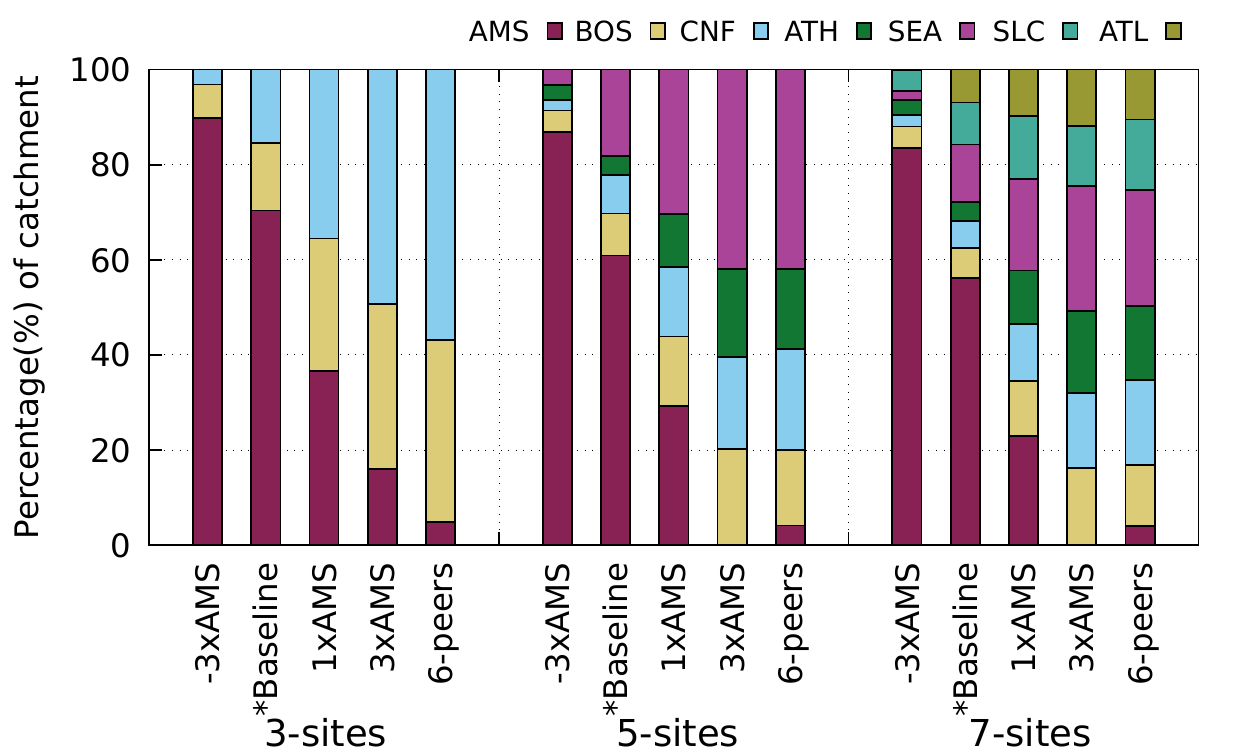}
1\vspace{-0.11in}
    \caption{\peering: Impacts of changing the number of anycast sites from 2020-04-07 to 2020-04-10.}
\label{fig:number_sites_peering}
\vspace{-0.2in}
\end{figure}

Next, we vary the \emph{number} of sites 
  and see how that changes control traffic.
We select 3, 5 and 7 sites from each testbed,
  and build a playbook to evaluate defense options.
\autoref{fig:number_sites_peering}
  shows selected configurations,
  grouped by number of sites.

\textbf{Baseline:}
With more sites, overall capacity increases
  and \emph{baseline load at each site falls}.
For example, in \autoref{fig:number_sites_peering},
  the baselines (with an asterisk*) at
  the largest site (AMS) shifts from 70\% of blocks with three sites
  to 61\% and 56\% with 5 and 7 sites.
Smaller sites shift less (BOS goes from 14\% to 6\% and 6\%,
  and CNF from 15\% to 8\% and 6\%).
Greater capacity and  distribution requires a larger and distributed attacker
  to exhaust the overall service.
We see similar results on our alternate testbed \tangled \ifisarxiv (\autoref{sec:more-tangled})\else,
  as described in the extended version of the paper~\cite{rizvi2020anycast}\fi.
\PostSubmission{IAD is not in Fig 13. I added some text in the appendix to somewhat solve this. ---asmrizvi 2020-06-02}


%

\textbf{Traffic flexibility:}
%
With more sites, \emph{the largest site usually shows the largest changes}
  and has the fewest catchment sizes.
Comparing  baseline to one prepending in  \autoref{fig:number_sites_peering},
  AMS shifts from 70\% to 37\% with three sites,
  from 61\% to 29\% with five,
  and from 56\% to 23\% with seven,
  always dropping by half.

Even with more sites,
  some blocks are often ``stuck'' at a particular site.
With three negative prepends, AMS gets most of the traffic,
  but it tops out at 90\% with three sites, and only 87\% and 84\% with five and seven.
We conclude that each site has its own set of ``stuck blocks'' that are captive to it
  and will not move with traffic engineering.

With more sites,
  the \emph{fine control of BGP communities becomes more important}
  because path-prepending becomes less sensitive.
For example, selective announcements with communities are need for
  AMS with 5 or 7 sites; prepending three times shifts all traffic.

\textbf{New sites:}
Adding more sites also shows how our playbook can help guide deployment of new sites.
Predicting traffic shifts for a new site is difficult,
  but experimenting with a test prefix can build a playbook pre-deployment.

\subsection{Playbook Stability Over Time}
	\label{sec:playbook_stability}

\reviewfix{SS22-A3}

A playbook has a limited use if routing changes immediately.  We know
routing changes when links fail, or when ISPs begin new peering or
purchase new transit. For how long is a playbook applicable?


\begin{table}
\centering
{\small 
\begin{tabular}{llll}
    \textbf{Months}  & \textbf{AMS(\%)} & \textbf{BOS(\%)} & \textbf{CNF(\%)} \\
  \hline
	2020-02  & 68.1 & 14.6 & 17.3\\ 
	2020-04  & 70.4  & 14.2 & 15.4 \\
	2020-06  & 65.3  & 14.1 & 20.6\\ 
\end{tabular}
}
\caption{Percent blocks in each catchment over time.}
\label{tab:catchment-time}
\vspace{-0.2in}
\end{table}

To answer this question,
  \autoref{tab:catchment-time}
  shows the fraction of /24 blocks going to each catchment over time for the baseline configuration.
We see that the fraction of blocks is generally quite stable,
  with only about 5\% of blocks shifting in or out of a site.
In addition, prior work has shown very strong anycast stability
  over hours to days~\cite{Levine06a,Wei17b}.
\reviewfix{SS22-F2, SS22-A2, SS22-A7, SS22-C4}
We checked the stability of \broot catchment.
We found that after two weeks 0.35\% prefixes,
  and after one month only 0.65\% prefixes
  changed their catchment (more on \autoref{sec:appendix_stability}).
While catchments are relatively stable,
  we expect operators will refresh playbooks periodically (perhaps weekly or monthly).


\section{Defenses at Work}
	\label{sec:fight_ddos}

\reviewfix{SS22-F1, SS22-F3, SS22-A9, SS22-B4, SS22-B5, SS22-B8, SS22-C3, SS22-D3, SS22-E1, SS22-E3}

In this section we describe four real-world attacks
  processing the traffic in our system.
We show that we can successfully respond to a different types of attacks in different ways.  

\textbf{Methodology:}
We use real-world attacks from 
  \broot server operator,
  \DutchScrubbingCenter,
  and from an anonymized enterprise network.
These events include polymorphic, adversarial, 
  and a volumetric attack.

We evaluate these events by
  simulating traffic rates against a three-site anycast network.
The first two events use \peering with our  
   AMS, BOS, CNF configuration from \autoref{sec:exploit_catchment}.
We vary this topology, using BOS, SEA, SLC from \autoref{sec:location_effects}
  in the last event.
We replay the traffic in simulation,
  assigning traffic to each anycast site
  based on catchments measured in our experiments.
We do not simulate the gradual route propagation,
  but instead have routing take effect 300\,s after a change
  (a conservative bound, most routing changes happen in half that time).
We then evaluate traffic levels at each site
  and compare that to 
  a target capacity.

\begin{figure*}
        \begin{subfigure}[t]{0.33\textwidth}
                \centering
                \includegraphics[width=.98\linewidth]{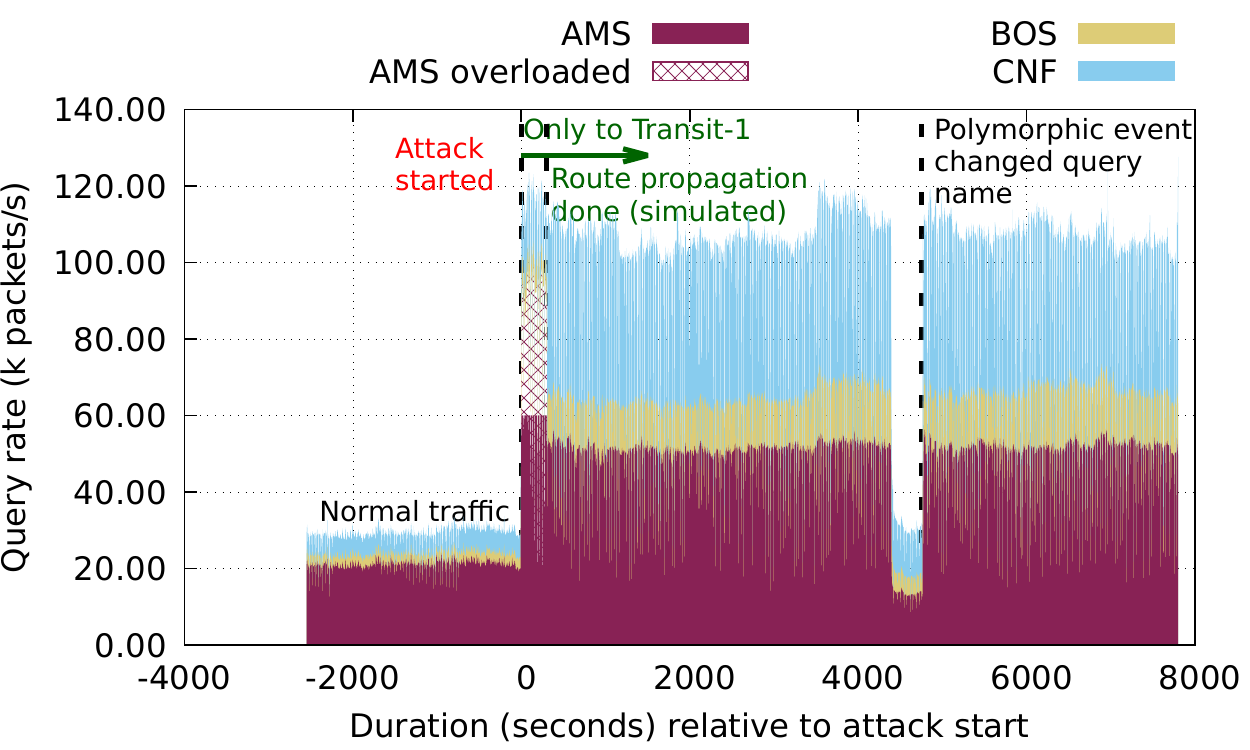}
                \caption{A polymorphic attack at \broot defended with community strings.}
                \label{fig:ddos-transit-1}
        \end{subfigure}
	\begin{subfigure}[t]{0.33\textwidth}
                \centering 
                \includegraphics[width=.98\linewidth]{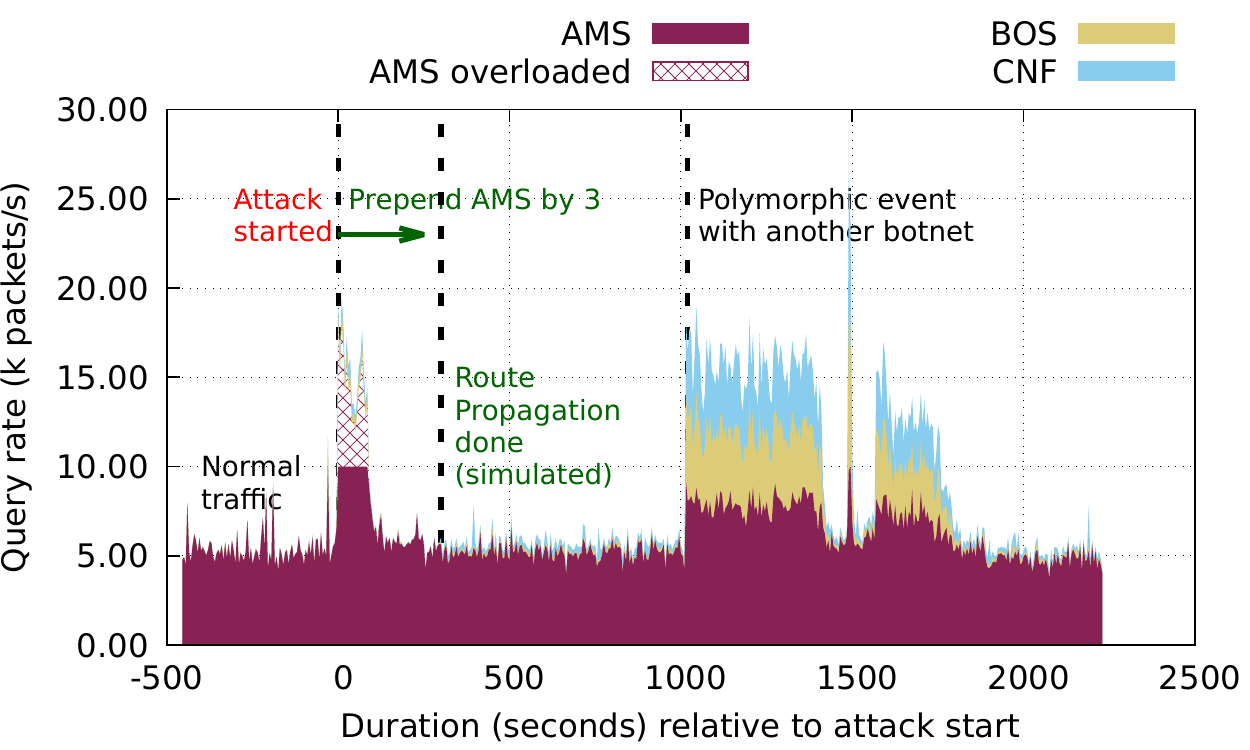}
                \caption{An  adversarial event at an enterprise mitigated using positive prepending.}
                \label{fig:ddos-ams-3}
        \end{subfigure}
        \begin{subfigure}[t]{0.33\textwidth}
                \centering
                \includegraphics[width=.98\linewidth]{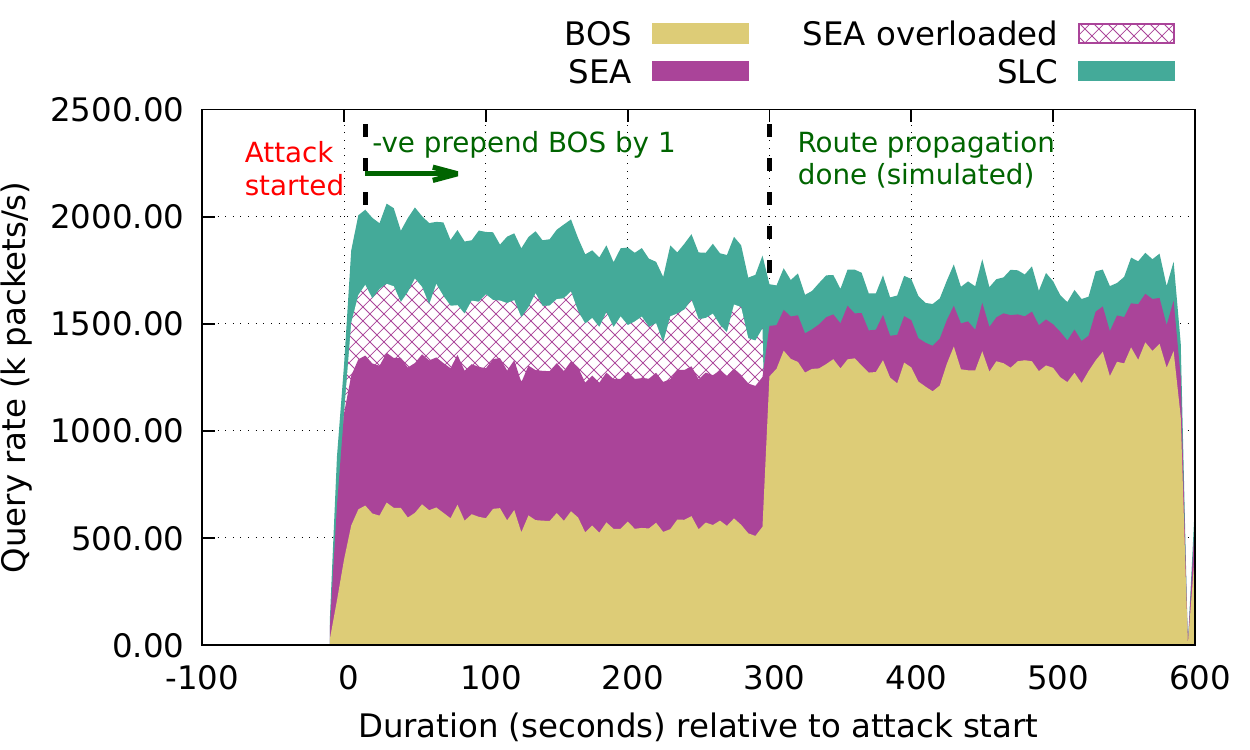}
                \caption{An event captured at \DutchScrubbingCenter defended using negative prepending.}
                \label{fig:ddos-super-site-bos}
        \end{subfigure}
        \vspace{-0.12in}
        \caption{Different attacks with various responses.}
        \label{fig:ddos_4_events} 
      \vspace{-0.18in}
\end{figure*}

For each attack we run our system in defense, estimating
  the attack size and selecting a pre-computed playbook response.
Since our playbook allows different responses:
  when we have choices we select different methods of defense:
  prepending, negative prepending,
  or community strings (\autoref{fig:ddos_4_events}).

\textbf{A 2017 polymorphic attack:}
Our first event 
  is a DNS flood from 2017-03-06 in 
  \broot\ifisanon~\cite{20170306Anon}\else~\cite{20170306}\fi\@
  (\autoref{fig:ddos-transit-1}).
This event was a volumetric polymorphic attack where the
  attack queries have common formats like
  {\it RANDOM.qycl520.com\textbackslash032} (from 0\,s) and 
  {\it RANDOM.cailing168.com\textbackslash032\textbackslash032} (changed at 4750\,s, so polymorphic in nature).
We assume 60k packets/s (30\,Mb/s) capacity at each anycast site.
The event was small enough that \broot was able to fully capture it
  across all active anycast sites at the time.
The event lasted about 5 \,hours,
  but we show only the first 2.25\,hours.
Services and attacks capacity today will both be much larger; we use a small
  attack, scaling the attack and capacity up would show similar results.

In \autoref{fig:ddos-transit-1} we can identity 
AMS site receives 100k packets/s traffic that
  is more than the capacity (shown as the maroon striped area).
Our system notices the attack from bitrate alerts.
It then estimates the AMS overload by computing the offered load
  using observed load and access fraction.
The system maps networks to number of packets to each site 
  using the pre-computed playbook (\autoref{tab:granular-ams-bos-cnf}).
Using this mapping our system/operator can then select a response.
From \autoref{fig:ddos-transit-1}, we can see
  the impact of the selected routing approach---announcing
  only to Transit-1 using community string.
After 300\,s, we can see no striped area which
  indicates the attack is mitigated.

The attacker changes the query names at 4750\,s,
  making this attack polymorphic.
Filtering on query names would need to react, but
  our routing changes can still
  mitigate the attack regardless of this type of change.

\textbf{A 2021 variable-length polymorphic attack:}
We next examine an HTTP-attack launched on an 
  enterprise network on 2021-09-05
  in \autoref{fig:ddos-ams-3}.
This polymorphic attack changes after each of three pauses.
The initial attack consists of millions of HTTP GETs (15k packets/s) launched
from an IoT botnet;
  it terminates when the enterprise's operator deploys IP-based filtering.
About 1000\,s later, a different botnet launched a
multi-vector attack combining HTTP GETs using random paths (to avoid
caching) and spoofed TCP ACKs.
We then see a lull, brief burst,
 another lull, and a burst to the end.

The initial attack at time 0 overloads one site (AMS),
  prompting our routing response.
After the estimation, we begin a route shift away from AMS,
  but the attack ends quickly (after 90\,s),
  while routes are still changing.

Since the normal traffic sources originate from Europe,
  most traffic went to AMS even after three prepends.
At 1020\,s the attack botnet changes, with more attack traffic
  from Asia and South America (based on IP geolocation from MaxMind)
Our route changes in response to the initial attack are still in place,
  and the renewed attack is successfully spread over all three sites,
  allowing AMS to tolerate the new attack.

Shifting attacks like this are common
  with more sophisticated
  adversaries.
Any approach (including ours) that defends with routing changes
  is limited by route propagation times,
  so the applicability of such defenses is limited for short-lived attacks like what occurred at 0\,s.
However, spreading traffic protects against many types of attack,
  as we see the renewed attacks after 1000\,s.
Varying attacks like this show the importance of reviewing
  defense effectiveness as the attack continues.

\textbf{An example attack on a different anycast topology:}
Finally, we consider an LDAP amplification attack,
  at \DutchScrubbingCenter on 2021-08-25.

In this case we simulate a super-site at BOS, capable of absorbing 1500k packets/s,
  while the other sites (SEA and SLC) support about half (700k packets/s).
In \autoref{fig:ddos-super-site-bos}, 
  the purple cross-hatched area shows
  how much the traffic will overwhelm SEA, a smaller site,
  but
  can be handled at the super-site.
We respond with negative prepending,
  with the traffic shift to BOS visible at 300\,s.
This response mitigates the attack (no striped area).

\textbf{Other attacks:}
We have assessed additional attacks,
  and describe them 
  in \autoref{sec:appendix_more_events}.
The additional polymorphic and volumetric attacks
  show that routing can successfully address attacks
  after routes propagate.

\section{Limitations and Future Work}
\label{sec:limitation}

\reviewfix{SS22-F1, SS22-F3, SS22-F5, SS22-B1, SS22-B3, SS22-B4, SS22-B5, SS22-C4, SS22-C6, SS22-D2, SS22-D4, SS22-D12, SS22-D14}

Our playbook of routing options (\autoref{sec:methodololgy})
  is effective against many attacks (\autoref{sec:fight_ddos} and \autoref{sec:appendix_more_events}).
However, like any defense, it is not impervious.
We next describe known limitations and areas of future work.

First, Internet routing is distributed, requiring time to converge.
The effects of routing defenses cannot be seen
  until convergence.
We do not make changes faster than 5 minutes.

Routing convergence time
  implies that routing changes
  will have limited applicability to short-lived attacks (less than 5 minutes).
Although routing changes will not hurt the service, their benefits may not occur
  until routing shifts.

In addition, routing convergence means that
  polymorphic attacks that shift traffic sources quickly
  will be more effective.
Routing changes are robust to polymorphic attacks that change method but
  take effect by traffic volume, they will spread load regardless of what it is,
  as we show in events in \autoref{sec:fight_ddos}.
However, when defending an attack where traffic shifts locations
  faster than routing converges,
  one must provision for the worst case volume to any site under
  the heaviest traffic it sees.
Rapid shifts make defense harder, but not impossible.


Finally, we assume the anycast catchments of the underlying service
  change slowly (over days).
We showed in \autoref{sec:playbook_stability} that
  this assumption generally holds.

Although we change routing during an attack to balance load across catchments,
  we do not explicitly attempt to locate attack origins.
As future work, we could use such information to
  improve defense selection.

\reviewfix{SS22-C2}
Attack response depends on human factors in service operators
  and attackers.
Explicitly studying such human factors is potential future research.
Our current work focused on the technical feasibility of our defenses.

\section{Conclusions}

This paper provides the first public evaluation of
  multiple anycast methods for DDoS defense.
Our system estimates attack size,
  selects a strategy from a pre-computed playbook,
  and automatically performs traffic engineering (TE) to rebalance load
  or to advise the operator.
Our contributions are attack-size estimation and
  playbook construction.
We experimentally evaluate TE mechanisms,
  showing that prepending is widely available but offers limited control,
  while BGP communities and path poisoning are the opposite.

\ifisanon
\else

{\vspace*{1ex}
\small
\noindent \textbf{Acknowledgments:}
ASM Rizvi and John Heidemann's work on this paper is supported, in part, by 
 the DHS HSARPA Cyber Security Division via contract number HSHQDC-17-R-B0004-TTA.02-0006-I.  
Joao Ceron and Leandro Bertholdo's work on this paper is supported by 
Netherlands Organisation for scientific research (4019020199), and 
European Union’s Horizon 2020 
research and innovation program 
(830927).
We would like to thank our anonymous reviewers for their valuable feedback.
We are also grateful to the \peering and \tangled admins
  who allowed us to run measurements.
We thank Dutch National Scrubbing Center for sharing DDoS data with us.  
\par
}
\fi

\makeatletter
    \renewcommand\@openbib@code{%
      \setlength{\itemsep}{0mm}
      \parsep \z@
      }%
\makeatother

\bibliographystyle{plain}
{\small
\bibliography{ref,johnh} 

\begin{thebibliography}{10}

\bibitem{ampath}
AMPATH.
\newblock Bgp resources.
\newblock \url{https://ampath.net/AMPATH_BGP_Policies.php}.
\newblock [Online; accessed 12-Oct-2021].

\bibitem{APNIC20a}
APNIC.
\newblock {BGP}-stats routing table report---{Japan} view.
\newblock
  \url{https://mailman.apnic.net/mailing-lists/bgp-stats/archive/2020/05/msg00001.html},
  May 1 2020.

\bibitem{bajpai2015lessons}
Vaibhav Bajpai, Steffie~Jacob Eravuchira, and J{\"u}rgen Sch{\"o}nw{\"a}lder.
\newblock Lessons learned from using the ripe atlas platform for measurement
  research.
\newblock {\em ACM SIGCOMM Computer Communication Review}, 45(3):35--42, 2015.

\bibitem{ballani2006measurement}
Hitesh Ballani, Paul Francis, and Sylvia Ratnasamy.
\newblock A measurement-based deployment proposal for {IP} anycast.
\newblock In {\em Proceedings of the 6th ACM SIGCOMM conference on Internet
  measurement}, pages 231--244, 2006.

\bibitem{basat2016heavyhitters}
Ran Ben-Basat, Gil Einziger, Roy Friedman, and Yaron Kassner.
\newblock Heavy hitters in streams and sliding windows.
\newblock In {\em IEEE INFOCOM 2016 - The 35th Annual IEEE International
  Conference on Computer Communications}, pages 1--9, 2016.

\bibitem{deter06}
Terry Benzel, Robert Braden, Dongho Kim, Cliford Neuman, Anthony Joseph, Keith
  Sklower, Ron Ostrenga, and Stephen Schwab.
\newblock Experience with deter: a testbed for security research.
\newblock In {\em 2nd International Conference on Testbeds and Research
  Infrastructures for the Development of Networks and Communities, 2006.
  TRIDENTCOM 2006.}, pages 10--pp. IEEE, 2006.

\bibitem{le2020tangled}
Leandro~M. Bertholdo, João~M. Ceron, Wouter B.~de Vries, Ricardo de~Oliveira
  Schmidt, Lisandro~Zambenedetti Granville, Roland~van Rijswijk-Deij, and Aiko
  Pras.
\newblock Tangled: A cooperative anycast testbed.
\newblock In {\em 2021 IFIP/IEEE International Symposium on Integrated Network
  Management (IM)}, pages 766--771, 2021.

\bibitem{Caesar05a}
Matthew Caesar and Jennifer Rexford.
\newblock {BGP} routing policies in {ISP} networks.
\newblock {\em {IEEE} Network Magazine}, 19(6):5--11, November 2005.

\bibitem{caida_rank}
CAIDA.
\newblock {AS} rank.
\newblock \url{https://asrank.caida.org/}, 2020.
\newblock [Online; accessed 12-Oct-2021].

\bibitem{caida-communities}
CAIDA.
\newblock {CAIDA UCSD BGP} community dictionary.
\newblock \url{https://www.caida.org/data/bgp-communities/}, 2020.
\newblock [Online; accessed 12-Oct-2021].

\bibitem{calder2015analyzing}
Matt Calder, Ashley Flavel, Ethan Katz-Bassett, Ratul Mahajan, and Jitendra
  Padhye.
\newblock Analyzing the performance of an anycast {CDN}.
\newblock In {\em Proceedings of the 2015 Internet Measurement Conference},
  pages 531--537, 2015.

\bibitem{carney2015method}
Mark~D Carney, Jeffrey~A Jackson, Andrew~L Bates, and Dante~J Pacella.
\newblock Method and apparatus for mitigating distributed denial of service
  attacks, November~24 2015.
\newblock US Patent 9,197,666.

\bibitem{rfc1997}
R.~Chandra, P.~Traina, and T.~Li.
\newblock {BGP} communities attribute.
\newblock Technical Report 1997, RFC Editor, 1996.

\bibitem{chang2005inbound}
Rocky~KC Chang and Michael Lo.
\newblock Inbound traffic engineering for multihomed {ASs} using {AS} path
  prepending.
\newblock {\em IEEE network}, 19(2):18--25, 2005.

\bibitem{Chiu15a}
Yi-Ching Chiu, Brandon Schlinker, Abhishek~Balaji Radhakrishnan, Ethan
  Katz-Bassett, and Ramesh Govindan.
\newblock Are we one hop away from a better {Internet}?
\newblock In {\em Proceedings of the ACM Internet Measurement Conference},
  pages 523--529, Tokyo, Japan, October 2015. {ACM}.

\bibitem{cicalese2015characterizing}
Danilo Cicalese, Jordan Aug{\'e}, Diana Joumblatt, Timur Friedman, and Dario
  Rossi.
\newblock Characterizing ipv4 anycast adoption and deployment.
\newblock In {\em Proceedings of the 11th ACM Conference on Emerging Networking
  Experiments and Technologies}, pages 1--13, 2015.

\bibitem{cicalese2018longitudinal}
Danilo Cicalese and Dario Rossi.
\newblock A longitudinal study of {IP} anycast.
\newblock {\em ACM SIGCOMM Computer Communication Review}, 48(1):10--18, 2018.

\bibitem{cloudflare-event}
Cloudflare.
\newblock Famous {DDoS} attacks | the largest {DDoS} attacks of all time.
\newblock \url{https://www.cloudflare.com/learning/ddos/famous-ddos-attacks/}.
\newblock [Online; accessed 12-Oct-2021].

\bibitem{de2011forecasting}
Alysha~M De~Livera, Rob~J Hyndman, and Ralph~D Snyder.
\newblock Forecasting time series with complex seasonal patterns using
  exponential smoothing.
\newblock {\em Journal of the American Statistical Association},
  106(496):1513--1527, 2011.

\bibitem{Vries17b}
Wouter~B. {de Vries}, Ricardo de~O.~Schmidt, Wes Hardaker, John Heidemann,
  Pieter-Tjerk de~Boer, and Aiko Pras.
\newblock Verfploeter: Broad and load-aware anycast mapping.
\newblock In {\em Proceedings of the ACM Internet Measurement Conference},
  London, UK, 2017.

\bibitem{dietzel2016blackholing}
Christoph Dietzel, Anja Feldmann, and Thomas King.
\newblock Blackholing at {IXPs}: On the effectiveness of {DDoS} mitigation in
  the wild.
\newblock In {\em International Conference on Passive and Active Network
  Measurement}, pages 319--332. Springer, 2016.

\bibitem{dousti2018automated}
Ramin~Ali Dousti, Frank Scalzo, and Suresh Bhogavilli.
\newblock Automated ddos attack mitigation via bgp messaging, March~22 2018.
\newblock US Patent App. 15/273,510.

\bibitem{fan2010selecting}
Xun Fan and John Heidemann.
\newblock Selecting representative ip addresses for internet topology studies.
\newblock In {\em Proceedings of the 10th ACM SIGCOMM conference on Internet
  measurement}, pages 411--423. ACM, 2010.

\bibitem{fan2013evaluating}
Xun Fan, John Heidemann, and Ramesh Govindan.
\newblock Evaluating anycast in the domain name system.
\newblock In {\em 2013 Proceedings IEEE INFOCOM}, pages 1681--1689. IEEE, 2013.

\bibitem{fayaz2015bohatei}
Seyed~K Fayaz, Yoshiaki Tobioka, Vyas Sekar, and Michael Bailey.
\newblock Bohatei: Flexible and elastic ddos defense.
\newblock In {\em 24th {USENIX} Security Symposium}, pages 817--832, 2015.

\bibitem{flavel2015fastroute}
Ashley Flavel, Pradeepkumar Mani, David Maltz, Nick Holt, Jie Liu, Yingying
  Chen, and Oleg Surmachev.
\newblock Fastroute: A scalable load-aware anycast routing architecture for
  modern {CDNs}.
\newblock In {\em 12th {USENIX} Symposium on Networked Systems Design and
  Implementation ({NSDI} 15)}, pages 381--394, 2015.

\bibitem{gao2005interdomain}
Ruomei Gao, Constantinos Dovrolis, and Ellen~W Zegura.
\newblock Interdomain ingress traffic engineering through optimized {AS}-path
  prepending.
\newblock In {\em International Conference on Research in Networking}, pages
  647--658. Springer, 2005.

\bibitem{giotsas2017inferring}
Vasileios Giotsas, Georgios Smaragdakis, Christoph Dietzel, Philipp Richter,
  Anja Feldmann, and Arthur Berger.
\newblock Inferring {BGP} blackholing activity in the internet.
\newblock In {\em Proceedings of the Internet Measurement Conference}, pages
  1--14. ACM, 2017.

\bibitem{rfc3258}
T.~Hardie.
\newblock Distributing authoritative name servers via shared unicast addresses.
\newblock Technical Report 3258, RFC Editor, 2002.

\bibitem{holloway2014mitigating}
Lee~Hahn Holloway, Srikanth~N Rao, Matthew~Browning Prince, Matthieu
  Philippe~Fran{\c{c}}ois Tourne, Ian~Gerald Pye, Ray~Raymond Bejjani, and
  Terry~Paul Rodery~Jr.
\newblock Mitigating a denial-of-service attack in a cloud-based proxy service,
  October~7 2014.
\newblock US Patent 8,856,924.

\bibitem{Hong18a}
Chi-Yao Hong, Subhasree Mandal, Mohammad Al-Fares, Min Zhu, Richard Alimi,
  Kondapa~Naidu B., Chandan Bhagat, Sourabh Jain, Jay Kaimal, Shiyu Liang,
  Kirill Mendelev, Steve Padgett, Faro Rabe, Saikat Ray, Malveeka Tewari, Matt
  Tierney, Monika Zahn, Jonathan Zolla, Joon Ong, and Amin Vahdat.
\newblock B4 and after: Managing hierarchy, partitioning, and asymmetry for
  availability and scale in {Google's} software-defined {WAN}.
\newblock In {\em Proceedings of the {ACM} SIGCOMM Conference}, Budapest,
  Hungary, August 2018. {ACM}.

\bibitem{Huston2018BGP}
Geoff Huston.
\newblock {BGP} in 2017.
\newblock \url{https://labs.apnic.net/?p=1102}, Jan 8 2018.
\newblock [Online; accessed 12-Oct-2021].

\bibitem{cymru-secure-bgp-template}
Team~Cymru Inc.
\newblock Secure {Cisco IOS BGP} template.
\newblock \url{https://www.team-cymru.com/secure-bgp-template.html}.
\newblock [Online; accessed 12-Oct-2021].

\bibitem{jia2014catch}
Quan Jia, Huangxin Wang, Dan Fleck, Fei Li, Angelos Stavrou, and Walter Powell.
\newblock Catch me if you can: A cloud-enabled ddos defense.
\newblock In {\em 2014 44th Annual IEEE/IFIP International Conference on
  Dependable Systems and Networks}, pages 264--275. IEEE, 2014.

\bibitem{Koch21a}
Thomas Koch, Ke~Li, Calvin Ardi, Ethan Katz-Bassett, Matt Calder, and John
  Heidemann.
\newblock Anycast in context: A tale of two systems.
\newblock In {\em Proceedings of the {ACM} SIGCOMM Conference}, Virtual, August
  2021. {ACM}.

\bibitem{krebs2016krebsonsecurity}
Brian Krebs.
\newblock Krebsonsecurity hit with record {DDoS}.
\newblock {\em KrebsOnSecurity, Sept}, 21, 2016.

\bibitem{kuipers17}
Jan~Harm Kuipers.
\newblock Anycast for {DDoS}.
\newblock \url{https://essay.utwente.nl/73795/1/Kuipers_MA_EWI.pdf}, 2017.
\newblock [Online; accessed 12-Oct-2021].

\bibitem{labovitz2000delayed}
Craig Labovitz, Abha Ahuja, Abhijit Bose, and Farnam Jahanian.
\newblock Delayed {Internet} routing convergence.
\newblock {\em ACM SIGCOMM Computer Communication Review}, 30(4):175--187,
  2000.

\bibitem{li2018internet}
Zhihao Li, Dave Levin, Neil Spring, and Bobby Bhattacharjee.
\newblock Internet anycast: Performance, problems, \& potential.
\newblock In {\em Proceedings of the 2018 Conference of the ACM Special
  Interest Group on Data Communication}, pages 59--73, 2018.

\bibitem{Li18a}
Zhihao Li, Dave Levin, Neil Spring, and Bobby Bhattacharjee.
\newblock Internet anycast: Performance, problems, and potential.
\newblock In {\em Proceedings of the {ACM} SIGCOMM Conference}, pages 59--73,
  Budapest, Hungary, August 2018. {ACM}.

\bibitem{liu2007two}
Ziqian Liu, Bradley Huffaker, Marina Fomenkov, Nevil Brownlee, et~al.
\newblock Two days in the life of the {DNS} anycast root servers.
\newblock In {\em International Conference on Passive and Active Network
  Measurement}, pages 125--134. Springer, 2007.

\bibitem{madory19a}
Doug Madory and Matt Prosser.
\newblock Excessive {BGP} {AS} path prepending is a self-inflicted
  vulnerability.
\newblock Presentation at RIPE 79, October 2019.

\bibitem{2018-memcrashed}
Marek Majkowski.
\newblock Memcrashed - major amplification attacks from {UDP} port 11211.
\newblock
  \url{https://blog.cloudflare.com/memcrashed-major-amplification-attacks-from-port-11211/},
  2018.
\newblock [Online; accessed 12-Oct-2021].

\bibitem{mcdaniel2020flexsealing}
Tyler McDaniel, Jared~M Smith, and Max Schuchard.
\newblock Flexsealing bgp against route leaks: peerlock active measurement and
  analysis.
\newblock {\em arXiv e-prints}, pages arXiv--2006, 2020.

\bibitem{mcquistin2019taming}
Stephen McQuistin, Sree~Priyanka Uppu, and Marcel Flores.
\newblock Taming anycast in the wild {Internet}.
\newblock In {\em Proceedings of the Internet Measurement Conference}, pages
  165--178, 2019.

\bibitem{Moura16b}
Giovane C.~M. Moura, Ricardo de~O.~Schmidt, John Heidemann, Wouter~B. {de
  Vries}, Moritz M{\"u}ller, Lan Wei, and Christian Hesselman.
\newblock Anycast vs {DDoS}: Evaluating the {November} 2015 root {DNS} event.
\newblock In {\em Proceedings of the ACM Internet Measurement Conference},
  November 2016.

\bibitem{nanda2009scalable}
Priyadarsi Nanda and AJ~Simmonds.
\newblock A scalable architecture supporting {QoS} guarantees using traffic
  engineering and policy based routing in the {Internet}.
\newblock {\em International Journal of Communications, Network and System
  Sciences}, 2009.

\bibitem{2015-event}
Root~Server Operators.
\newblock Events of 2015-11-30.
\newblock \url{https://root-servers.org/media/news/events-of-20151130.txt},
  2015.
\newblock [Online; accessed 12-Oct-2021].

\bibitem{2016-event}
Root~Server Operators.
\newblock Events of 2016-06-25.
\newblock \url{https://root-servers.org/media/news/events-of-20160625.txt},
  2016.
\newblock [Online; accessed 12-Oct-2021].

\bibitem{rfc1546}
Craig Partridge, Trevor Mendez, and Walter Milliken.
\newblock Host anycasting service.
\newblock Technical Report 1546, RFC Editor, 1993.

\bibitem{voipms12021}
The~Canadian Press.
\newblock Canadian communications company voip.ms hit by cyber attack.
\newblock
  \url{https://www.thestar.com/business/2021/09/21/canadian-communications-company-voipms-hit-by-cyber-attack.html/},
  09 2021.

\bibitem{20170306}
LANDER project.
\newblock Lander:b root anomaly-20170306.
\newblock
  \url{https://ant.isi.edu/datasets/readmes/B_Root_Anomaly-20170306.README.txt},
  2019.
\newblock [Online; accessed 12-Oct-2021].

\bibitem{quoitin2005performance}
Bruno Quoitin, Cristel Pelsser, Olivier Bonaventure, and Steve Uhlig.
\newblock A performance evaluation of {BGP}-based traffic engineering.
\newblock {\em International journal of network management}, 15(3):177--191,
  2005.

\bibitem{quoitin2003interdomain}
Bruno Quoitin, Cristel Pelsser, Louis Swinnen, Olivier Bonaventure, and Steve
  Uhlig.
\newblock Interdomain traffic engineering with {BGP}.
\newblock {\em IEEE Communications magazine}, 41(5):122--128, 2003.

\bibitem{ripe-dns-built-in}
RIPE.
\newblock Measurements.
\newblock \url{https://atlas.ripe.net/measurements/10310/}.
\newblock [Online; accessed 12-Oct-2021].

\bibitem{RIPE21a}
RIPE.
\newblock Root dns observations.
\newblock Measurement ID 1009 (A-Root), 1010 (B-Root), etc., 2021.

\bibitem{riperis}
{RIPE Network Coordination Centre}.
\newblock {RIPE - Routing Information Service (RIS)}.
\newblock
  \url{https://https://www.ripe.net/analyse/internet-measurements/routing-information-service-ris},
  2020.

\bibitem{Rizvi19a}
{ASM} Rizvi, John Heidemann, and Jelena Mirkovic.
\newblock Dynamically selecting defenses to {DDoS} for {DNS} (extended).
\newblock Technical Report ISI-TR-736, USC/Information Sciences Institute, May
  2019.

\bibitem{sarat2006use}
Sandeep Sarat, Vasileios Pappas, and Andreas Terzis.
\newblock On the use of anycast in {DNS}.
\newblock In {\em Proceedings of 15th International Conference on Computer
  Communications and Networks}, pages 71--78. IEEE, 2006.

\bibitem{schlinker19peering}
Brandon Schlinker, Todd Arnold, Italo Cunha, and Ethan Katz-Bassett.
\newblock {PEERING: Virtualizing BGP at the Edge for Research}.
\newblock In {\em Proc. ACM CoNEXT}, Orlando, FL, December 2019.

\bibitem{Schlinker17a}
Brandon Schlinker, Hyojeong Kim, Timothy Cui, Ethan Katz-Bassett, Harsha~V.
  Madhyastha, Italo Cunha, James Quinn, Saif Hasan, Petr Lapukhov, and Hongyi
  Zeng.
\newblock Engineering egress with {Edge} {Fabric}: Steering oceans of content
  to the world.
\newblock In {\em Proceedings of the {ACM} SIGCOMM Conference}, pages 418--431,
  Los Angeles, CA, USA, August 2017. {ACM}.

\bibitem{Schmidt17a}
Ricardo de~O. Schmidt, John Heidemann, and Jan~Harm Kuipers.
\newblock Anycast latency: How many sites are enough?
\newblock In {\em International Conference on Passive and Active Network
  Measurement}, pages 188--200, Sydney, Australia, March 2017.

\bibitem{scholl2015methods}
Thomas~Bradley Scholl.
\newblock Methods and apparatus for distributed backbone internet ddos
  mitigation via transit providers, February~3 2015.
\newblock US Patent 8,949,459.

\bibitem{Shaikh2001dnsselection}
A.~{Shaikh}, R.~{Tewari}, and M.~{Agrawal}.
\newblock On the effectiveness of {DNS}-based server selection.
\newblock In {\em Proceedings IEEE INFOCOM 2001. Conference on Computer
  Communications. Twentieth Annual Joint Conference of the IEEE Computer and
  Communications Society (Cat. No.01CH37213)}, volume~3, pages 1801--1810
  vol.3, 2001.

\bibitem{voipms2021}
AX~Sharma.
\newblock Phone calls disrupted by ongoing ddos cyber attack on voip.ms.
\newblock
  \url{https://arstechnica.com/gadgets/2021/09/canadian-voip-provider-hit-by-ddos-attack-phone-calls-disrupted/},
  09 2021.

\bibitem{aws2020cldap}
AWS Shield.
\newblock Aws shield - threat landscape report – q1 2020.
\newblock
  \url{https://aws-shield-tlr.s3.amazonaws.com/2020-Q1_AWS_Shield_TLR.pdf}, 08
  2020.

\bibitem{silva2017bgpdelay}
R.~B.~da Silva and E.~Souza Mota.
\newblock A survey on approaches to reduce {BGP} interdomain routing
  convergence delay on the {Internet}.
\newblock {\em IEEE Communications Surveys \& Tutorials}, 19(4):2949--2984,
  2017.

\bibitem{smith2017ddosaas}
Daniel Smith.
\newblock The growth of {DDoS}-as-a-service: Stresser services.
\newblock
  \url{https://blog.radware.com/security/2017/09/growth-of-ddos-as-a-service-stresser-services/},
  2017.
\newblock [Online; accessed 12-Oct-2021].

\bibitem{smith2016network}
Donald~J Smith, Michael Glenn, John~A Schiel, and Christopher~L Garner.
\newblock Network traffic data scrubbing with services offered via anycasted
  addresses, May~24 2016.
\newblock US Patent 9,350,706.

\bibitem{smith2018routing}
Jared~M. Smith and Max Schuchard.
\newblock Routing around congestion: Defeating {DDoS} attacks and adverse
  network conditions via reactive {BGP} routing.
\newblock In {\em 2018 IEEE Symposium on Security and Privacy (SP)}, pages
  599--617. IEEE, 2018.

\bibitem{snijders2016practical}
Job Snijders.
\newblock Practical everyday bgp filtering with as\_path filters: Peer locking.
\newblock {\em NANOG-67, Chicago, June}, 2016.

\bibitem{spatscheck2013multi}
Oliver Spatscheck, Zakaria Al-Qudah, Seunjoon Lee, Michael Rabinovich, and
  Jacobus Van Der~Merwe.
\newblock Multi-autonomous system anycast content delivery network, December~10
  2013.
\newblock US Patent 8,607,014.

\bibitem{staff2015ripe}
RIPE~NCC Staff.
\newblock Ripe atlas: A global internet measurement network.
\newblock {\em Internet Protocol Journal}, 18(3), 2015.

\bibitem{comm_list}
One Step.
\newblock {BGP} community guides.
\newblock \url{https://onestep.net/communities/}.
\newblock [Online; accessed 12-Oct-2021].

\bibitem{swildens2011global}
Eric Sven-Johan Swildens, Zaide Liu, and Richard~David Day.
\newblock Global traffic management system using {IP} anycast routing and
  dynamic load-balancing, March~8 2011.
\newblock US Patent 7,904,541.

\bibitem{teixeira2007bgp}
Renata Teixeira, Steve Uhlig, and Christophe Diot.
\newblock {BGP} route propagation between neighboring domains.
\newblock In {\em International Conference on Passive and Active Network
  Measurement}, pages 11--21. Springer, 2007.

\bibitem{routeviewsproject}
{University of Oregon}.
\newblock {Route Views Project}.
\newblock \url{http://www.routeviews.org/routeviews/}, 2021.

\bibitem{b-dataset}
USC/ISI.
\newblock Usc/isi ant datasets.
\newblock \url{https://ant.isi.edu/datasets/all.html}, 2019.
\newblock [Online; accessed 12-Oct-2021].

\bibitem{Wei17b}
Lan Wei and John Heidemann.
\newblock Does anycast hang up on you?
\newblock In {\em 2017 Network Traffic Measurement and Analysis Conference
  (TMA)}, pages 1--9, Dublin, Ireland, July 2017. {IEEE}.

\bibitem{weiden2010anycast}
Fernanda Weiden and Peter Frost.
\newblock Anycast as a load balancing feature.
\newblock In {\em Proceedings of the 24th international conference on Large
  installation system administration}, pages 1--6. USENIX Association, 2010.

\bibitem{wilson2012tools}
Curt Wilson.
\newblock Attack of the {Shuriken}: Many hands, many weapons.
\newblock \url{https://www.arbornetworks.com/blog/asert/ddos-tools/}, 2012.
\newblock [Online; accessed 12-Oct-2021].

\end{thebibliography}
}

\begin{appendices}

\ifisarxiv
\section{Anycast and BGP Background}
\label{sec:anycast_bgp_background}

\reviewfix{SS22-A6}

IP anycast is a routing method used to route incoming requests to
different locations (sites). Each site uses the same IP address,
but at 
different geographic locations.
Anycast then uses Internet routing with BGP
to determine how to associate users to different sites---that is known as
  site's anycast \emph{catchment}.
BGP has a standard path selection algorithm
  that considers routing policy and approximate distance~\cite{Caesar05a}.

\autoref{fig:anycast-bgp-client} shows a conceptual
version of one of our three-site anycast deployments.
Clients are splitted
  into three sites from three continents.
Although we illustrate catchments by continent here,
  in practice they follow BGP routing rules and not geography,
  with users intermixed.

Although BGP is not perfect, in practice it often does
  a reasonably good job at associating users to nearby sites
  and thereby minimizing service latency~\cite{Li18a,Koch21a}.
Moreover, anycast increases the service resiliency since
it spreads traffic over multiple sites.
If one site goes offline, perhaps due to maintenance,
  that site withdraws its BGP route and routing
  automatically redistributes users previously going to that site
  to other sites.
Thus, anycast avoids service interruptions
  in addition to capacity expansion.

Operators can influence the routing decisions process using different
traffic engineering techniques (TE) to manipulate BGP.  We describe
TE techniques in \autoref{sec:route_manipulation} and how they can
be used to rebalance the load during a DDoS attack.

\begin{figure}[ht]
\centering
\includegraphics[width=0.9\linewidth]{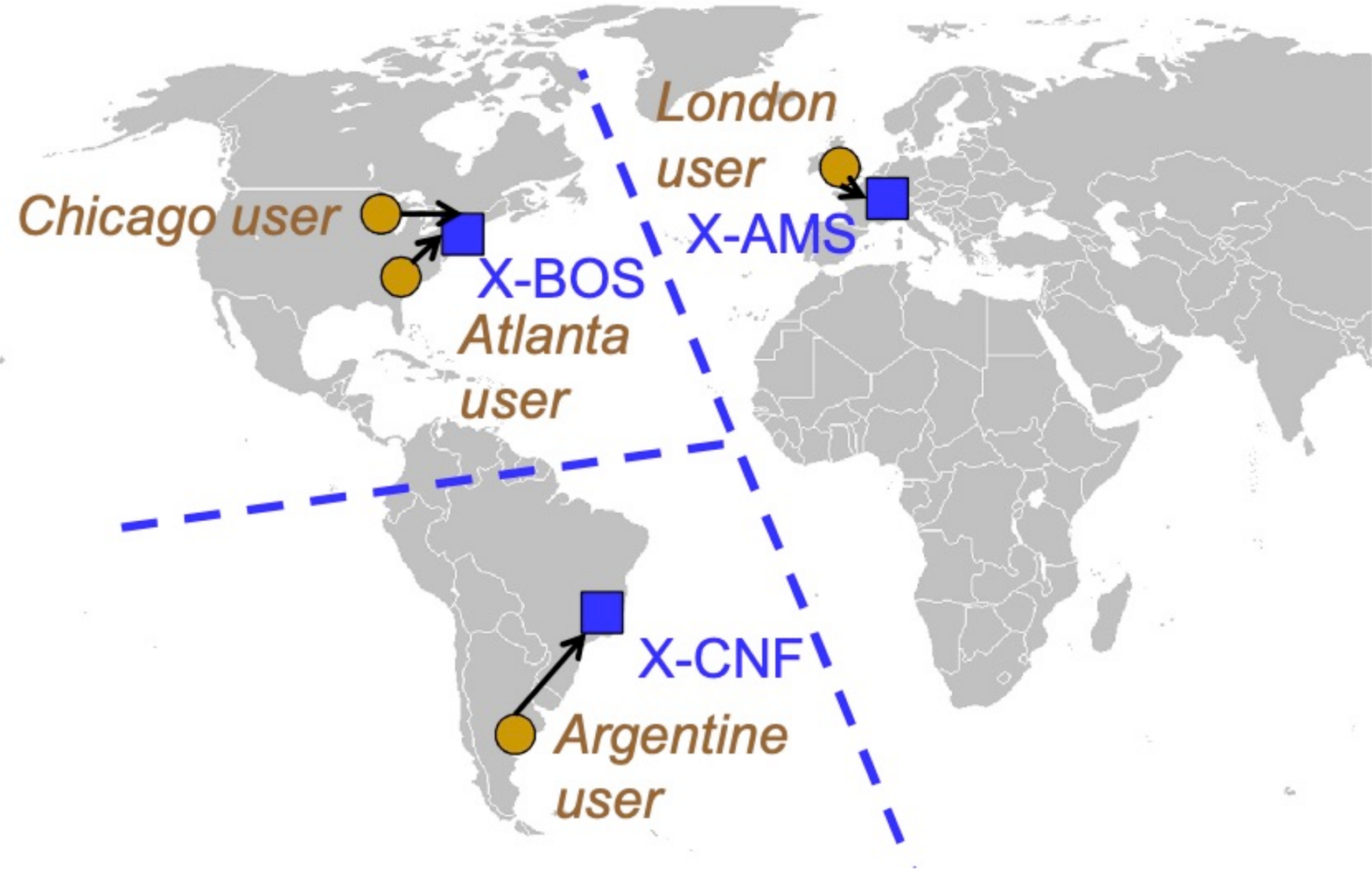}
    \caption{An example of a three-site Anycast deployment with possible catchments.}
\label{fig:anycast-bgp-client}
  \vspace{-0.5em}
\end{figure}

\else

\section{Anycast and Verfploeter}
\label{sec:anycast_bgp_verfploeter}

\reviewfix{SS22-A6}

IP anycast is a routing method used to route incoming requests to
different locations (sites). Each site uses the same IP address,
but at 
different geographic locations.
Anycast then uses Internet routing with BGP
to determine how to associate users to different sites---that is known as
  site's anycast \emph{catchment}.
BGP has a standard path selection algorithm
  that considers routing policy and approximate distance~\cite{Caesar05a}.

Operators can influence the routing decisions process using different
traffic engineering techniques (TE) to manipulate BGP.  We describe
TE techniques in \autoref{sec:route_manipulation} and how they can
be used to rebalance the load during a DDoS attack.

We use Verfploeter~\cite{Vries17b} to find out the client
  to anycast site mapping.
Using Verfploeter we build our BGP playbook
  with various BGP changes (\autoref{sec:playbook}).
The main intuition behind Verfploeter is to
  send pings to millions of address blocks\ifisanon~\cite{anon-b-dataset, fan2010selecting} \else~\cite{b-dataset, fan2010selecting}\fi, 
  using an anycast prefix as source address.
The replies to these pings will be routed to the nearest anycast
  site by the inter-domain routing system from which
  we can map address blocks to the anycast sites.

\fi

\ifisarxiv  
\section{Verfploeter Mapping}
	\label{sec:appendix_verfploeter}

We use Verfploeter~\cite{Vries17b} to find out the client
  to anycast site mapping.
Using Verfploeter we build our BGP playbook
  with various BGP changes (\autoref{sec:playbook}).

The main intuition behind Verfploeter is to
send pings using an anycast prefix as source address.
The replies to these pings will be routed to the nearest anycast
site by the inter-domain routing system. \autoref{fig:vp-pinger} shows
how this works. One of the sites (green) of the anycast service runs a
packet generator that sends pings (ICMP Echo Requests) to a hitlist of
IP addresses. The replies (ICMP Echo Replies) from these
IPs are then routed to the ``closest'' (in terms of routing distance) anycast
site. The catchment of the site is thus determined by the IP prefixes
from which ping replies arrive at that particular site.

While in principle Verfploeter can work with any type of IP hitlist,
  for our measurement we use a publicly available hitlist\ifisanon~\cite{anon-b-dataset} \else~\cite{b-dataset} \fi based on the
  Fan et, al.~\cite{fan2010selecting} methodology.
This hitlist includes the IP addresses that are most likely to respond to pings
  for each /24 prefix in the IPv4 address space. 

\begin{figure}[ht]
   \centering
  \includegraphics[width=0.8\columnwidth] {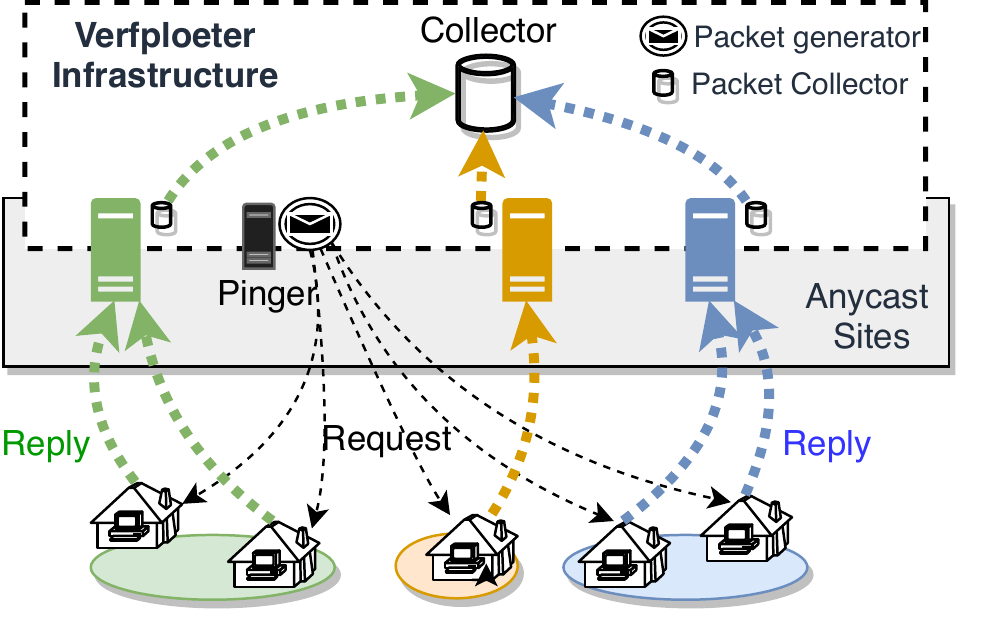}
  \caption{Overview of the Verfploeter approach (from~\cite{le2020tangled}).}
  \label{fig:vp-pinger}
  \vspace{-0.15in}
\end{figure}

\fi

\section{Operator Assistance System}
\label{sec:operator_assistance}

To assist operators (\autoref{sec:op-assistance-system}), we provide an interface for defense. To react and reconfigure the anycast network, the operators can use a web interface
similar to an equalizer, choosing the percentage of load to be increased or dropped at an anycast site.
The possible ranges of slider positions are based on the playbook alternatives or presets of routing policies.
This process hides the playbook complexity from the operator,
making the process less error-prone and more intuitive, but still giving the operator a full control of the BGP routing.

In \autoref{fig:user-interface} we can visualize a snapshot of this interface. Each
slider represents an anycast site and each site has predetermined settings indicated
by “notches". The positions of the "notches" are the results of all the
measurements obtained to create our playbook. The bar graph shows the results
of the measurement process, indicating how many networks will be attracted to
each anycast site. The operators can visualize the forecasted traffic to each
position and then apply the configuration on the production network.

\begin{figure}
\centering
\includegraphics[width=0.8\linewidth]{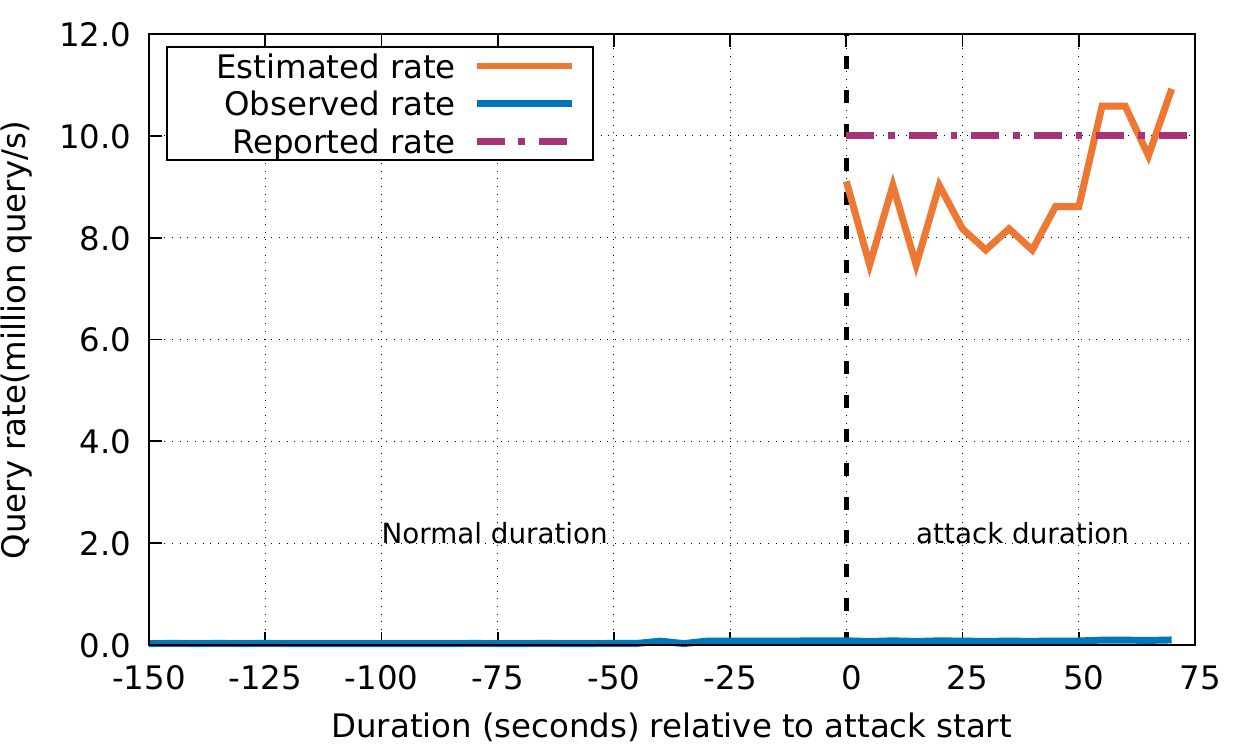}
\vspace{-0.07in}
    \caption{Estimating real-world attack events: estimating June 2016 event with 0.91\% access fraction.}
\label{fig:est-2016}
    \vspace{-0.1in}
\end{figure}

\ifisarxiv
\begin{figure}
\centering
 \includegraphics[clip,width=1\columnwidth]{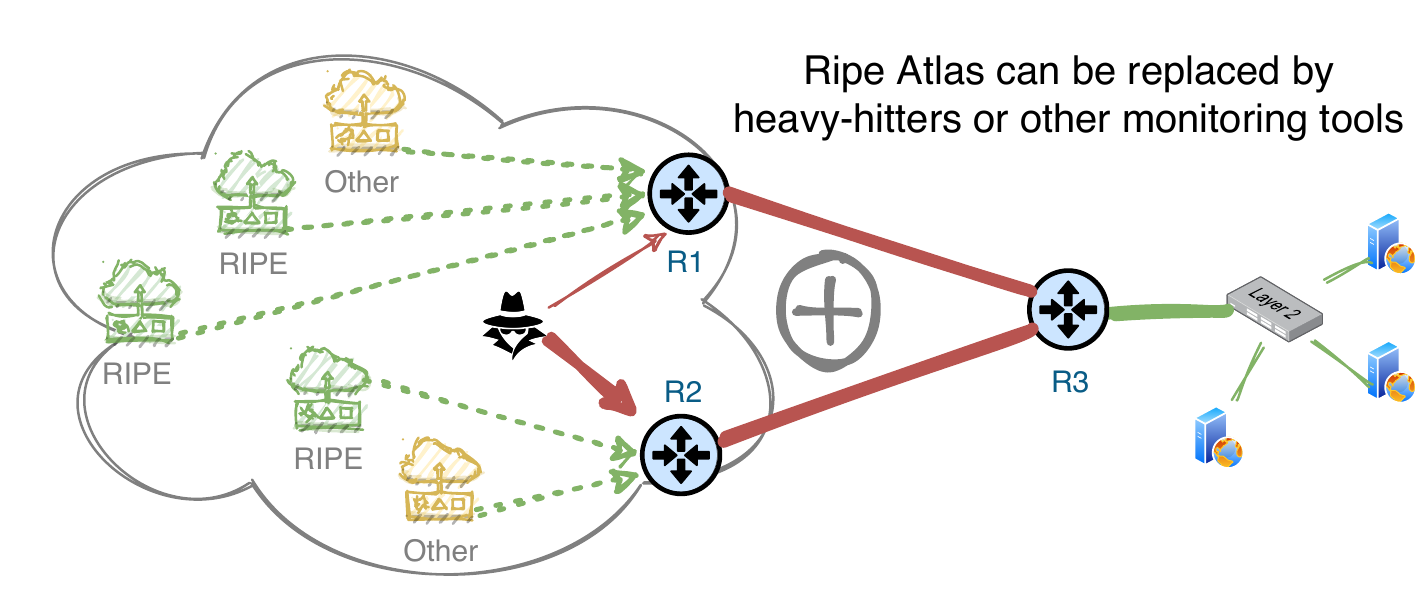}
\caption{Topology with two upstream providers.}
\label{fig:topology}
\vspace{-0.15in}
\end{figure}
\fi

\begin{figure*}
\centering
  \begin{minipage}{0.3\textwidth}
\begin{center}
   \vspace{-0.1in}
    \includegraphics[width=\textwidth]{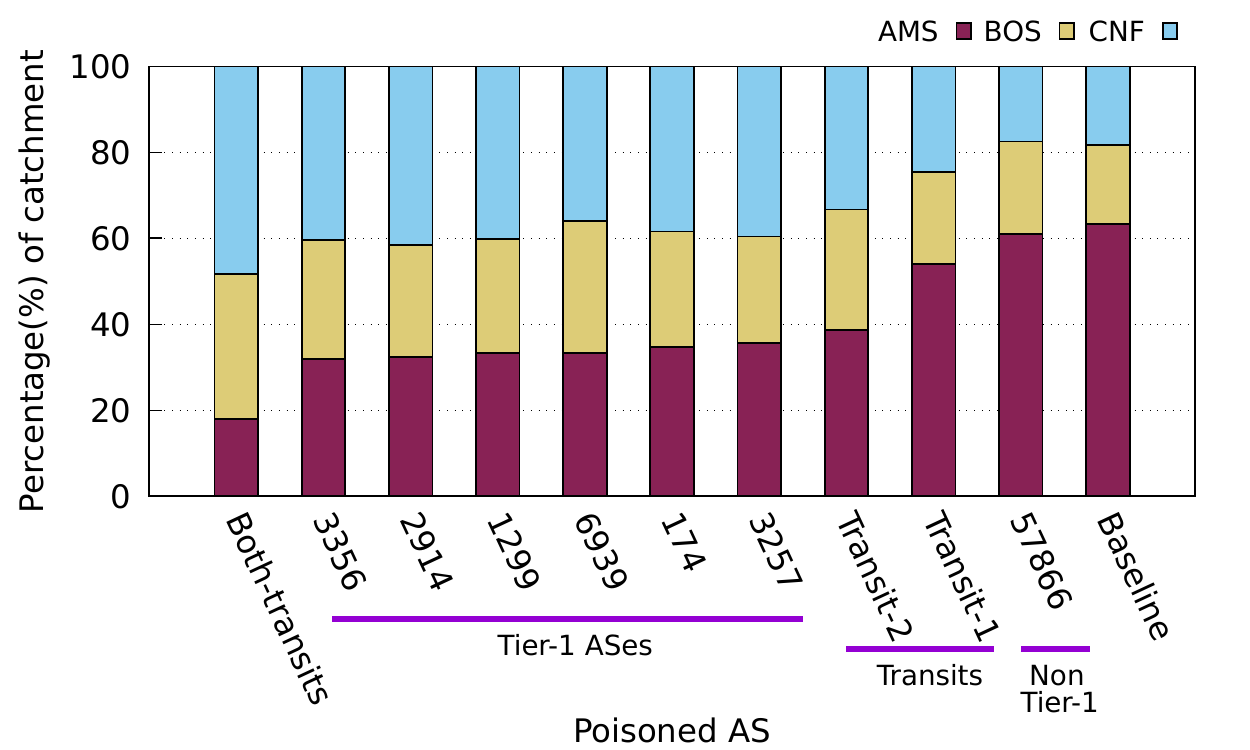}
\end{center}
     \vspace{-0.1in}
  \caption{\peering: Impact of path poisoning (from AMS on 2021-04-09).}
          \label{fig:poison-peering-2021-04-15}
  \end{minipage}\hfill
  \begin{minipage}{0.3\textwidth}
\begin{center}
    \includegraphics[width=\textwidth]{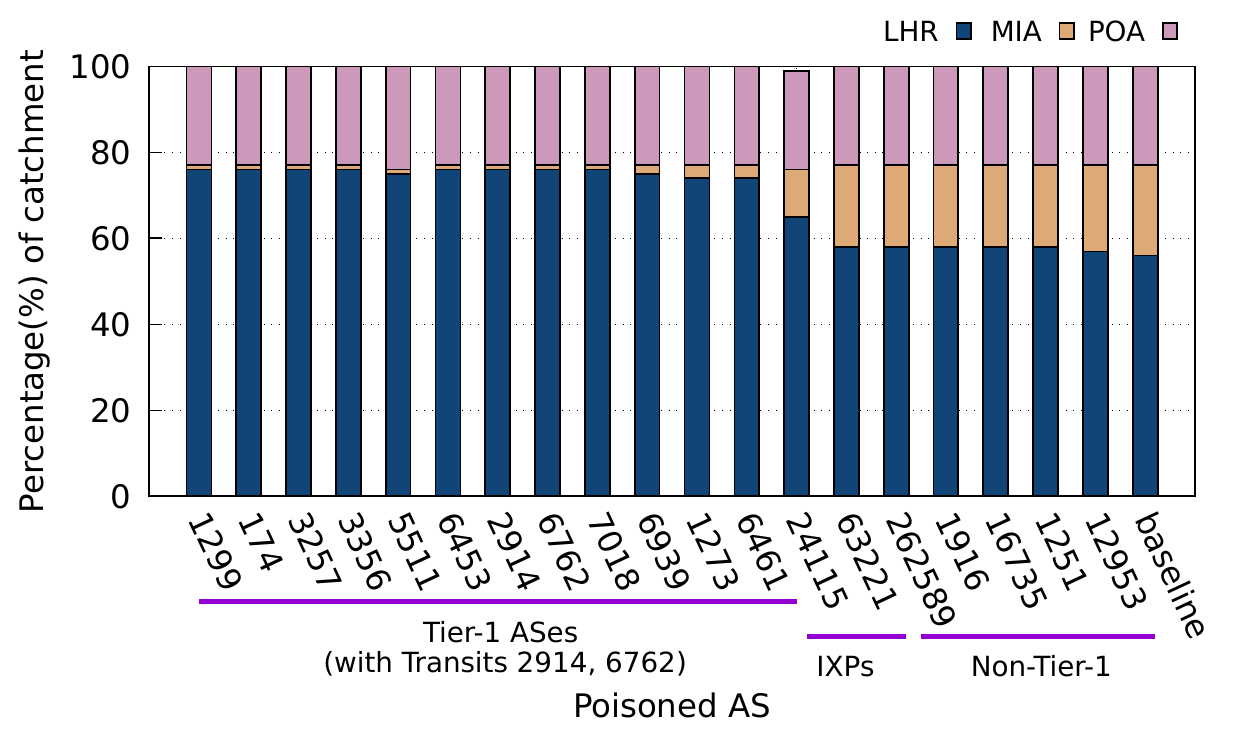}
\end{center}
       \vspace{-0.1in}
  \caption{\tangled: Impact of path poisoning (from MIA on 2021-04-11).}
          \label{fig:poison-tangled-2021-04-15}
  \end{minipage}\hfill
  \begin{minipage}{0.3\textwidth}

\begin{center}
\includegraphics[width=\textwidth]{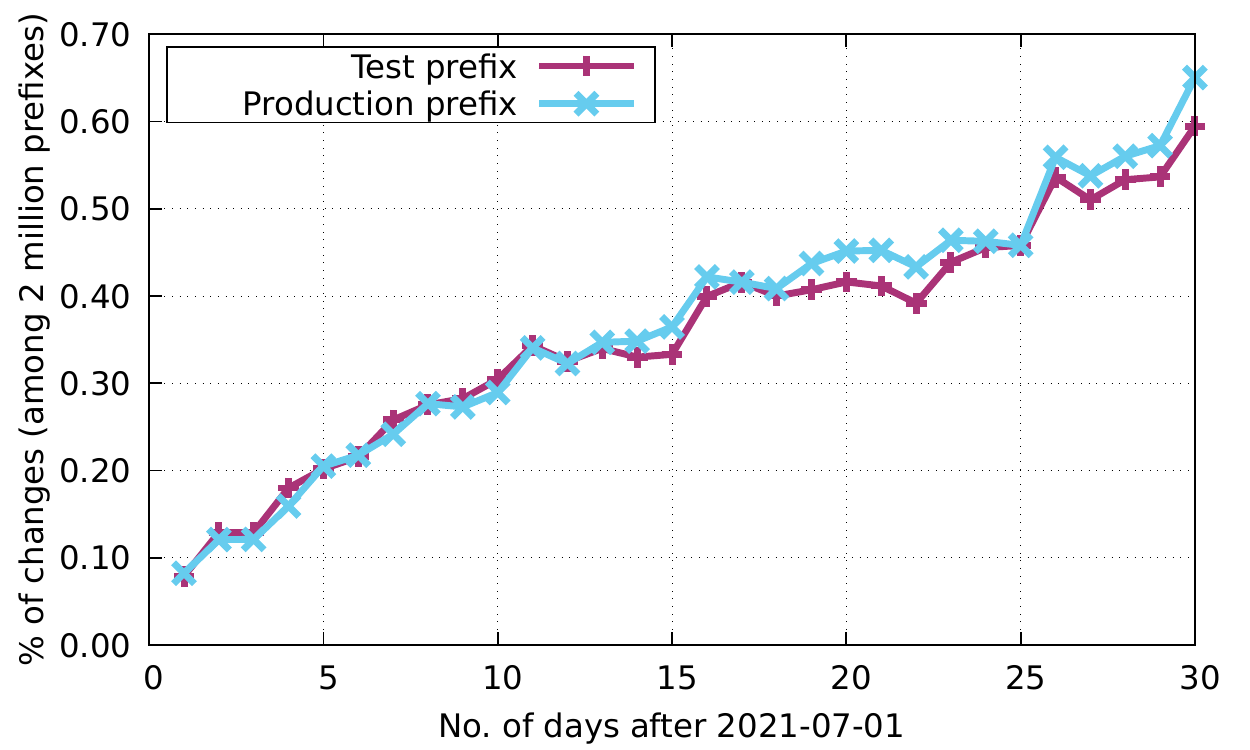}
\end{center}

     \vspace{-0.1in}
\caption{One month of catchment stability in \broot.}
\label{fig:broot-stability}

  \end{minipage}
\end{figure*}

\begin{table*}
\resizebox{\textwidth}{!}{\begin{tabular}{ll|llll|llll|llll}
\rowcolor[HTML]{EAECF0} 
\multicolumn{2}{c}{\cellcolor[HTML]{EAECF0}{\color[HTML]{202122} }} &
\multicolumn{4}{c}{\cellcolor[HTML]{EAECF0}{\color[HTML]{202122} \textbf{AMS(\%)}}} &
\multicolumn{4}{c}{\cellcolor[HTML]{EAECF0}{\color[HTML]{202122} \textbf{BOS(\%)}}} &
\multicolumn{4}{c}{\cellcolor[HTML]{EAECF0}{\color[HTML]{202122} \textbf{CNF(\%)}}} \\
\rowcolor[HTML]{F8F9FA} 
\multicolumn{2}{c}{\multirow{-2}{*}{\cellcolor[HTML]{EAECF0}{\color[HTML]{202122} \textbf{Policy / Day}}}} &
            {\color[HTML]{202122} 00 GMT} &
            {\color[HTML]{202122} 06 GMT} &
            {\color[HTML]{202122} 12 GMT} &
            {\color[HTML]{202122} 18 GMT} &
            {\color[HTML]{202122} 00 GMT} &
            {\color[HTML]{202122} 06 GMT} &
            {\color[HTML]{202122} 12 GMT} &
            {\color[HTML]{202122} 18 GMT} &
            {\color[HTML]{202122} 00 GMT} &
            {\color[HTML]{202122} 06 GMT} &
            {\color[HTML]{202122} 12 GMT} &
            {\color[HTML]{202122} 18 GMT} \\ 
\rowcolor[HTML]{F8F9FA} 
\cellcolor[HTML]{F8F9FA}{\color[HTML]{202122} } &
          {\color[HTML]{202122} Day-1 load} &
          {\color[HTML]{202122} 77} &
          {\color[HTML]{202122} 84} &
          {\color[HTML]{202122} 84} &
          {\color[HTML]{202122} 84} &
          {\color[HTML]{202122} 10} &
          {\color[HTML]{202122} 8} &
          {\color[HTML]{202122} 8} &
          {\color[HTML]{202122} 7} &
          {\color[HTML]{202122} 13} &
          {\color[HTML]{202122} 8} &
          {\color[HTML]{202122} 8} &
          {\color[HTML]{202122} 9}  \\

\rowcolor[HTML]{F8F9FA} 
\cellcolor[HTML]{F8F9FA}{\color[HTML]{202122} } &
         {\color[HTML]{202122} Day-2 load} &
         {\color[HTML]{202122} 77} &
         {\color[HTML]{202122} 84} &
         {\color[HTML]{202122} 84} &
         {\color[HTML]{202122} 80} &
         {\color[HTML]{202122} 10} &
         {\color[HTML]{202122} 8} &
         {\color[HTML]{202122} 7} &
         {\color[HTML]{202122} 9} &
         {\color[HTML]{202122} 13} &
         {\color[HTML]{202122} 8} &
         {\color[HTML]{202122} 9} &
         {\color[HTML]{202122} 11} \\

\rowcolor[HTML]{F8F9FA} 
\multirow{-3}{*}{\cellcolor[HTML]{F8F9FA}{\color[HTML]{202122} Baseline}} &
         {\color[HTML]{202122} Catchment} &
         \multicolumn{4}{c|}{\color[HTML]{202122} 68} &
         \multicolumn{4}{c|}{\color[HTML]{202122} 15} &
         \multicolumn{4}{c}{\color[HTML]{202122} 17} \\ \hline

\rowcolor[HTML]{F8F9FA} 
\cellcolor[HTML]{F8F9FA}{\color[HTML]{202122} } &
          {\color[HTML]{202122} Day-1 load} &
          {\color[HTML]{202122} 43} &
          {\color[HTML]{202122} 49} &
          {\color[HTML]{202122} 49} &
          {\color[HTML]{202122} 58} &
          {\color[HTML]{202122} 18} &
          {\color[HTML]{202122} 20} &
          {\color[HTML]{202122} 18} &
          {\color[HTML]{202122} 13} &
          {\color[HTML]{202122} 39} &
          {\color[HTML]{202122} 32} &
          {\color[HTML]{202122} 33} &
          {\color[HTML]{202122} 29}  \\

\rowcolor[HTML]{F8F9FA} 
\cellcolor[HTML]{F8F9FA}{\color[HTML]{202122} } &
         {\color[HTML]{202122} Day-2 load} &
         {\color[HTML]{202122} 43} &
         {\color[HTML]{202122} 46} &
         {\color[HTML]{202122} 46} &
         {\color[HTML]{202122} 50} &
         {\color[HTML]{202122} 18} &
         {\color[HTML]{202122} 18} &
         {\color[HTML]{202122} 18} &
         {\color[HTML]{202122} 18} &
         {\color[HTML]{202122} 39} &
         {\color[HTML]{202122} 36} &
         {\color[HTML]{202122} 36} &
         {\color[HTML]{202122} 32} \\
          
\rowcolor[HTML]{F8F9FA} 
\multirow{-3}{*}{\cellcolor[HTML]{F8F9FA}{\color[HTML]{202122} 1xPrepend AMS}} &
         {\color[HTML]{202122} Catchment} &
         \multicolumn{4}{c|}{\color[HTML]{202122} 37} &
         \multicolumn{4}{c|}{\color[HTML]{202122} 25} &
         \multicolumn{4}{c}{\color[HTML]{202122} 38} \\ \hline

\rowcolor[HTML]{F8F9FA} 
\cellcolor[HTML]{F8F9FA}{\color[HTML]{202122} } &
          {\color[HTML]{202122} Day-1 load} &
          {\color[HTML]{202122} 78} &
          {\color[HTML]{202122} 85} &
          {\color[HTML]{202122} 83} &
          {\color[HTML]{202122} 83} &
          {\color[HTML]{202122} 4} &
          {\color[HTML]{202122} 3} &
          {\color[HTML]{202122} 4} &
          {\color[HTML]{202122} 3} &
          {\color[HTML]{202122} 18} &
          {\color[HTML]{202122} 12} &
          {\color[HTML]{202122} 13} &
          {\color[HTML]{202122} 14}  \\

\rowcolor[HTML]{F8F9FA} 
\cellcolor[HTML]{F8F9FA}{\color[HTML]{202122} } &
         {\color[HTML]{202122} Day-2 load} &
         {\color[HTML]{202122} 78} &
         {\color[HTML]{202122} 85} &
         {\color[HTML]{202122} 83} &
         {\color[HTML]{202122} 79} &
         {\color[HTML]{202122} 4} &
         {\color[HTML]{202122} 4} &
         {\color[HTML]{202122} 4} &
         {\color[HTML]{202122} 4} &
         {\color[HTML]{202122} 18} &
         {\color[HTML]{202122} 12} &
         {\color[HTML]{202122} 13} &
         {\color[HTML]{202122} 16} \\
          
\rowcolor[HTML]{F8F9FA} 
\multirow{-3}{*}{\cellcolor[HTML]{F8F9FA}{\color[HTML]{202122} 1xPrepend BOS}} &
         {\color[HTML]{202122} Catchment} &
         \multicolumn{4}{c|}{\color[HTML]{202122} 70} &
         \multicolumn{4}{c|}{\color[HTML]{202122} 7} &
         \multicolumn{4}{c}{\color[HTML]{202122} 23} \\ \hline

\rowcolor[HTML]{F8F9FA} 
\cellcolor[HTML]{F8F9FA}{\color[HTML]{202122} } &
          {\color[HTML]{202122} Day-1 load} &
          {\color[HTML]{202122} 83} &
          {\color[HTML]{202122} 88} &
          {\color[HTML]{202122} 87} &
          {\color[HTML]{202122} 87} &
          {\color[HTML]{202122} 11} &
          {\color[HTML]{202122} 10} &
          {\color[HTML]{202122} 9} &
          {\color[HTML]{202122} 8} &
          {\color[HTML]{202122} 6} &
          {\color[HTML]{202122} 2} &
          {\color[HTML]{202122} 3} &
          {\color[HTML]{202122} 5}  \\

\rowcolor[HTML]{F8F9FA} 
\cellcolor[HTML]{F8F9FA}{\color[HTML]{202122} } &
         {\color[HTML]{202122} Day-2 load} &
         {\color[HTML]{202122} 83} &
         {\color[HTML]{202122} 89} &
         {\color[HTML]{202122} 87} &
         {\color[HTML]{202122} 85} &
         {\color[HTML]{202122} 11} &
         {\color[HTML]{202122} 9} &
         {\color[HTML]{202122} 9} &
         {\color[HTML]{202122} 10} &
         {\color[HTML]{202122} 6} &
         {\color[HTML]{202122} 2} &
         {\color[HTML]{202122} 4} &
         {\color[HTML]{202122} 5} \\
          
\rowcolor[HTML]{F8F9FA} 
\multirow{-3}{*}{\cellcolor[HTML]{F8F9FA}{\color[HTML]{202122} 1xPrepend CNF}} &
         {\color[HTML]{202122} Catchment} &
         \multicolumn{4}{c|}{\color[HTML]{202122} 77} &
         \multicolumn{4}{c|}{\color[HTML]{202122} 19} &
         \multicolumn{4}{c}{\color[HTML]{202122} 4} \\ \hline

\rowcolor[HTML]{F8F9FA} 
\cellcolor[HTML]{F8F9FA}{\color[HTML]{202122} } &
          {\color[HTML]{202122} Day-1 load} &
          {\color[HTML]{202122} 88} &
          {\color[HTML]{202122} 93} &
          {\color[HTML]{202122} 92} &
          {\color[HTML]{202122} 91} &
          {\color[HTML]{202122} 5} &
          {\color[HTML]{202122} 4} &
          {\color[HTML]{202122} 5} &
          {\color[HTML]{202122} 3} &
          {\color[HTML]{202122} 6} &
          {\color[HTML]{202122} 3} &
          {\color[HTML]{202122} 4} &
          {\color[HTML]{202122} 5}  \\

\rowcolor[HTML]{F8F9FA}
\cellcolor[HTML]{F8F9FA}{\color[HTML]{202122} } &
         {\color[HTML]{202122} Day-2 load} &
         {\color[HTML]{202122} 88} &
         {\color[HTML]{202122} 93} &
         {\color[HTML]{202122} 92} &
         {\color[HTML]{202122} 90} &
         {\color[HTML]{202122} 5} &
         {\color[HTML]{202122} 4} &
         {\color[HTML]{202122} 4} &
         {\color[HTML]{202122} 5} &
         {\color[HTML]{202122} 7} &
         {\color[HTML]{202122} 2} &
         {\color[HTML]{202122} 4} &
         {\color[HTML]{202122} 6} \\

\ifisarxiv
\rowcolor[HTML]{F8F9FA} 
\multirow{-3}{*}{\cellcolor[HTML]{F8F9FA}{\color[HTML]{202122} -1xPrepend AMS}} &

         {\color[HTML]{202122} Catchment} &
         \multicolumn{4}{c|}{\color[HTML]{202122} 87} &
         \multicolumn{4}{c|}{\color[HTML]{202122} 8} &
         \multicolumn{4}{c}{\color[HTML]{202122} 5} \\ \hline

\rowcolor[HTML]{F8F9FA} 
\cellcolor[HTML]{F8F9FA}{\color[HTML]{202122} } &
          {\color[HTML]{202122} Day-1 load} &
          {\color[HTML]{202122} 58} &
          {\color[HTML]{202122} 65} &
          {\color[HTML]{202122} 63} &
          {\color[HTML]{202122} 69} &
          {\color[HTML]{202122} 33} &
          {\color[HTML]{202122} 30} &
          {\color[HTML]{202122} 31} &
          {\color[HTML]{202122} 24} &
          {\color[HTML]{202122} 9} &
          {\color[HTML]{202122} 5} &
          {\color[HTML]{202122} 6} &
          {\color[HTML]{202122} 7}  \\

\rowcolor[HTML]{F8F9FA} 
\cellcolor[HTML]{F8F9FA}{\color[HTML]{202122} } &
         {\color[HTML]{202122} Day-2 load} &
         {\color[HTML]{202122} 54} &
         {\color[HTML]{202122} 60} &
         {\color[HTML]{202122} 62} &
         {\color[HTML]{202122} 60} &
         {\color[HTML]{202122} 37} &
         {\color[HTML]{202122} 35} &
         {\color[HTML]{202122} 32} &
         {\color[HTML]{202122} 32} &
         {\color[HTML]{202122} 9} &
         {\color[HTML]{202122} 5} &
         {\color[HTML]{202122} 6} &
         {\color[HTML]{202122} 8} \\

\rowcolor[HTML]{F8F9FA} 
\multirow{-3}{*}{\cellcolor[HTML]{F8F9FA}{\color[HTML]{202122} -1xPrepend BOS}} &
         {\color[HTML]{202122} Catchment} &
         \multicolumn{4}{c|}{\color[HTML]{202122} 42} &
         \multicolumn{4}{c|}{\color[HTML]{202122} 49} &
         \multicolumn{4}{c}{\color[HTML]{202122} 9} \\  \hline

\rowcolor[HTML]{F8F9FA} 
\cellcolor[HTML]{F8F9FA}{\color[HTML]{202122} } &
          {\color[HTML]{202122} Day-1 load} &
          {\color[HTML]{202122} 45} &
          {\color[HTML]{202122} 51} &
          {\color[HTML]{202122} 51} &
          {\color[HTML]{202122} 58} &
          {\color[HTML]{202122} 6} &
          {\color[HTML]{202122} 4} &
          {\color[HTML]{202122} 4} &
          {\color[HTML]{202122} 4} &
          {\color[HTML]{202122} 49} &
          {\color[HTML]{202122} 45} &
          {\color[HTML]{202122} 45} &
          {\color[HTML]{202122} 38}  \\

\rowcolor[HTML]{F8F9FA} 
\cellcolor[HTML]{F8F9FA}{\color[HTML]{202122} } &
         {\color[HTML]{202122} Day-2 load} &
         {\color[HTML]{202122} 45} &
         {\color[HTML]{202122} 48} &
         {\color[HTML]{202122} 48} &
         {\color[HTML]{202122} 52} &
         {\color[HTML]{202122} 5} &
         {\color[HTML]{202122} 4} &
         {\color[HTML]{202122} 4} &
         {\color[HTML]{202122} 5} &
         {\color[HTML]{202122} 50} &
         {\color[HTML]{202122} 47} &
         {\color[HTML]{202122} 48} &
         {\color[HTML]{202122} 43} \\
          
\rowcolor[HTML]{F8F9FA} 
\multirow{-3}{*}{\cellcolor[HTML]{F8F9FA}{\color[HTML]{202122} -1xPrepend CNF}} &
         {\color[HTML]{202122} Catchment} &
         \multicolumn{4}{c|}{\color[HTML]{202122} 42} &
         \multicolumn{4}{c|}{\color[HTML]{202122} 9} &
         \multicolumn{4}{c}{\color[HTML]{202122} 49} \\ \hline

\rowcolor[HTML]{F8F9FA} 
\cellcolor[HTML]{F8F9FA}{\color[HTML]{202122} } &
          {\color[HTML]{202122} Day-1 load} &
          {\color[HTML]{202122} 41} &
          {\color[HTML]{202122} 57} &
          {\color[HTML]{202122} 55} &
          {\color[HTML]{202122} 55} &
          {\color[HTML]{202122} 22} &
          {\color[HTML]{202122} 22} &
          {\color[HTML]{202122} 23} &
          {\color[HTML]{202122} 23} &
          {\color[HTML]{202122} 31} &
          {\color[HTML]{202122} 21} &
          {\color[HTML]{202122} 22} &
          {\color[HTML]{202122} 22}  \\

\rowcolor[HTML]{F8F9FA} 
\cellcolor[HTML]{F8F9FA}{\color[HTML]{202122} } &
         {\color[HTML]{202122} Day-2 load} &
         {\color[HTML]{202122} 48} &
         {\color[HTML]{202122} 59} &
         {\color[HTML]{202122} 57} &
         {\color[HTML]{202122} 48} &
         {\color[HTML]{202122} 21} &
         {\color[HTML]{202122} 18} &
         {\color[HTML]{202122} 21} &
         {\color[HTML]{202122} 26} &
         {\color[HTML]{202122} 41} &
         {\color[HTML]{202122} 23} &
         {\color[HTML]{202122} 22} &
         {\color[HTML]{202122} 26} \\        
\fi          
          
\rowcolor[HTML]{F8F9FA} 
\multirow{-3}{*}{\cellcolor[HTML]{F8F9FA}{\color[HTML]{202122} Transit-1}} &
         {\color[HTML]{202122} Catchment} &
         \multicolumn{4}{c|}{\color[HTML]{202122} 38} &
         \multicolumn{4}{c|}{\color[HTML]{202122} 24} &
         \multicolumn{4}{c}{\color[HTML]{202122} 38} \\ \hline
\ifisarxiv
\rowcolor[HTML]{F8F9FA} 
\cellcolor[HTML]{F8F9FA}{\color[HTML]{202122} } &
          {\color[HTML]{202122} Day-1 load} &
          {\color[HTML]{202122} 64} &
          {\color[HTML]{202122} 72} &
          {\color[HTML]{202122} 73} &
          {\color[HTML]{202122} 75} &
          {\color[HTML]{202122} 13} &
          {\color[HTML]{202122} 10} &
          {\color[HTML]{202122} 10} &
          {\color[HTML]{202122} 9} &
          {\color[HTML]{202122} 23} &
          {\color[HTML]{202122} 18} &
          {\color[HTML]{202122} 17} &
          {\color[HTML]{202122} 16}  \\

\rowcolor[HTML]{F8F9FA} 
\cellcolor[HTML]{F8F9FA}{\color[HTML]{202122} } &
         {\color[HTML]{202122} Day-2 load} &
         {\color[HTML]{202122} 64} &
         {\color[HTML]{202122} 72} &
         {\color[HTML]{202122} 73} &
         {\color[HTML]{202122} 70} &
         {\color[HTML]{202122} 12} &
         {\color[HTML]{202122} 11} &
         {\color[HTML]{202122} 9} &
         {\color[HTML]{202122} 11} &
         {\color[HTML]{202122} 23} &
         {\color[HTML]{202122} 18} &
         {\color[HTML]{202122} 18} &
         {\color[HTML]{202122} 19}  \\
          
\rowcolor[HTML]{F8F9FA} 
\multirow{-3}{*}{\cellcolor[HTML]{F8F9FA}{\color[HTML]{202122} Transit-2}} &
         {\color[HTML]{202122} Catchment} &
         \multicolumn{4}{c|}{\color[HTML]{202122} 53} &
         \multicolumn{4}{c|}{\color[HTML]{202122} 19} &
         \multicolumn{4}{c}{\color[HTML]{202122} 28} \\ \hline
\fi         
\end{tabular}}
         \caption{Load distribution with \peering catchment and \broot load. Catchment: 2020-02-24, Load: 2020-02-25 and 2020-02-26 (only showing selected policies). Catchment distribution remains similar over the course of the day showing by a single value.}
       \label{tab:load-distribution-hours}
      \vspace{-0.18in}
\end{table*}
        
\ifisarxiv

\begin{figure*}
        \begin{subfigure}[t]{0.33\textwidth}
                \centering
                \includegraphics[width=.98\linewidth]{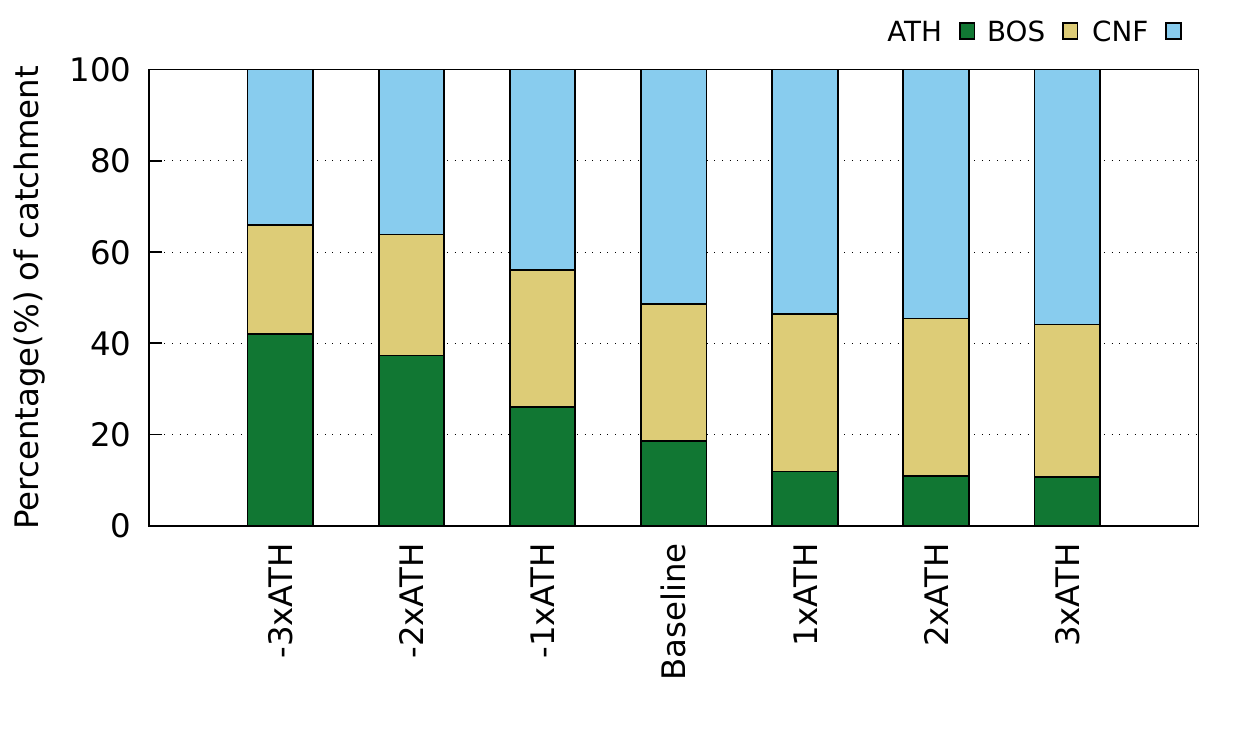}
                \caption{ATH site.}
                \label{fig:ath-only-site-2020-05-30}
        \end{subfigure}
	\begin{subfigure}[t]{0.33\textwidth}
                \centering 
                \includegraphics[width=.98\linewidth]{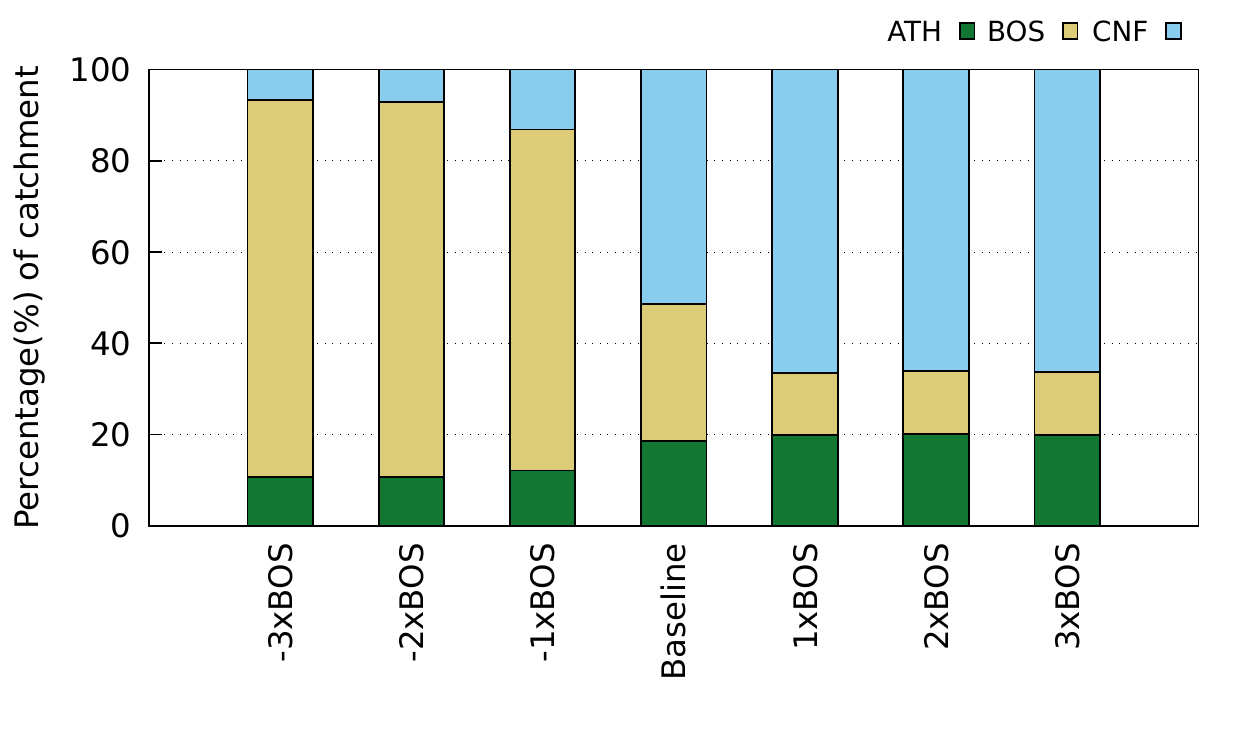}
                \caption{BOS site.}
                \label{fig:bos-only-site-2020-05-30}
        \end{subfigure}
        \begin{subfigure}[t]{0.33\textwidth}
                \centering
                \includegraphics[width=.98\linewidth]{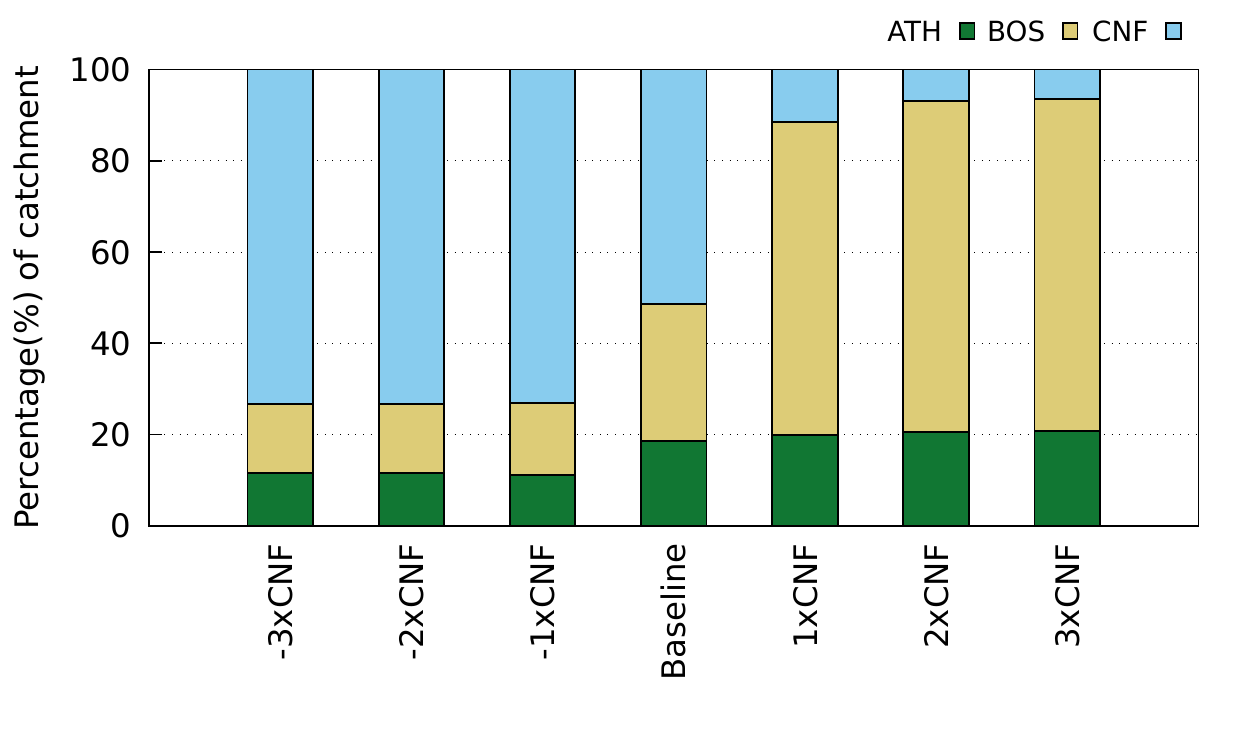}
                \caption{CNF site.}
                \label{fig:cnf-only-site-2020-05-30}
        \end{subfigure}
       \caption{\peering: Impact of path prepending in catchment distribution with ATH, BOS and CNF sites on 2020-05-30.}
       \label{fig:ath-bos-cnf-only-site-2020-05-30}
      \vspace{-0.15in}
\end{figure*}

\begin{figure*}
        \begin{subfigure}[t]{0.33\textwidth}
                \centering
                \includegraphics[width=.98\linewidth]{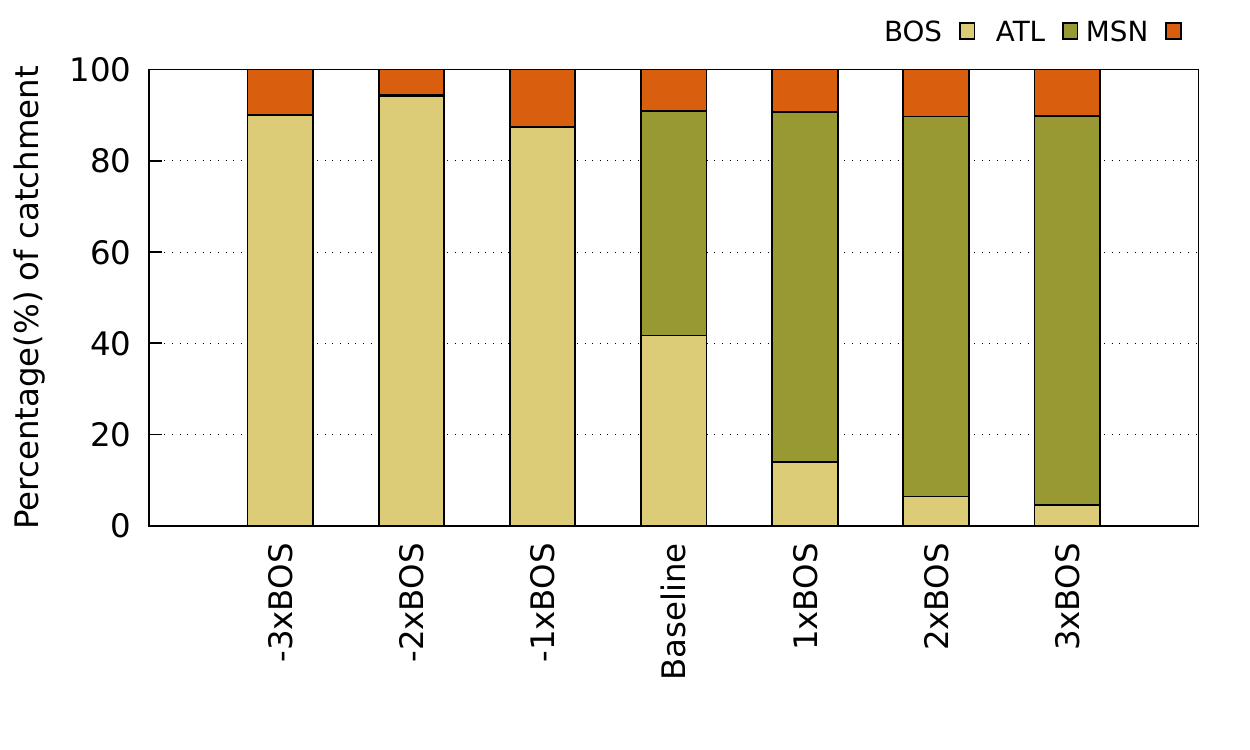}
                \caption{BOS site.}
                \label{fig:bos-only-site-2020-05-29}
        \end{subfigure}
	\begin{subfigure}[t]{0.33\textwidth}
                \centering 
                \includegraphics[width=.98\linewidth]{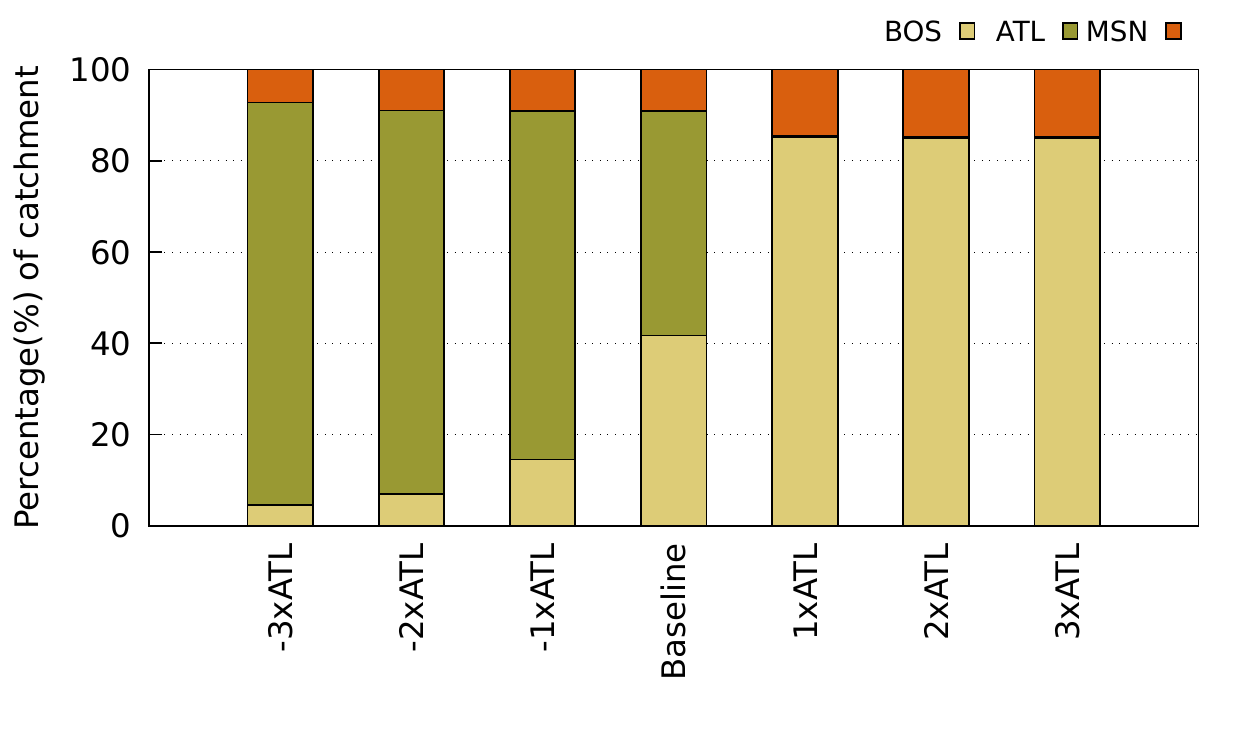}
                \caption{ATL site.}
                \label{fig:atl-only-site-2020-05-29}
        \end{subfigure}
        \begin{subfigure}[t]{0.33\textwidth}
                \centering
                \includegraphics[width=.98\linewidth]{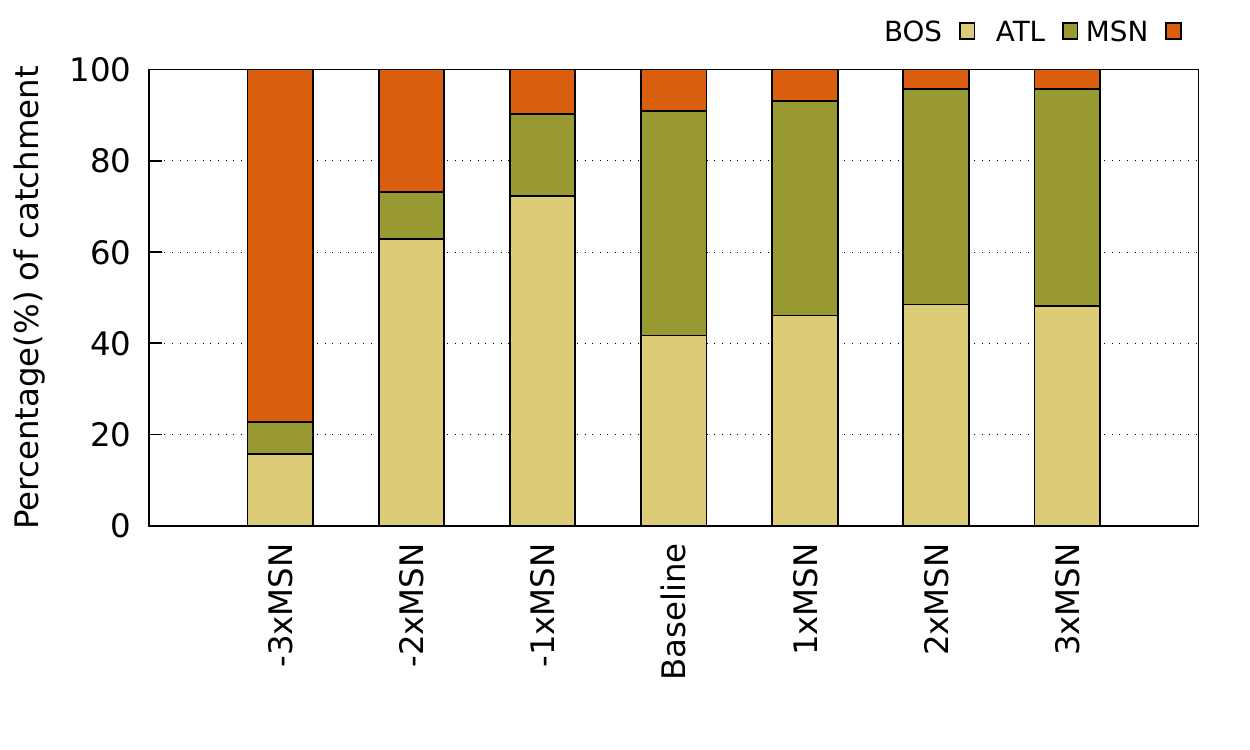}
                \caption{MSN site.}
                \label{fig:msn-only-site-2020-05-29}
        \end{subfigure}
       \caption{\peering: Impact of path prepending in catchment distribution with BOS, ATL and MSN sites on 2020-05-29.}
       \label{fig:bos-atl-msn-only-site-2020-05-29}
	\vspace{-0.15in}
\end{figure*}

\begin{figure}
\centering
\includegraphics[width=1.0\linewidth]{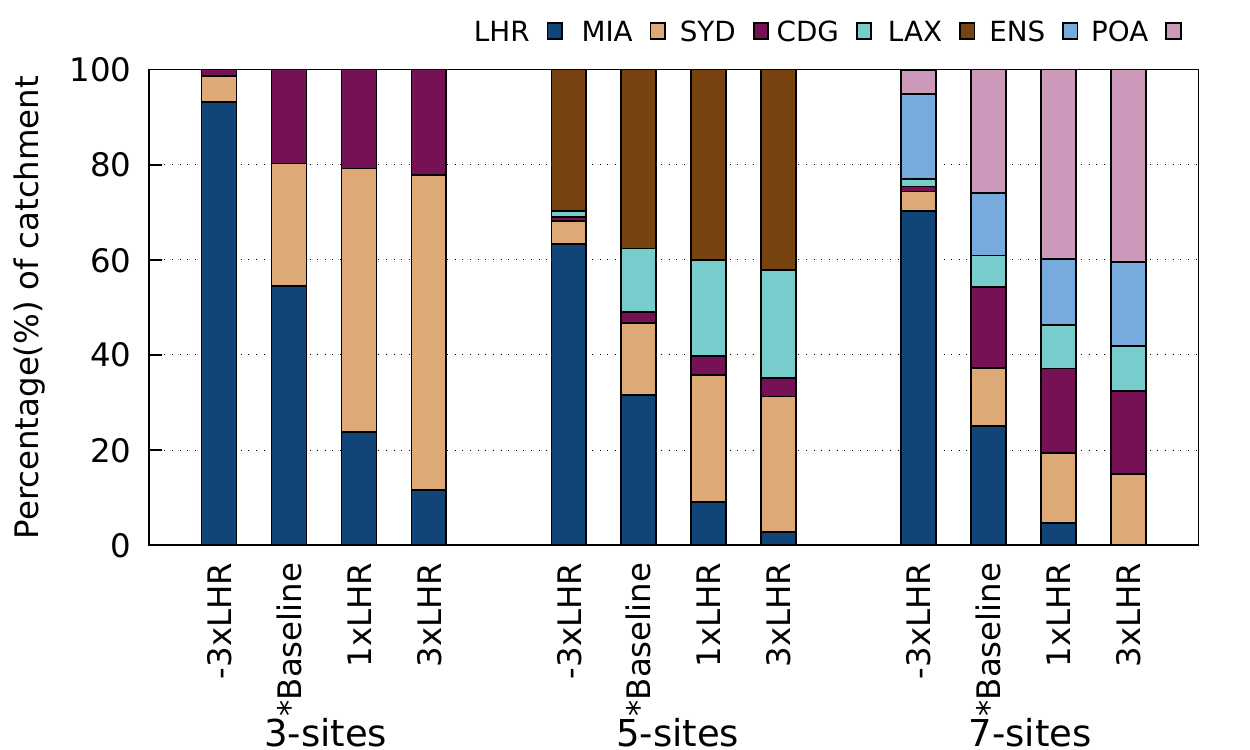}
\vspace{-0.25in}
    \caption{\tangled: Impacts of changing the number of anycast sites.}
\label{fig:number_sites_tangled}
\vspace{-0.15in}
\end{figure}

\begin{figure*}[ht]
\centering
\begin{tabular}{cccc}

          \begin{subfigure}[t]{0.33\textwidth}
                \centering
                \includegraphics[width=.98\linewidth]{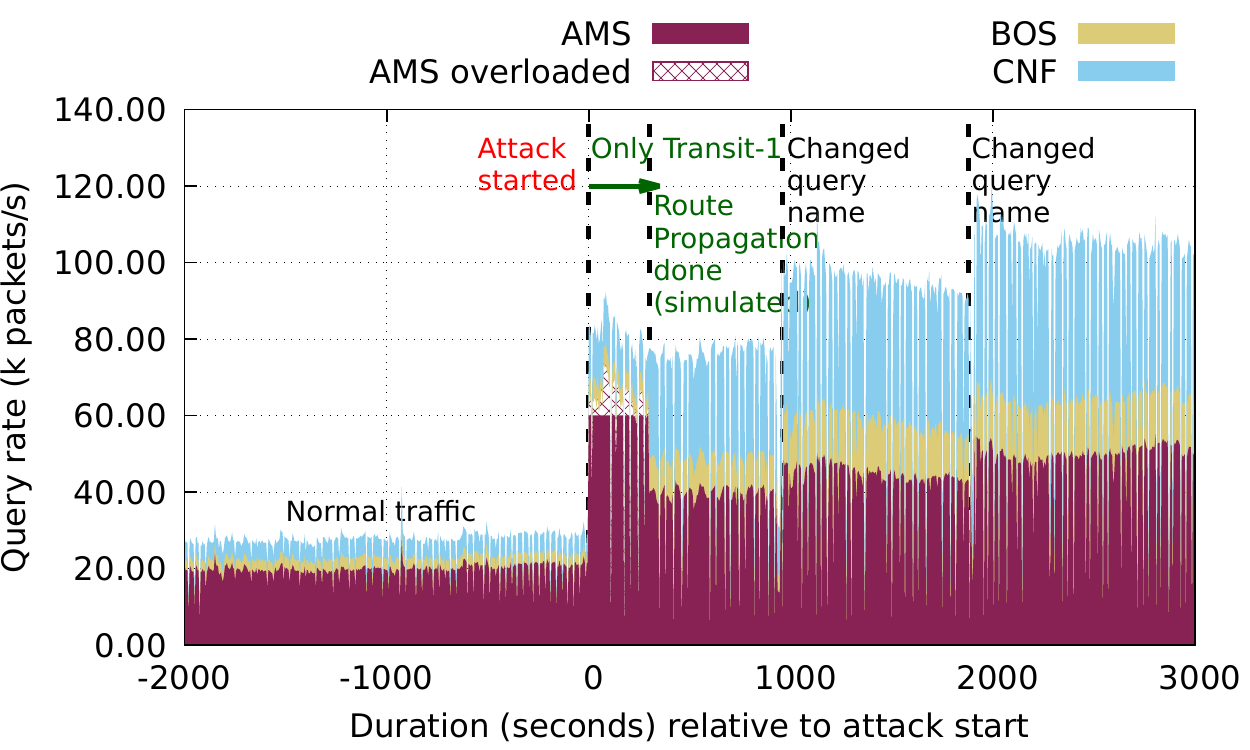}
                \caption{A 2017 event at \broot mitigated using community strings.}
                \label{fig:transit-1-20170221}
          \end{subfigure} &

          \begin{subfigure}[t]{0.33\textwidth}
            \centering
                \includegraphics[width=.93\linewidth]{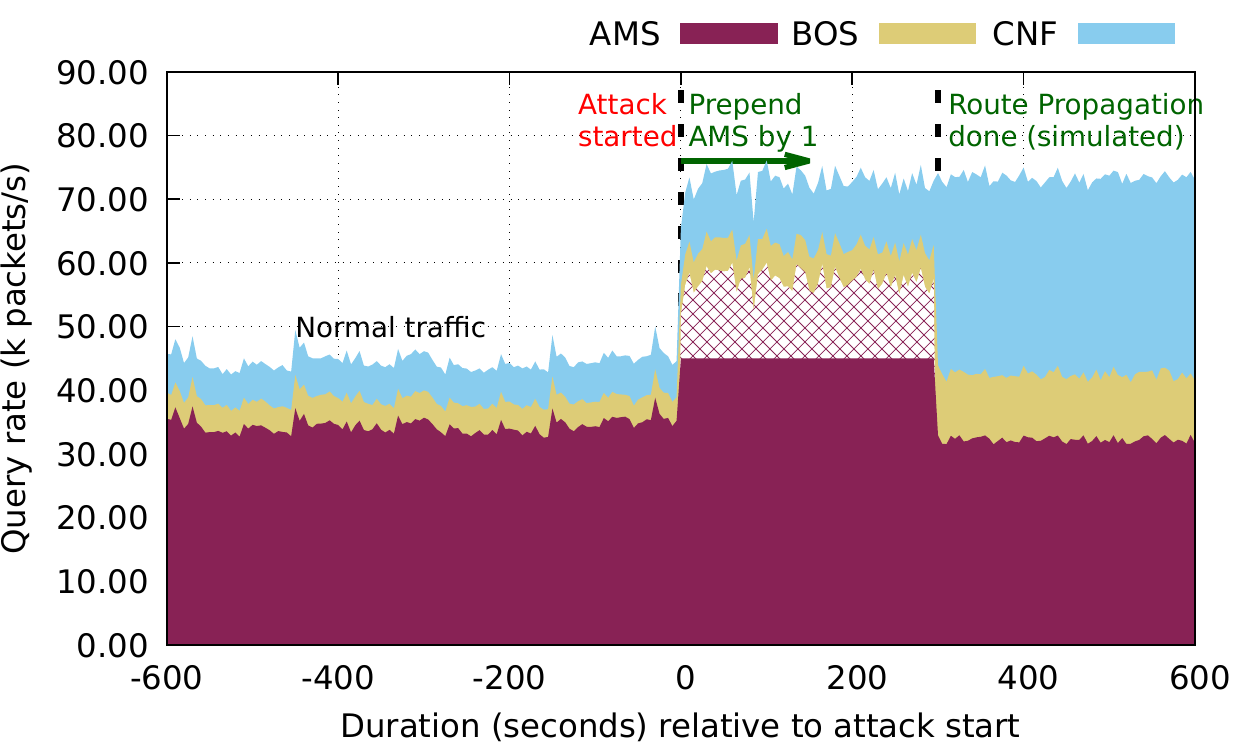}
                \caption{A 2020 event at \broot defended using positive prepending.}
                \label{fig:prepend-1-20200214}
          \end{subfigure} &

          \begin{subfigure}[t]{0.33\textwidth}
                \centering
                \includegraphics[width=.98\linewidth]{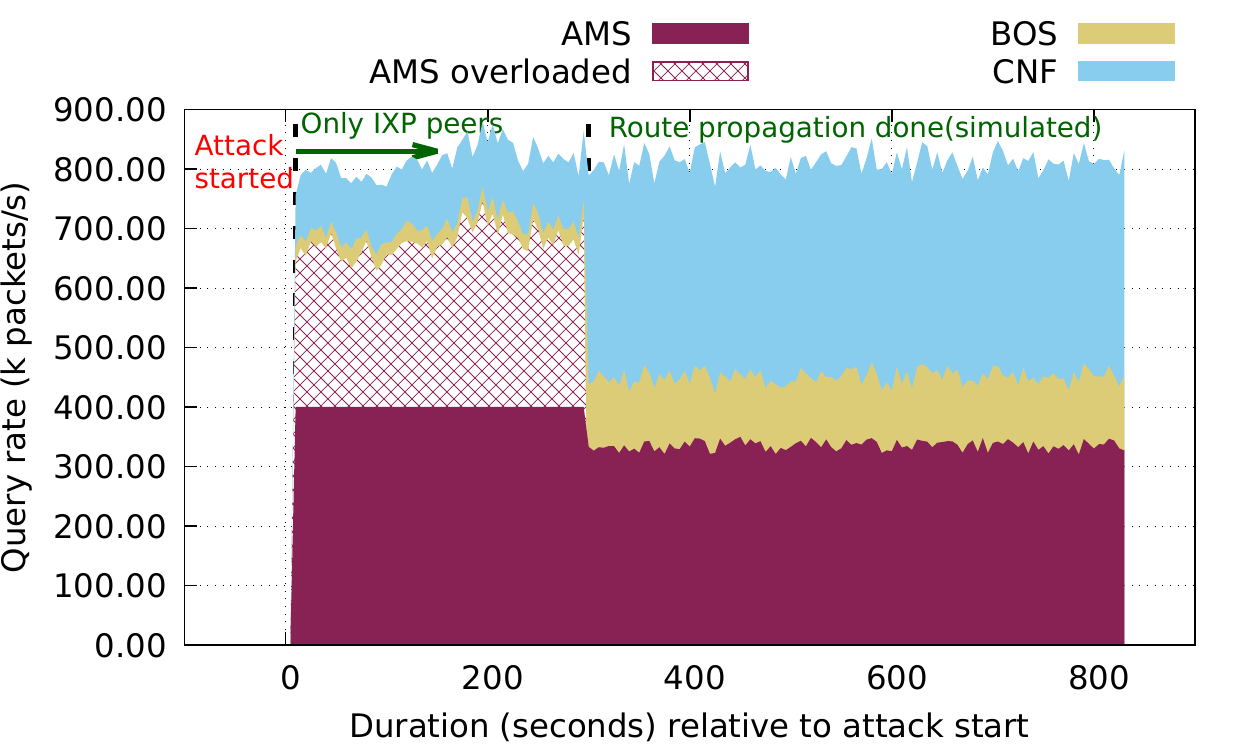}
                \caption{A 2021 event captured at \DutchScrubbingCenter mitigated using community strings.}
                \label{fig:ixp-20210822}
          \end{subfigure}
          
\end{tabular}
\begin{tabular}{cccc}
   \begin{subfigure}[t]{0.33\textwidth}
     \centering
                \includegraphics[width=.90\linewidth]{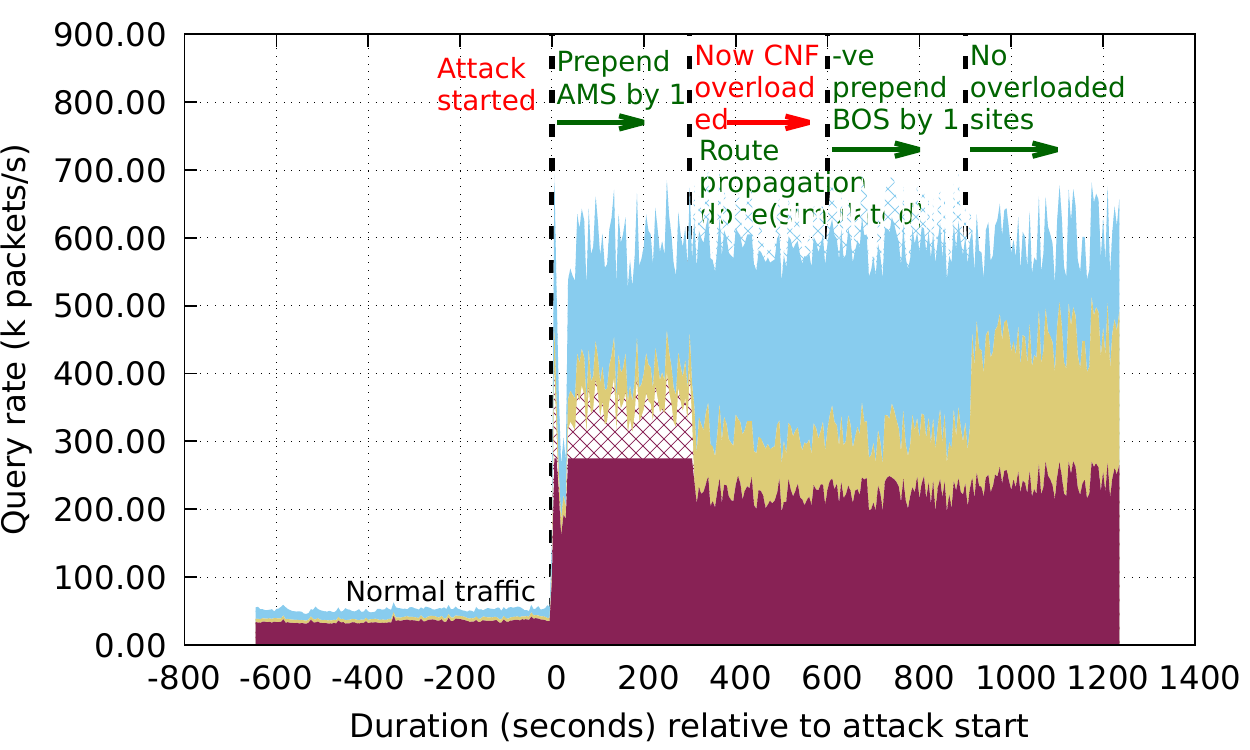}
                \caption{A 2021 event at \broot defended using negative prepending.}
                \label{fig:ams1-bos-1-20210528}
    \end{subfigure} &
    \begin{subfigure}[t]{0.33\textwidth}
      \centering

                \includegraphics[width=.98\linewidth]{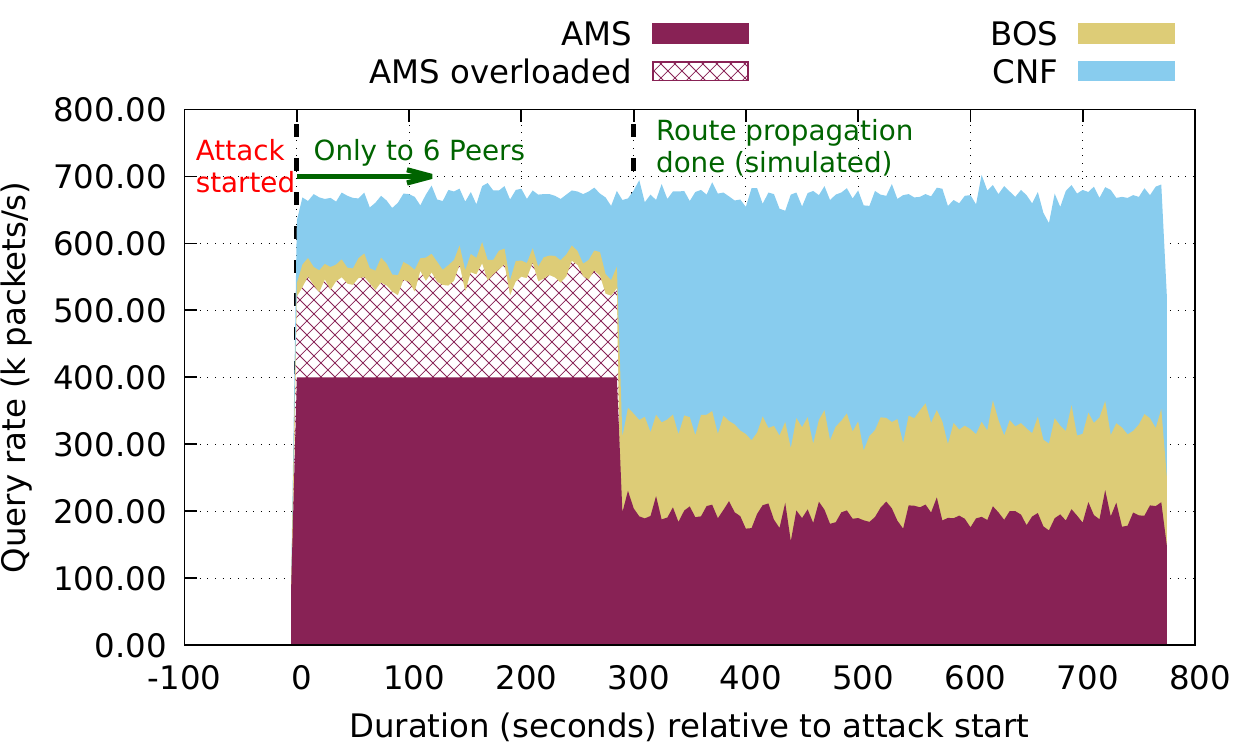}
                \caption{A 2021 event at \DutchScrubbingCenter mitigated using community strings.}
                \label{fig:ddos-6-peers}
     \end{subfigure}
\end{tabular}
\caption{Different attacks with various responses (extended).}
\label{fig:Name}
\vspace{-0.15in}
\end{figure*}

\else

\begin{figure*}[ht]
       \begin{subfigure}[t]{0.33\textwidth}
            \centering
                \includegraphics[width=.98\linewidth]{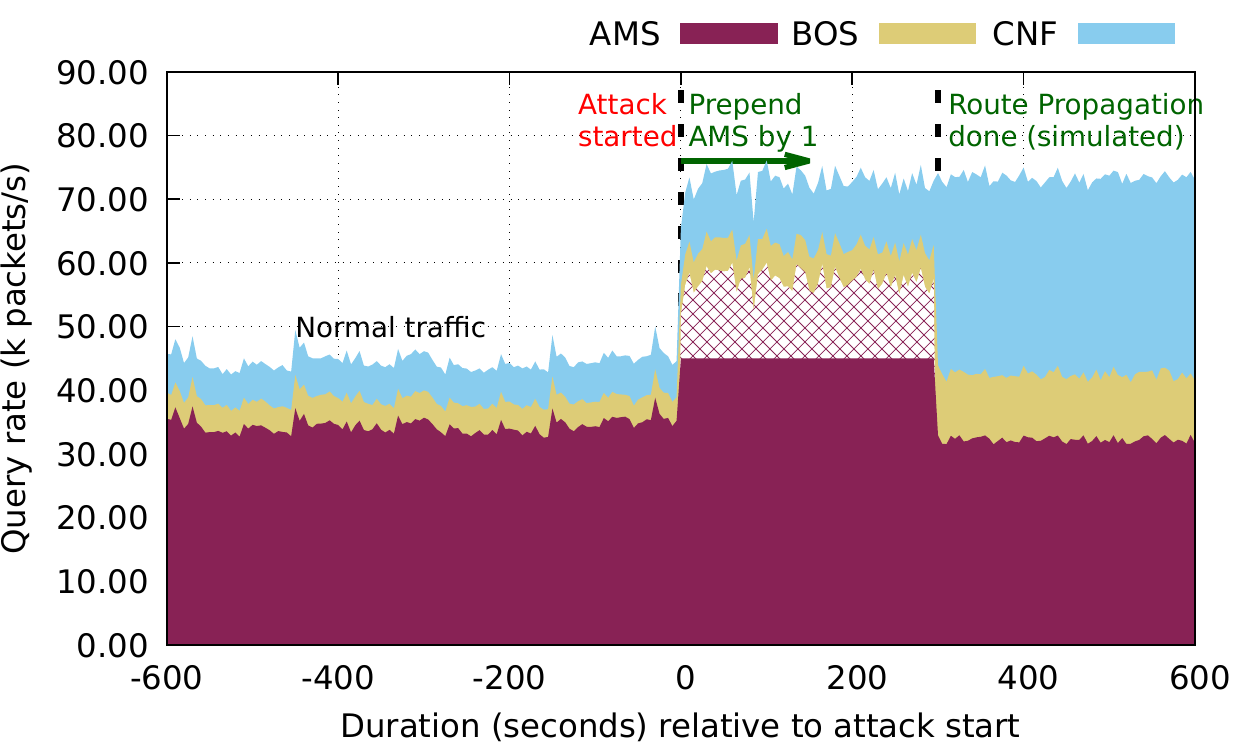}
                \caption{A 2020 event at \broot defended using positive prepending.}
                \label{fig:prepend-1-20200214}
        \end{subfigure} 
	\begin{subfigure}[t]{0.33\textwidth}
            \centering
                \includegraphics[width=.98\linewidth]{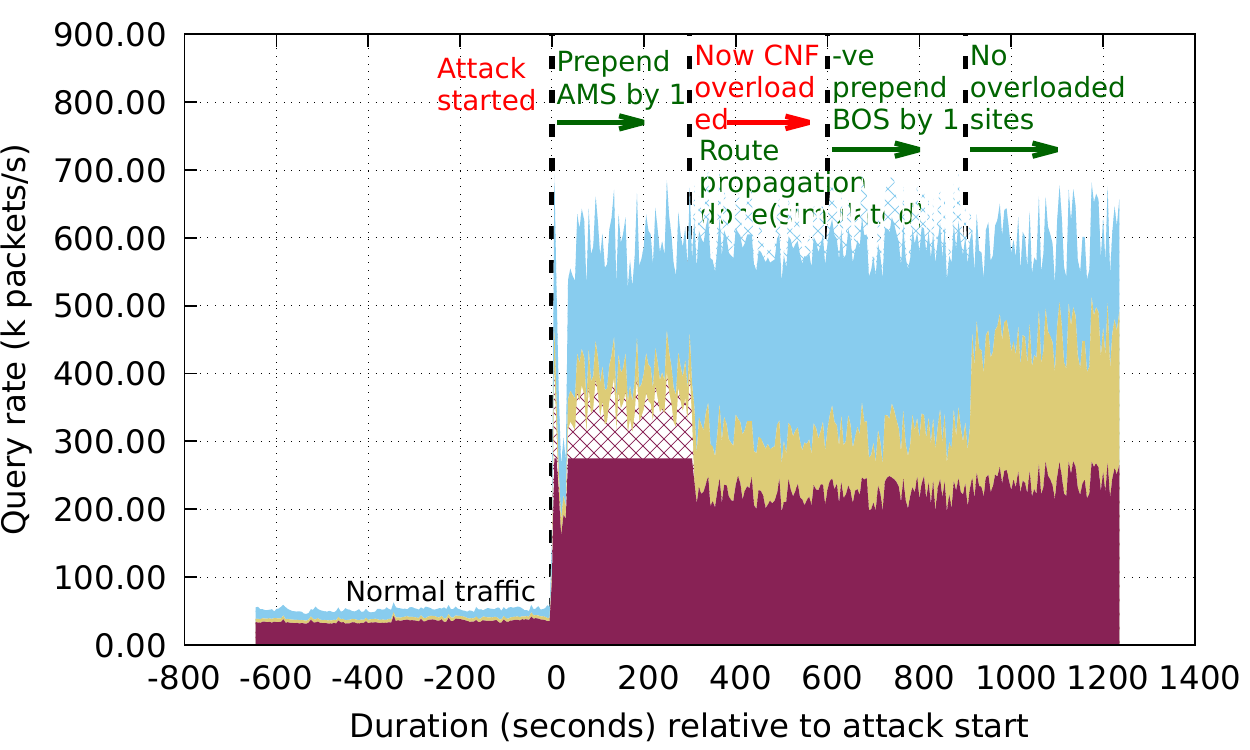}
                \caption{A 2021 event at \broot defended using negative prepending.}
                \label{fig:ams1-bos-1-20210528}
        \end{subfigure}
        \begin{subfigure}[t]{0.33\textwidth}
      \centering

                \includegraphics[width=.98\linewidth]{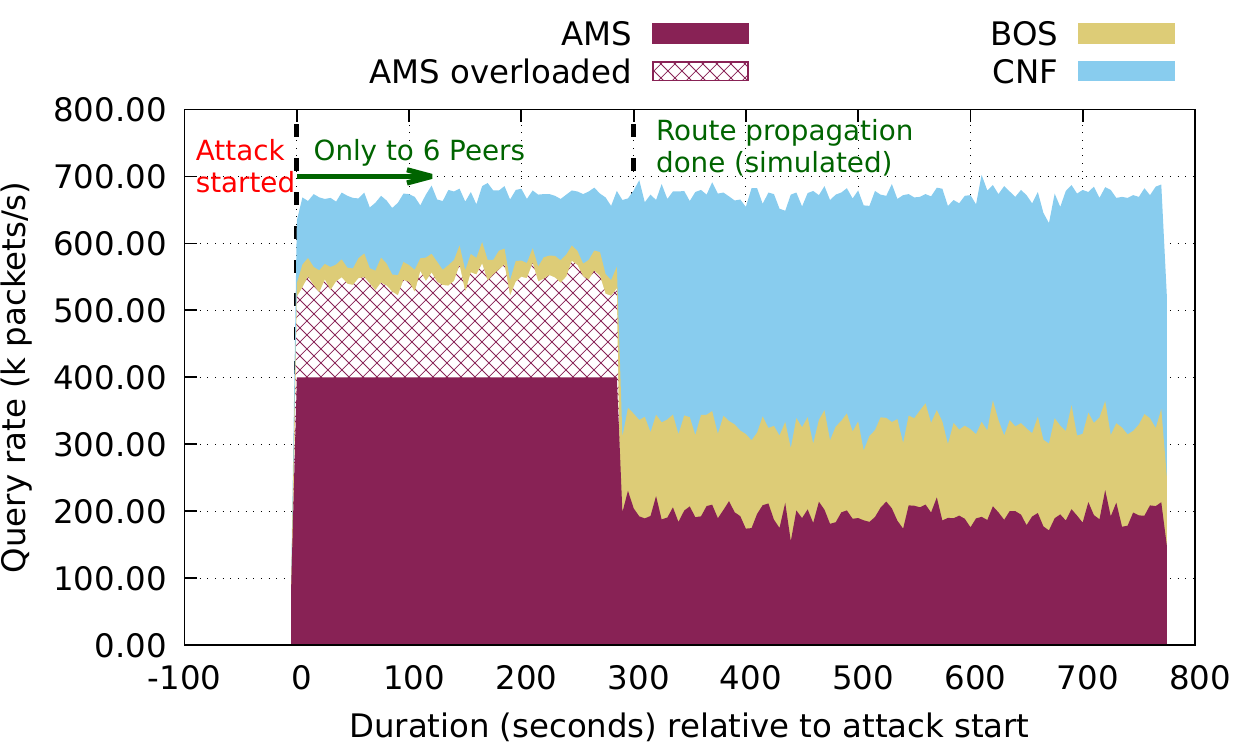}
                \caption{A 2021 event at \DutchScrubbingCenter mitigated using community strings.}
                \label{fig:ddos-6-peers}
     \end{subfigure}
        \caption{Different attacks with various responses (extended).}
        \vspace{-0.18in}
       \label{fig:Name}
\end{figure*}

\fi

\section{Detailed Attack Size Estimation}
\label{sec:testbed-experiment-details}

We validate our attack estimation with both testbed experiments and
  real-world events.

\subsection{Testbed Experiment}
\label{sec:testbed-experiment}

We validate our model with experiments in a testbed
  (DETER~\cite{deter06})
  where we can control all factors,
  where actual offered load
  is estimated and topology is fixed.
  
\ifisarxiv
We consider a simple topology (\autoref{fig:topology})
  where two access links from R1 and R2 towards R3 has 
  a capacity of 100Mb/s each.
We assume the link from the service router (R3) to the servers
  has 1\,Gb/s capacity, so the internal network is never a bottleneck.

Here we use a slightly unequal legitimate traffic--40\,Mb/s from R1 to R3 and
  60\,Mb/s from R2 to R3.
As part of legitimate traffic
  we generate known-good traffic
  that we use for estimation (\autoref{sec:estimating}):
  2\,Mb/s on R1-R3 link and 3\,Mb/s on R2-R3.

The attack consists of 250\,Mb/s of traffic
  following the distribution of 100\,Mb/s on R1-R3 and 150\,Mb/s on R2-R3.
Offered load is therefore  
   140\,Mb/s (1.4$\times$ link capacity) and 210\,Mb/s (2.1$\times$ link capacity) on the two links,
   for a total offered load of 350\,Mb/s (29.2\,k queries/second).

We show our estimation works well
  with the testbed experiment in \autoref{tab:events-estimation}.
Our estimation needs to know the access fraction or $\alpha$ (how much traffic
  arrives at the system during an attack).
We observe the rate of known-good traffic at the server
  to find the $\alpha$.
\autoref{tab:events-estimation} shows the expected typical known good traffic (``normal''),
  the observed rate under attack (``observed'') and the computed $\alpha$.

Using the observed offered load (``observed'') of 16.3\,kq/s at the server, and
  $\alpha$ of 0.49, we estimate 33.2\,kq/s offered load (``estimated''),
  which is close to the actual reported 29.2\,kq/s (``reported'').
Normal offered load (``normal'') is 8.5\,kq/s,
  which is lower than what we can observe during an attack.
Our observation is significantly lower than the reported or estimated rate,
  which tells us the importance of estimation.

We carried out other tested experiments, varying topology,
  traffic ratios, and distribution of attackers.
In general, we find our approach works well
  unless attack traffic is highly unbalanced (one or few sources, not a distributed DoS). 
\else

We provide the details of our testbed experiment
in the extended version of this paper~\cite{rizvi2020anycast}.

\fi
 
\subsection{Case Studies: 2016-06-25 Event}
\label{sec:case-study-2016}

We showed real-world case studies in \autoref{sec:estimation_case_studies}.
Here we show that our approach works for
  another event from 2016-06-25 (\autoref{fig:est-2016}).
We observe our estimation (varying orange line)
  is close to the reported line (dashed purple line).
We can also see that our observation is
  only a tiny fraction of the true offered load (bottom blue line).

Both these results show
  the effectiveness of our approach with
  both testbeds and real-world events.

\section{BGP Poisoning Granularity}

\label{sec:granularity-poison-detail}


With poisoning coverage limited by filters (\autoref{sec:coverage-poisoning}),
  we next examine what granularity control it provides.
We expect to see limited range since we cannot poison \tier ASes,
  and small ASes carry little traffic.

We test path poisoning in both \peering and \tangled
  using three sites from each testbed.
As expected, we observe the same traffic distribution
  when we poison any \tier AS---30-35\% load at AMS (\peering in \autoref{fig:poison-peering-2021-04-15}) and 1-3\% load at MIA (\tangled in \autoref{fig:poison-tangled-2021-04-15}).

When we poison a non-\tier AS that is more than one hop away,
  we observe a small change in the traffic distribution.
In \peering, we can see that poisoning AS57866
  reduces a small fraction of traffic from AMS (\autoref{fig:poison-peering-2021-04-15}).
We observe a similar outcome
  in \tangled (\autoref{fig:poison-tangled-2021-04-15}).

Our results prove that poisoning \tier ASes
  is limited by the filters, and poisoning non-\tier ASes
  that are multi-hops away
  can change only a small fraction of traffic.

Poisoning an immediate upstream is equivalent to
  not announcing to them, so we do not consider that case here.

\section{Catchment Stability}
\label{sec:appendix_stability}

\reviewfix{SS22-F2, SS22-A7, SS22-C4, SS22-E4}

Our insight is that we can use a playbook
  for several days or weeks since the catchment
  remains stable over the time (\autoref{sec:playbook_stability}).
To test this we use one month of \broot catchment mapping
  with test and production prefixes.
We observe the stability in \broot catchment.

From \autoref{fig:broot-stability}, we can see that
  the catchment remains stable over time.
In two weeks, only 0.35\% prefixes,
  and in one month, only 0.65\% prefixes
  changed their catchment
  when we compare the catchment with day 1
  considering $\sim$2 million prefixes.
This shows only a tiny fraction of prefixes
  changes the catchment even after a month
  irrespective of the changes made by the ASes.
Hence, building the playbook once every week/month
  should be sufficient.

We also make the catchment mapping at different times of the day.
We found catchment distribution remains similar
  at different times of the day.

\section{Load Distribution}
	\label{sec:appendix_playbook_load}

        \reviewfix{SS22-F2, SS22-A7}

Our playbook with catchment (\autoref{sec:playbook}) distribution
  gives an adequate prediction of traffic distribution
  which we successfully apply in \autoref{sec:fight_ddos}.
Since services care about load, we want to see
  how the load is distributed in different routing changes.
An operator can simply make the load playbook
  based on the already computed catchment mapping
  without making additional BGP announcements.

In \autoref{tab:load-distribution-hours},
  we can see different routing changes and
  their impacts over load distribution in different
  times of the day.
Load changes over the day---fewer load at 00 GMT
  in AMS site since most Europe sleeps at that time.
BOS and CNF receive more load
  at 00 GMT as that is a busy hour for these two regions.
We can also observe that some prefixes contribute
  more load due to the difference in number of clients
  behind each prefix.
For this reason, BOS prefixes (mostly North American prefixes)
  contribute less load compared to the prefixes at other two sites.
We can also see that load remains stable at the same time
  of different days (varies within 5\% most of the time).

We can also see that the relative catchment distribution follows
  the load distribution,
  however, it is not exactly the same.
Decisions will be even better when
  an operator considers different load playbooks
  at different times of the day.
Building multiple load playbooks is simple
  since we can just use the same catchment mapping (catchment mapping remains stable (\autoref{sec:appendix_stability})).

\ifisarxiv  
\section{BGP in Other Anycast Set-up }
	\label{sec:other_bgp}

In this section, we present the complementary results from the
experiments performed in both testbeds. Here we present data used to
generalize our findings.

\subsection{\peering: A Small Site in Europe}
\label{sec:small-eu}

AMS in \peering is well-connected with two transits,
  and several IXP peers. 
Next, instead of AMS, we take ATH in Europe which is connected through
  a research network in Greece.
Our goal is to see whether the findings from \autoref{sec:exploit_catchment}
  are still valid in this anycast setup.
Like the previous setup, we also take BOS and CNF.

\autoref{fig:ath-bos-cnf-only-site-2020-05-30} shows the catchment distribution.
In the baseline case, as ATH is not connected like AMS,
  it gets only 20\% traffic.
CNF serves almost 50\% traffic in this anycast setup.
So, the traffic distribution is still skewed in this setup.
Even if both AMS and ATH are in Europe,
  we see a different catchment control which
  indicates the importance of site's connectivity.

Prepending works similarly in this setup. 
BOS and CNF can cut most of their traffic after first prepend.
However, ATH can only shift 7\% traffic after first prepend,
  and for more prepends it does not show any effect.
In all these sites, we can also see that some blocks
  are always ``stuck'' to a particular site.
Using negative prepending, we can push most of
  the traffic to BOS and CNF\@.
However, we can only push 40\% traffic to ATH site.

\subsection{\peering: Sites in Nearby Location}
\label{sec:nearby-sites}

Next, we take three sites that are in nearby location
  and have similar connectivity.
We select sites from Boston (BOS), Atlanta (ATL) and Wisconsin (MSN)
  in \peering.
All these sites are located within the eastern half
  of the U.S., and they are connected through education network---BOS
  with Northeastern University, ATL with Georgia Institute of Technology, and
  MSN with University of Wisconsin - Madison.

When the sites are located in nearby geo-location, and
  connected by similar network,
  path prepending can result in an ``all or none'' outcome.
When we prepend from ATL or BOS,
  most traffic goes away from these sites.
Prepending one time from ATL leaves no traffic in ATL,
  and prepending one time from BOS cuts traffic from 42\% to 14\%.
Even with negative prepending,
  BOS can get over 90\% traffic,
  and ATL can get nearly 90\% traffic.
So, BOS and ATL can cut or gain almost all the traffic
  with positive and negative prepending.

MSN receives a small fraction of traffic in the baseline, and
  some blocks are always ``stuck'' at MSN\@.
With two negative prepends,
  MSN receives only 27\% catchment,
  however, with the third negative prepend,
  MSN receives almost 80\% catchment.
This slow and sudden increase in the catchment
  suggests us why we need a BGP ``playbook'' for anycast setup.

\subsection{More Sites in \tangled}
        \label{sec:more-tangled}
        
We want to confirm that increasing the
  number of sites in \tangled shows the similar
  results that we get in \autoref{sec:effects_no_sites}.

Like \peering, we take 3, 5 and 7 sites in \tangled testbed (\autoref{fig:number_sites_tangled}).
As the overall capacity increases with more number of sites,
  in \tangled also we can see the baseline traffic is reduced
  in each site as the traffic spreads out.
LHR gets 55\%, 31\% and 25\% traffic when there are 3, 5 and 7 sites respectively.

With more sites, as there are more capacity in other sites,
  one site can cut almost all of its traffic.
For example,
  LHR can cut all of its traffic when there 7 sites
  which is not possible when there are 3 or 5 sites.

\PostSubmission{Leandro, I added this part here. ---asmrizvi 2020-06-02}
We can see a ``shadowing'' instance when we have a 7-site testbed.
After adding ENS and POA, we can see that
  all traffic from LAX site disappears (\autoref{fig:number_sites_tangled}).
We believe Cogent traffic now shifts from LAX to POA,
  and academic traffic shifts from LAX to ENS\@.
We explain this issue in \autoref{sec:effects_no_sites} when
  IAD shadows LAX.

Like \peering, adding more sites can create new
  options where the shifted traffic goes.
For example, with 3 sites, LHR traffic goes to MIA
  when we make prepending.
But when we have 7 sites, a significant amount of LHR traffic goes to POA---POA
  traffic increases from 26\% to 40\% when we prepend from LHR\@.
Hence, it is necessary to keep a ``playbook'' to see
  the traffic distribution after a BGP change.
\PostSubmission{ for John: what is the difference beetwen ``playmate'' and
``shadow'' ?
case figure 13: LAX traffic (5 sites) goes to (POA+ENS) ant 7 sites
LAX lose cogent traffic to POA and academic traffic to SURFNET/ENSCHEDE} 
\PostSubmission{Leandro, I realized we did not use the term playmate anywhere else. So, I am deleting it from here. ---asmrizvi 2020-06-02}
\PostSubmission {add graph about shadow example (WAS,IAD) Los Nettos case or CDGxLHR case}
\PostSubmission{johnh to Leandro: I'm not fond of the term ``platemate'' (it can have negative connotations in American English) ---johnh 2020-06-10}
  
\fi

\section{More Attacks And Mitigation}
	\label{sec:appendix_more_events}

        \reviewfix{SS22-F1, SS22-F3, SS22-A9, SS22-B4, SS22-B5, SS22-B8, SS22-C3, SS22-D3, SS22-E1}


We evaluate more attack events captured at \broot and
  at \DutchScrubbingCenter.
We follow the same methodology mentioned in \autoref{sec:fight_ddos}.
We use the same playbook built with AMS, BOS and CNF sites (\autoref{sec:exploit_catchment}) from \peering.

\ifisarxiv
\textbf{A polymorphic event at \broot:}
We observed a polymorphic event at \broot on 2017-02-21
  where the attackers used three
  different query names---RANDOM.phone.tianxintv.cn\textbackslash032, RANDOMclgc88.com\textbackslash032, and RANDOM.jiang.com\textbackslash032.
The total offered load at AMS site
  exceeds the capacity of 60k packets/s (striped area).

Our system announces only to Transit-1
  using community strings to mitigate this attack (\autoref{fig:transit-1-20170221}).
We can see that there is no striped area
  after the deployment of the new routing policy.
Also, when the attackers change pattern,
  our system does not need to make any routing changes.
This proves the applicability of TE approaches
  in polymorphic events.
\fi

\textbf{A 2020 volumetric attack at \broot:}
We observed an ephemeral volumetric event at \broot on 2020-02-14
  where the attackers used a single query name---peacecorps.gov.
This event lasted very briefly for 3 minutes.
In practice, no routing approach can work
  against such short-lived attacks due
  to the propagation delay of BGP\@.
We stretched the event with similar traffic rate
  so that we can see the impact if the attack continues for more time.

In this event also, AMS is overloaded
  with 60k packets/s when the assumed capacity is 40k packets/s (\autoref{fig:prepend-1-20200214}).
We prepend AMS by 1 so that the traffic shifts away from AMS\@.
After 300\,s, we can see no overloaded striped area in AMS\@.

These volumetric attacks are common at root servers.
Routing based approaches can defend against such attacks.

\ifisarxiv
\textbf{A DNS amplification attack:}
We evaluate another DNS amplification attack
  collected at \DutchScrubbingCenter on 2021-08-22.
In this event too, AMS site receives a
  huge traffic exceeding its capacity (large striped area).
Announcing only to IXP peers,
  our system can mitigate this attack event (\autoref{fig:ixp-20210822}).
This is another example where
  a community string helps us to mitigate the attack.
\fi

\textbf{A 2021 \broot event where our system iterates:}
We evaluate another event at \broot occurred on 2021-05-28.
In this event, the queries were IP fragmented (large packet size),
  and the common query name was pizzaseo.com (we stretched the event since it was short-lived).
When the attack started, our system finds AMS site overloaded (\autoref{fig:ams1-bos-1-20210528}).
Our system finds prepending from AMS
  is the best approach to reduce traffic from AMS\@.
However, after prepending AMS by 1, CNF site gets
  the most redirected traffic, and becomes overloaded.
Redirected attack sources prefer CNF over BOS.
When our system finds CNF site overloaded,
  it deploys an approach that will reduce traffic
  from CNF since it is now overloaded.
Our system deploys negative prepending to push more
  traffic towards BOS site.
After 900\,s, we can see there is no overloaded site.
This event shows how our system can gradually find out
  the best routing approach.

\textbf{Defending with community strings:}
We next consider an attack observed at
  \DutchScrubbingCenter on 2021-08-27.
This attack was a  volumetric
  DNS amplification.

In this attack, AMS is overloaded.
Consulting the playbook,
  we select a response using community strings
  to shift traffic,
  retaining six IXP peers at AMS,
  while dropping all other peers and transits.
The impact of this change is visible at 300\,s in \autoref{fig:ddos-6-peers},
  as the attack is successfully spread across all sites.

This example shows how different community strings
provide control over traffic distribution.
\ifisarxiv
\else
We show more events in the extended version of this paper~\cite{rizvi2020anycast}.
\fi

\end{appendices}

\end{document}
